\documentclass[aps,prd,reprint,nofootinbib,longbibliography]{revtex4-2}
\usepackage[utf8]{inputenc}
\usepackage{amsmath,amssymb,mathrsfs}
\usepackage{microtype}
\usepackage[hidelinks]{hyperref}
\makeatletter
\newenvironment{manuscriptwide}[1][22]{%
  \par\ignorespaces
  \setbox\widetext@top\vbox{\hb@xt@\hsize{\leaders\hrule\hfil\vrule\@height6\p@}}%
  \setbox\widetext@bot\hb@xt@\hsize{\vrule\@depth6\p@\leaders\hrule\hfil}%
  \onecolumngrid
  \vskip#1\p@
  \dimen@\ht\widetext@top\advance\dimen@\dp\widetext@top
  \cleaders\box\widetext@top\vskip\dimen@
  \vskip6\p@
  \prep@math@patch
}{%
  \par\vskip6\p@
  \setbox\widetext@bot\vbox{\hb@xt@\hsize{\hfil\box\widetext@bot}}%
  \dimen@\ht\widetext@bot\advance\dimen@\dp\widetext@bot
  \cleaders\box\widetext@bot\vskip\dimen@
  \vskip8.5\p@
  \twocolumngrid\global\@ignoretrue\@endpetrue
}
\makeatother
\newsavebox{\equationmeasure}
\newcommand{\eqanchor}[1]{\hypertarget{eq#1}{}}
\newcommand{\crosspdflink}[3]{\href{#1\##2}{#3}}
\hypersetup{pdftitle={Pinched Multi Affine Geometry and Confinement: Describing the Yang-Mills Mass Gap},pdfauthor={Shoshauna Gauvin},pdfsubject={Version 62; Pinched multi affine geometry, confinement and the Yang-Mills mass-gap problem}}
\begin{document}

\title{Pinched Multi Affine Geometry and Confinement:\\Describing the Yang-Mills Mass Gap}

\author{Shoshauna Gauvin}

\affiliation{Department of Physics and Astronomy, University of Waterloo, Waterloo, Ontario, Canada N2L 3G1}

\date[]{}

\begin{abstract}We give physical gap certificates for regulated Yang-Mills Hamiltonians and conditional continuum-transfer criteria. At fixed spatial cutoff, the physical $SU(3)$ Hamiltonian has a volume-independent gap of at least $8a/3$ for magnetic/electric ratio $b/a\le1/648$, with a thermodynamic ground state and a surviving finite-energy Wilson excitation. Its centered two-point autocorrelations have positive spectral representations; on the stated periodic cubes, every positive magnetic coupling gives the elementary plaquette at least two active energies. Quantum relative entropy identifies the physical energy form, while exact elimination retains the dynamical memory of discarded modes. We derive volume- and exterior-uniform vacuum score bounds, construct compatible finite hierarchies, and give summability and spectral-tightness criteria for limiting dynamics and surviving observables. Pinching motivates a conditional tube-energy balance; its use for an infrared gap requires uniform comparison with the complete vacuum-subtracted quantum energy after relaxation, including control of local low modes and localization errors. The remaining inputs include uniform infrared and shell estimates, local vacuum comparison rates, and relativistic field reconstruction. The interacting four-dimensional construction, nontrivial renormalized local fields and a uniform physical mass gap remain unproved.\end{abstract}

\maketitle

\hypertarget{sec1}{}\section{Introduction}

A positive mass gap separates the vacuum from the spectrum of physical excitations. The open four-dimensional Yang-Mills problem asks for a nontrivial quantum field theory on $\mathbb R^4$, for each compact simple gauge group, satisfying the standard axioms and possessing such a gap. We study how entropy and curvature estimates can contribute to this question through their relation to the physical Hamiltonian.\par

Our geometric input is the entropy-geometric framework of [\hyperlink{ref2}{1}]. On the faithful states of a supplied qutrit (three-level) Hermitian carrier, Umegaki relative entropy induces the Bogoliubov-Kubo-Mori (BKM) metric. Normalized affine-invariant transport and a chosen parallel Cartan form define a $\mathrm{PSU}(3)$ reduction of the eight-dimensional tangent bundle. The associated orthogonal $SU(3)$-module splitting is $\mathfrak{so}(8)=\mathfrak h_8\oplus\mathfrak m_{20}$, with $\mathfrak h_8=\operatorname{ad}\mathfrak{su}(3)$ and $(\mathfrak m_{20})_{\mathbb C}=\mathbf{10}\oplus\overline{\mathbf{10}}$. Relative to this reduction, the BKM Levi-Civita connection decomposes as $D^\Psi=D^{\rm col}+\Phi_{20}$: $D^{\rm col}$ preserves the transported Cartan form and comparison cubic, while the complementary one-form $\Phi_{20}$ measures the failure of $D^\Psi$ to preserve them [\hyperlink{ref2}{1}, Corollary 30 and Theorem 17]. This identifies color-preserving transport and its defect, providing geometrically specified variables for the curvature and energy comparisons in Section \hyperlink{sec5}{V}.\par

We take the cited decomposition, connection-defect identities and invariant-tensor normalizations as explicit assumptions, within their stated carrier, reduction and norm conventions; their entropy-geometric derivation is not reproved here. Section \hyperlink{sec5}{V} and Proposition \crosspdflink{Entropy_Geometry_and_Lattice_Gaps_Supplement.pdf}{res6.44}{6.30} rely on these inputs, including the cubic normalization in [\hyperlink{ref2}{1}, Theorem 18]. A spacetime realization, action coefficients and comparison with the actual quantum energy remain separate requirements. The regulated Hamiltonian results in Sections \hyperlink{sec2}{II}-\hyperlink{sec4}{IV} and the remaining operator and continuum-transfer results in Section \hyperlink{sec6}{VI} do not depend on these imported qutrit inputs. The channel-information-loss Hessian used in Section \hyperlink{sec3}{III} is distinct from this connection defect and is proved directly in the Supplement, Appendix \crosspdflink{Entropy_Geometry_and_Lattice_Gaps_Supplement.pdf}{appA.28}{A.29}.\par

The distinction between state geometry and physical dynamics is essential. The physical quantum Hamiltonian acts on gauge-invariant field states and determines the gap. When $\sigma$ is its actual Gibbs state, the calibrated modular Hamiltonian $-\beta^{-1}\log\sigma$ reproduces that evolution. BKM mechanics instead generates cotangent paths of faithful entropy states, while an endpoint generator reconstructs an assigned map from its samples. These map generators do not identify the physical quantum Hamiltonian. Once that Hamiltonian and its vacuum are specified, quantum relative entropy gives a direct characterization of its energy form.\par

The methods include established strong-coupling stability [\hyperlink{ref37}{2}, \hyperlink{ref58}{3}, \hyperlink{ref59}{4}], Feynman-Kac derivative formulas [\hyperlink{ref96}{5}], shorting and Schur/Feshbach elimination [\hyperlink{ref90}{6}, \hyperlink{ref101}{7}, \hyperlink{ref104}{8}], spectral representations and functional calculus. The contributions demonstrated here are the explicit gauge-sector gap constants and surviving Wilson excitation (Section \hyperlink{sec2}{II}); the elementary-plaquette result at every positive magnetic coupling on the stated periodic cubes (Proposition \hyperlink{res2.16}{2.8}); the calibrated entropy-energy and balanced-record identities (Theorems \hyperlink{res3.1}{3.1} and \hyperlink{res3.2}{3.2}); and the actual-vacuum score and comparison estimates, quantitative relaxation bounds, conditional hierarchy and transfer criteria, and lost-direction counterexample (Section \hyperlink{sec6}{VI}). Pinching motivates Section \hyperlink{sec5}{V}'s conditional flux-tube certificate and classical curvature-feedback test, including the obstruction to direct energy domination. The interacting four-dimensional continuum Yang-Mills construction and its required uniform physical decay estimate remain unproved.\par

Section \hyperlink{sec6}{VI} follows the proof route from actual-vacuum bounds and exact reduction through regulator comparisons, observable survival and spectral transfer. Section \hyperlink{sec7}{VII} identifies the remaining Yang-Mills estimates.\par

Proofs are collected in the companion Supplemental Material [\hyperlink{ref79}{9}]: Appendix \crosspdflink{Entropy_Geometry_and_Lattice_Gaps_Supplement.pdf}{appA}{A} covers physical electric, entropy and response estimates, Appendix \crosspdflink{Entropy_Geometry_and_Lattice_Gaps_Supplement.pdf}{appB}{B} screening and exceptional configurations, and Appendix \crosspdflink{Entropy_Geometry_and_Lattice_Gaps_Supplement.pdf}{appC}{C} compatible dynamics, regulator comparisons and spectral transfer.\par

\hypertarget{sec2}{}\section{Physical Hamiltonians and the fixed-cutoff result}

\hypertarget{sec2.1}{}\subsection{Gauge-invariant electric spectrum}

We first isolate the electric contribution to the physical excitation energy at a fixed regulator. Let $G$ be a nontrivial compact connected Lie group with a fixed bi-invariant Riemannian metric and normalized Haar measure. For an irreducible unitary representation $\pi$, write $d_\pi$, $\chi_\pi$, and $c_\pi$ for its dimension, character, and Casimir eigenvalue. The nonpositive Laplace-Beltrami operator $\Delta_G$ satisfies $-\Delta_G\pi_{ij}=c_\pi\pi_{ij}$. Connectedness and compact elliptic theory ensure that the following minimum is positive and attained:
\sbox{\equationmeasure}{$\displaystyle\begin{gathered}c_G:=\min_{\pi\ne\mathbf 1}c_\pi>0.\end{gathered}$}
\ifdim\wd\equationmeasure>\dimexpr\columnwidth-36pt\relax
\typeout{MANUSCRIPT-WIDE 2.1}\begin{manuscriptwide}\eqanchor{2.1}\begin{equation*}\tag{2.1}\begin{gathered}c_G:=\min_{\pi\ne\mathbf 1}c_\pi>0.\end{gathered}\end{equation*}\end{manuscriptwide}
\else \eqanchor{2.1}\begin{equation*}\tag{2.1}\begin{gathered}c_G:=\min_{\pi\ne\mathbf 1}c_\pi>0.\end{gathered}\end{equation*}\fi

Let $\Gamma=(\mathsf V,\mathsf E)$ be a finite connected simple graph containing a cycle. Orient each edge once, write $s(e),t(e)$ for its endpoints, and choose weights $a_e>0$. Set
\sbox{\equationmeasure}{$\displaystyle\begin{gathered}M=G^{\mathsf E},\\  dm=\prod_{e\in\mathsf E}dg_e,\\  K=G^{\mathsf V},\\  (k\cdot U)_e=k_{s(e)}U_e k_{t(e)}^{-1}.\end{gathered}$}
\ifdim\wd\equationmeasure>\dimexpr\columnwidth-36pt\relax
\typeout{MANUSCRIPT-WIDE 2.2}\begin{manuscriptwide}\eqanchor{2.2}\begin{equation*}\tag{2.2}\begin{gathered}M=G^{\mathsf E},\quad dm=\prod_{e\in\mathsf E}dg_e,\\ K=G^{\mathsf V},\quad (k\cdot U)_e=k_{s(e)}U_e k_{t(e)}^{-1}.\end{gathered}\end{equation*}\end{manuscriptwide}
\else \eqanchor{2.2}\begin{equation*}\tag{2.2}\begin{gathered}M=G^{\mathsf E},\\  dm=\prod_{e\in\mathsf E}dg_e,\\  K=G^{\mathsf V},\\  (k\cdot U)_e=k_{s(e)}U_e k_{t(e)}^{-1}.\end{gathered}\end{equation*}\fi

Gauge transformations act at every vertex, including the boundary; there are no external charges or matter fields. Define
\sbox{\equationmeasure}{$\displaystyle\begin{gathered}\mathcal H_{\rm phys}=L^2(M,dm)^K,\\  H_{\rm el}=-\sum_e a_e\Delta_e,\\  \mathcal E_m(f)=\int_M\sum_e a_e|\nabla_e f|^2\,dm.\end{gathered}$}
\ifdim\wd\equationmeasure>\dimexpr\columnwidth-36pt\relax
\typeout{MANUSCRIPT-WIDE 2.3}\begin{manuscriptwide}\eqanchor{2.3}\begin{equation*}\tag{2.3}\begin{gathered}\mathcal H_{\rm phys}=L^2(M,dm)^K,\quad H_{\rm el}=-\sum_e a_e\Delta_e,\\ \mathcal E_m(f)=\int_M\sum_e a_e|\nabla_e f|^2\,dm.\end{gathered}\end{equation*}\end{manuscriptwide}
\else \eqanchor{2.3}\begin{equation*}\tag{2.3}\begin{gathered}\mathcal H_{\rm phys}=L^2(M,dm)^K,\\  H_{\rm el}=-\sum_e a_e\Delta_e,\\  \mathcal E_m(f)=\int_M\sum_e a_e|\nabla_e f|^2\,dm.\end{gathered}\end{equation*}\fi

The form domain is $H^1(M)\cap\mathcal H_{\rm phys}$. Bi-invariance makes gauge transformations commute with the operator and form. The restricted operator is self-adjoint with compact resolvent; smooth approximation followed by compact gauge averaging gives invariant smooth operator and form cores. Its kernel consists of constants because $M$ is connected. Define
\sbox{\equationmeasure}{$\displaystyle\begin{gathered}g_a(\Gamma)=\min_{C\ {\rm simple\ cycle}}\sum_{e\in C}a_e.\end{gathered}$}
\ifdim\wd\equationmeasure>\dimexpr\columnwidth-36pt\relax
\typeout{MANUSCRIPT-WIDE 2.4}\begin{manuscriptwide}\eqanchor{2.4}\begin{equation*}\tag{2.4}\begin{gathered}g_a(\Gamma)=\min_{C\ {\rm simple\ cycle}}\sum_{e\in C}a_e.\end{gathered}\end{equation*}\end{manuscriptwide}
\else \eqanchor{2.4}\begin{equation*}\tag{2.4}\begin{gathered}g_a(\Gamma)=\min_{C\ {\rm simple\ cycle}}\sum_{e\in C}a_e.\end{gathered}\end{equation*}\fi

\hypertarget{res2.1}{}\noindent \textbf{Theorem \hyperlink{res2.1}{2.1} (exact electric gap).} The first nonzero eigenvalue of the physical electric Hamiltonian is
\sbox{\equationmeasure}{$\displaystyle\begin{gathered}\delta_{\rm el}=c_G g_a(\Gamma).\end{gathered}$}
\ifdim\wd\equationmeasure>\dimexpr\columnwidth-36pt\relax
\typeout{MANUSCRIPT-WIDE 2.5}\begin{manuscriptwide}\eqanchor{2.5}\begin{equation*}\tag{2.5}\begin{gathered}\delta_{\rm el}=c_G g_a(\Gamma).\end{gathered}\end{equation*}\end{manuscriptwide}
\else \eqanchor{2.5}\begin{equation*}\tag{2.5}\begin{gathered}\delta_{\rm el}=c_G g_a(\Gamma).\end{gathered}\end{equation*}\fi

Equivalently, $\mathcal E_m(f)\ge\delta_{\rm el}\operatorname{Var}_m(f)$ on the physical form domain, with optimal constant $\delta_{\rm el}$.\par

Proof in the Supplement, Appendix \crosspdflink{Entropy_Geometry_and_Lattice_Gaps_Supplement.pdf}{appA.1}{A.1}.\par

\noindent \textbf{Graph and boundary scope.} Support can include bridges joining cyclic pieces; the proof only requires a cycle. On a forest, leaf removal makes every physical label trivial, leaving a one-dimensional physical space with no finite excited eigenvalue. Fixing boundary gauge transformations or adding charges can admit open flux lines and changes Theorem \hyperlink{res2.1}{2.1}. Short periodic lattices require their actual graph girth; the simple-graph hypotheses exclude self-loops and parallel edges.\par

\hypertarget{sec2.2}{}\subsection{Actual ground law}

Let $V\in C^\infty(M;\mathbb R)^K$ and $H=H_{\rm el}+V$. On $L^2(M,dm)$, $H$ is lower bounded and self-adjoint on the electric domain, with compact resolvent; its invariant restriction is the physical Hamiltonian. Write the physical eigenvalues with multiplicity as $E_0\le E_1\le\cdots$ and its gap as $\Delta(H)=E_1-E_0$. Set
\sbox{\equationmeasure}{$\displaystyle\begin{gathered}\operatorname{osc}V=\max_M V-\min_M V,\\  \Gamma_a(f,g)=\sum_e a_e\langle\nabla_e f,\nabla_e g\rangle.\end{gathered}$}
\ifdim\wd\equationmeasure>\dimexpr\columnwidth-36pt\relax
\typeout{MANUSCRIPT-WIDE 2.6}\begin{manuscriptwide}\eqanchor{2.6}\begin{equation*}\tag{2.6}\begin{gathered}\operatorname{osc}V=\max_M V-\min_M V,\quad \Gamma_a(f,g)=\sum_e a_e\langle\nabla_e f,\nabla_e g\rangle.\end{gathered}\end{equation*}\end{manuscriptwide}
\else \eqanchor{2.6}\begin{equation*}\tag{2.6}\begin{gathered}\operatorname{osc}V=\max_M V-\min_M V,\\  \Gamma_a(f,g)=\sum_e a_e\langle\nabla_e f,\nabla_e g\rangle.\end{gathered}\end{equation*}\fi

For complex functions, the Hermitian extension is conjugate-linear in its first argument. Weights and potentials share a fixed energy normalization; the semigroup parameter $t$ has inverse-energy units.\par

\hypertarget{res2.2}{}\noindent \textbf{Theorem \hyperlink{res2.2}{2.2} (ground-state realization).} The full Hamiltonian has a unique normalized strictly positive ground state $\Omega$, with eigenvalue $E_0$. It is smooth and gauge invariant. With
\sbox{\equationmeasure}{$\displaystyle\begin{gathered}d\mu=\Omega^2dm,\\  Wf=\Omega f,\\  \mathcal E_\mu(f)=\int_M\Gamma_a(f,f)\,d\mu,\end{gathered}$}
\ifdim\wd\equationmeasure>\dimexpr\columnwidth-36pt\relax
\typeout{MANUSCRIPT-WIDE 2.7}\begin{manuscriptwide}\eqanchor{2.7}\begin{equation*}\tag{2.7}\begin{gathered}d\mu=\Omega^2dm,\qquad Wf=\Omega f,\qquad \mathcal E_\mu(f)=\int_M\Gamma_a(f,f)\,d\mu,\end{gathered}\end{equation*}\end{manuscriptwide}
\else \eqanchor{2.7}\begin{equation*}\tag{2.7}\begin{gathered}d\mu=\Omega^2dm,\\  Wf=\Omega f,\\  \mathcal E_\mu(f)=\int_M\Gamma_a(f,f)\,d\mu,\end{gathered}\end{equation*}\fi

multiplication by $\Omega$ is unitary from $L^2(\mu)^K$ to $\mathcal H_{\rm phys}$. On smooth invariant functions, and by closure on the corresponding form domains,
\sbox{\equationmeasure}{$\displaystyle\begin{gathered}\big\|(H-E_0)^{1/2}Wf\big\|^2=\mathcal E_\mu(f).\end{gathered}$}
\ifdim\wd\equationmeasure>\dimexpr\columnwidth-36pt\relax
\typeout{MANUSCRIPT-WIDE 2.8}\begin{manuscriptwide}\eqanchor{2.8}\begin{equation*}\tag{2.8}\begin{gathered}\big\|(H-E_0)^{1/2}Wf\big\|^2=\mathcal E_\mu(f).\end{gathered}\end{equation*}\end{manuscriptwide}
\else \eqanchor{2.8}\begin{equation*}\tag{2.8}\begin{gathered}\big\|(H-E_0)^{1/2}Wf\big\|^2=\mathcal E_\mu(f).\end{gathered}\end{equation*}\fi

In particular, with the infimum taken over real smooth invariant functions of positive variance,
\sbox{\equationmeasure}{$\displaystyle\begin{gathered}\Delta(H)=\inf_f\frac{\mathcal E_\mu(f)}{\operatorname{Var}_\mu(f)}.\end{gathered}$}
\ifdim\wd\equationmeasure>\dimexpr\columnwidth-36pt\relax
\typeout{MANUSCRIPT-WIDE 2.9}\begin{manuscriptwide}\eqanchor{2.9}\begin{equation*}\tag{2.9}\begin{gathered}\Delta(H)=\inf_f\frac{\mathcal E_\mu(f)}{\operatorname{Var}_\mu(f)}.\end{gathered}\end{equation*}\end{manuscriptwide}
\else \eqanchor{2.9}\begin{equation*}\tag{2.9}\begin{gathered}\Delta(H)=\inf_f\frac{\mathcal E_\mu(f)}{\operatorname{Var}_\mu(f)}.\end{gathered}\end{equation*}\fi

Complex functions give the same infimum by real and imaginary decomposition.\par

Proof in the Supplement, Appendix \crosspdflink{Entropy_Geometry_and_Lattice_Gaps_Supplement.pdf}{appA.2}{A.2}.\par

For a collection $\mathsf P$ of simple cycles, a nontrivial irreducible representation $R$, and coefficients $b_p\ge0$, define
\sbox{\equationmeasure}{$\displaystyle\begin{gathered}V_W(U)=\sum_{p\in\mathsf P}b_p\left(1-\frac{\operatorname{Re}\chi_R(U_p)}{d_R}\right),\\  B=\sum_pb_p.\end{gathered}$}
\ifdim\wd\equationmeasure>\dimexpr\columnwidth-36pt\relax
\typeout{MANUSCRIPT-WIDE 2.10}\begin{manuscriptwide}\eqanchor{2.10}\begin{equation*}\tag{2.10}\begin{gathered}V_W(U)=\sum_{p\in\mathsf P}b_p\left(1-\frac{\operatorname{Re}\chi_R(U_p)}{d_R}\right),\qquad B=\sum_pb_p.\end{gathered}\end{equation*}\end{manuscriptwide}
\else \eqanchor{2.10}\begin{equation*}\tag{2.10}\begin{gathered}V_W(U)=\sum_{p\in\mathsf P}b_p\left(1-\frac{\operatorname{Re}\chi_R(U_p)}{d_R}\right),\\  B=\sum_pb_p.\end{gathered}\end{equation*}\fi

Each summand lies between zero and $2b_p$ and vanishes at the identity configuration. Each cycle holonomy is Haar under $m$, hence $\int V_Wdm=B$; plaquette independence is unnecessary.\par

At each fixed regulator, Proposition \crosspdflink{Entropy_Geometry_and_Lattice_Gaps_Supplement.pdf}{res2.14}{2.16} constructs a stationary reflection-positive continuous-path law with exactly Theorem \hyperlink{res2.2}{2.2}'s Hilbert space and time generator, and proves strict concavity of the vacuum energy in its coupling for every nonconstant smooth invariant interaction. At nonzero coupling the unfixed link law cannot factorize, although each single-link marginal is Haar; this is not a decomposition into physical subregions. Time is inverse-energy Euclidean time, and no spatial-cutoff-uniform estimate follows.\par

Proposition \crosspdflink{Entropy_Geometry_and_Lattice_Gaps_Supplement.pdf}{res2.8}{2.10}, Theorem \crosspdflink{Entropy_Geometry_and_Lattice_Gaps_Supplement.pdf}{res2.9}{2.11} and Corollaries \crosspdflink{Entropy_Geometry_and_Lattice_Gaps_Supplement.pdf}{res2.10}{2.12} and \crosspdflink{Entropy_Geometry_and_Lattice_Gaps_Supplement.pdf}{res2.11}{2.13} give finite-volume certificates whose constants depend on the graph and potential. The following fixed-cutoff result has volume-independent constants.\par

\hypertarget{sec2.3}{}\subsection{Fixed-cutoff interacting result}

The strong-coupling construction gives explicit volume-independent constants for the physical gap and a surviving local Wilson excitation. It removes the auxiliary electric spectral cutoff; space remains a cubic lattice and Hamiltonian time is continuous.\par

Yarotsky's theorem gives a uniform gap, thermodynamic ground state and clustering for infinite-dimensional local Hilbert spaces under its smallness hypotheses [\hyperlink{ref58}{3}, Theorem 1]. Osterwalder and Seiler established lattice physical positivity, strong-coupling infinite-volume analyticity and Wilson confinement [\hyperlink{ref59}{4}]. The proof below supplies the constants $8a/3$ and $1/648$ through its gauge-sector estimates and cutoff removal in the stated normalization.\par

In the Hamiltonian lattice framework [\hyperlink{ref4}{10}], let each positively oriented spatial link carry $L^2(SU(3))$ with normalized Haar measure. Write $L_e=-\Delta_e$ in the normalization $\operatorname{tr}(T^AT^B)=\delta^{AB}/2$, so every nontrivial representation has electric Casimir at least $4/3$. Count each elementary spatial plaquette once and set
\sbox{\equationmeasure}{$\displaystyle\begin{gathered}H_\Lambda=a\sum_eL_e+b\sum_p(1-F_p),\\  F_p=\frac13\operatorname{Re}\operatorname{tr}U_p,\\  a>0,\\  b\ge0.\end{gathered}$}
\ifdim\wd\equationmeasure>\dimexpr\columnwidth-36pt\relax
\typeout{MANUSCRIPT-WIDE 2.11}\begin{manuscriptwide}\eqanchor{2.11}\begin{equation*}\tag{2.11}\begin{gathered}H_\Lambda=a\sum_eL_e+b\sum_p(1-F_p),\qquad F_p=\frac13\operatorname{Re}\operatorname{tr}U_p,\qquad a>0,\quad b\ge0.\end{gathered}\end{equation*}\end{manuscriptwide}
\else \eqanchor{2.11}\begin{equation*}\tag{2.11}\begin{gathered}H_\Lambda=a\sum_eL_e+b\sum_p(1-F_p),\\  F_p=\frac13\operatorname{Re}\operatorname{tr}U_p,\\  a>0,\\  b\ge0.\end{gathered}\end{equation*}\fi

Here $a$ and $b$ are energy coefficients; $a$ is not lattice spacing. The finite graphs are open cubic boxes, finite outgoing-link block restrictions, or periodic cubic volumes of side at least four. Each plaquette has four distinct links, and every endpoint gauge transformation is imposed. For a block restriction, retain the three outgoing links at each selected vertex and every chosen plaquette whose actual three-block support is retained. Its endpoint graph is a subgraph of the cubic lattice. This convention permits electric boundary links and defines the finite restrictions used below without identifying different link variables.\par

Lemmas \crosspdflink{Entropy_Geometry_and_Lattice_Gaps_Supplement.pdf}{res2.12}{2.14} and \crosspdflink{Entropy_Geometry_and_Lattice_Gaps_Supplement.pdf}{res2.13}{2.15} adapt the creation-operator method [\hyperlink{ref37}{2}, Sections 2.1-2.4] to gauge rotors; the Supplement proves its volume-independent constants and electric-cutoff removal. For the proof strategy below, set $\delta=4a/3$ and $J=\max_u\sum_{X\ni u}\|V_X\|$ for the shifted interaction $V=\sum_X V_X$ on outgoing-link blocks, counting repeated supports as in Lemma \crosspdflink{Entropy_Geometry_and_Lattice_Gaps_Supplement.pdf}{res2.13}{2.15}.\par

\hypertarget{res2.3}{}\noindent \textbf{Theorem \hyperlink{res2.3}{2.3} (explicit finite-volume physical gap).} For every finite graph specified above, if $b/a\le1/648$, the actual Hamiltonian \hyperlink{eq2.11}{(2.11)} has a unique normalized positive ground vector $\Psi_\Lambda$. It is physical, and
\sbox{\equationmeasure}{$\displaystyle\begin{gathered}H_{\Lambda,\mathrm{phys}}-E_\Lambda\ \ge\ \frac{8a}{3}\bigl(I-|\Psi_\Lambda\rangle\langle\Psi_\Lambda|\bigr).\end{gathered}$}
\ifdim\wd\equationmeasure>\dimexpr\columnwidth-36pt\relax
\typeout{MANUSCRIPT-WIDE 2.12}\begin{manuscriptwide}\eqanchor{2.12}\begin{equation*}\tag{2.12}\begin{gathered}H_{\Lambda,\mathrm{phys}}-E_\Lambda\ \ge\ \frac{8a}{3}\bigl(I-|\Psi_\Lambda\rangle\langle\Psi_\Lambda|\bigr).\end{gathered}\end{equation*}\end{manuscriptwide}
\else \eqanchor{2.12}\begin{equation*}\tag{2.12}\begin{gathered}H_{\Lambda,\mathrm{phys}}-E_\Lambda\ \ge\ \frac{8a}{3}\bigl(I-|\Psi_\Lambda\rangle\langle\Psi_\Lambda|\bigr).\end{gathered}\end{equation*}\fi

The claim concerns the physical spectrum above the full ground energy $E_\Lambda$; the same stronger lower bound is not asserted for nonphysical excitations.\par

\noindent \textbf{Proof strategy.} The cycle bound \crosspdflink{Entropy_Geometry_and_Lattice_Gaps_Supplement.pdf}{eq2.25}{(2.25)} makes Lemma \crosspdflink{Entropy_Geometry_and_Lattice_Gaps_Supplement.pdf}{res2.13}{2.15} independent of volume and electric cutoff. At $J/\delta\le1/128$, the creation equation contracts the radius-$1/24$ ball; its shifted equation has Lipschitz constant below $47/48$ for energy differences below $2\delta$. Continuity selects the physical ground branch, and gauge-invariant form cores and min-max remove the cutoff. The Wilson estimate $J\le27b/4$ makes $b/a\le1/648$ sufficient; positivity identifies the full ground state.\par

Proof in the Supplement, Appendix \crosspdflink{Entropy_Geometry_and_Lattice_Gaps_Supplement.pdf}{appA.8}{A.8}. Proposition \hyperlink{res2.4}{2.4} supplies local energy compactness and a surviving plaquette fluctuation. Theorem \hyperlink{res2.5}{2.5} then passes the spectral bound to the complete invariant GNS sector by local normality, gauge averaging and convergence of physical correlation functions.\par

\hypertarget{res2.4}{}\noindent \textbf{Proposition \hyperlink{res2.4}{2.4} (local survival and finite excitation energy).} Let $\omega_\Lambda$ be the actual finite-volume ground state of \hyperlink{eq2.11}{(2.11)}, and let $F$ be any retained plaquette. At every finite $b/a$ and every retained link $e$,
\sbox{\equationmeasure}{$\displaystyle\begin{gathered}a\,\omega_\Lambda(L_e)\le4b.\end{gathered}$}
\ifdim\wd\equationmeasure>\dimexpr\columnwidth-36pt\relax
\typeout{MANUSCRIPT-WIDE 2.13}\begin{manuscriptwide}\eqanchor{2.13}\begin{equation*}\tag{2.13}\begin{gathered}a\,\omega_\Lambda(L_e)\le4b.\end{gathered}\end{equation*}\end{manuscriptwide}
\else \eqanchor{2.13}\begin{equation*}\tag{2.13}\begin{gathered}a\,\omega_\Lambda(L_e)\le4b.\end{gathered}\end{equation*}\fi

If $b/a\le1/648$, set $v_{F,\Lambda}=(F-\omega_\Lambda(F))\Psi_\Lambda$ and $K_\Lambda=H_\Lambda-E_\Lambda$. Then
\sbox{\equationmeasure}{$\displaystyle\begin{gathered}s_{F,\Lambda}:=\|v_{F,\Lambda}\|^2\ge\eta_0:=\frac{511-18\sqrt{430}}{7776}>\frac1{60},\\  \|K_\Lambda^{1/2}v_{F,\Lambda}\|^2\le\frac{2a}{3}.\end{gathered}$}
\ifdim\wd\equationmeasure>\dimexpr\columnwidth-36pt\relax
\typeout{MANUSCRIPT-WIDE 2.14}\begin{manuscriptwide}\eqanchor{2.14}\begin{equation*}\tag{2.14}\begin{gathered}s_{F,\Lambda}:=\|v_{F,\Lambda}\|^2\ge\eta_0:=\frac{511-18\sqrt{430}}{7776}>\frac1{60},\qquad \|K_\Lambda^{1/2}v_{F,\Lambda}\|^2\le\frac{2a}{3}.\end{gathered}\end{equation*}\end{manuscriptwide}
\else \eqanchor{2.14}\begin{equation*}\tag{2.14}\begin{gathered}s_{F,\Lambda}:=\|v_{F,\Lambda}\|^2\ge\eta_0:=\frac{511-18\sqrt{430}}{7776}>\frac1{60},\\  \|K_\Lambda^{1/2}v_{F,\Lambda}\|^2\le\frac{2a}{3}.\end{gathered}\end{equation*}\fi

Proof in the Supplement, Appendix \crosspdflink{Entropy_Geometry_and_Lattice_Gaps_Supplement.pdf}{appA.9}{A.9}.\par

\hypertarget{sec2.4}{}\subsection{Infinite volume and physical correlations}

Let $\mathcal A$ be the norm closure of the union of the full bounded local link algebras $\mathcal A_S=\mathcal B(L^2(SU(3)^S))$, where $S$ ranges over finite sets of spatial links and inclusions act as the identity on additional links. For each limiting state on $\mathcal A$ below, write $(\pi,\mathscr H,\Omega)$ for its Gelfand-Naimark-Segal (GNS) representation, with cyclic vacuum vector $\Omega$.\par

\hypertarget{res2.5}{}\noindent \textbf{Theorem \hyperlink{res2.5}{2.5} (thermodynamic Hamiltonian and complete physical gap).} Use exhausting native restrictions of the fixed interaction \hyperlink{eq2.11}{(2.11)}, with every fixed local link and plaquette eventually retained, or the periodic sequence described in the Supplement, Appendix \crosspdflink{Entropy_Geometry_and_Lattice_Gaps_Supplement.pdf}{appA.10}{A.10}. At each fixed finite $b/a$, finite-volume ground states have locally normal thermodynamic subsequences. Their GNS representations possess a strongly continuous Hamiltonian time evolution with nonnegative generator $K$. If $b/a\le1/648$, the common fixed-vector space of all vertex gauge transformations reduces $K$ and satisfies
\sbox{\equationmeasure}{$\displaystyle\begin{gathered}\begin{gathered}
\mathscr H_{\mathrm{phys}}=
\overline{\left\{\pi(A)\Omega:
\begin{gathered}A\text{ bounded, local,}\\\text{gauge invariant}\end{gathered}
\right\}},\\
K_{\mathrm{phys}}\ge\frac{8a}{3}(I-P_\Omega).
\end{gathered}\end{gathered}$}
\ifdim\wd\equationmeasure>\dimexpr\columnwidth-36pt\relax
\typeout{MANUSCRIPT-WIDE 2.15}\begin{manuscriptwide}[6]\eqanchor{2.15}\begin{equation*}\tag{2.15}\begin{gathered}\begin{gathered}
\mathscr H_{\mathrm{phys}}=
\overline{\left\{\pi(A)\Omega:
\begin{gathered}A\text{ bounded, local,}\\\text{gauge invariant}\end{gathered}
\right\}},\\
K_{\mathrm{phys}}\ge\frac{8a}{3}(I-P_\Omega).
\end{gathered}\end{gathered}\end{equation*}\end{manuscriptwide}
\else \eqanchor{2.15}\begin{equation*}\tag{2.15}\begin{gathered}\begin{gathered}
\mathscr H_{\mathrm{phys}}=
\overline{\left\{\pi(A)\Omega:
\begin{gathered}A\text{ bounded, local,}\\\text{gauge invariant}\end{gathered}
\right\}},\\
K_{\mathrm{phys}}\ge\frac{8a}{3}(I-P_\Omega).
\end{gathered}\end{gathered}\end{equation*}\fi

A translation-invariant such state can be selected. Vacuum uniqueness in \hyperlink{eq2.15}{(2.15)} is within this physical representation; uniqueness among all infinite-volume states or independence of boundary conditions is not asserted.\par

Proof in the Supplement, Appendix \crosspdflink{Entropy_Geometry_and_Lattice_Gaps_Supplement.pdf}{appA.10}{A.10}.\par

\hypertarget{res2.6}{}\noindent \textbf{Corollary \hyperlink{res2.6}{2.6} (two-sided Wilson correlations and finite spectral edge).} In any representation of Theorem \hyperlink{res2.5}{2.5} with $b/a\le1/648$, let $v_F=(\pi(F)-\omega(F))\Omega$ for an elementary spatial plaquette and $s_F=\|v_F\|^2$. Then $s_F\ge\eta_0>1/60$, $v_F$ is in the form domain of $K$, and
\sbox{\equationmeasure}{$\displaystyle\begin{gathered}\|K^{1/2}v_F\|^2\le\frac{2a}{3},\\  \frac1{60}e^{-40at}\le\langle v_F,e^{-tK}v_F\rangle\le s_Fe^{-8at/3},\\  t\ge0.\end{gathered}$}
\ifdim\wd\equationmeasure>\dimexpr\columnwidth-36pt\relax
\typeout{MANUSCRIPT-WIDE 2.16}\begin{manuscriptwide}\eqanchor{2.16}\begin{equation*}\tag{2.16}\begin{gathered}\|K^{1/2}v_F\|^2\le\frac{2a}{3},\qquad \frac1{60}e^{-40at}\le\langle v_F,e^{-tK}v_F\rangle\le s_Fe^{-8at/3},\quad t\ge0.\end{gathered}\end{equation*}\end{manuscriptwide}
\else \eqanchor{2.16}\begin{equation*}\tag{2.16}\begin{gathered}\|K^{1/2}v_F\|^2\le\frac{2a}{3},\\  \frac1{60}e^{-40at}\le\langle v_F,e^{-tK}v_F\rangle\le s_Fe^{-8at/3},\\  t\ge0.\end{gathered}\end{equation*}\fi

In particular the physical sector is nontrivial, and its positive spectral edge obeys
\sbox{\equationmeasure}{$\displaystyle\begin{gathered}\frac{8a}{3}\le\inf\sigma\!\left(K\big|_{\mathscr H_{\mathrm{phys}}\cap\Omega^\perp}\right)\le40a.\end{gathered}$}
\ifdim\wd\equationmeasure>\dimexpr\columnwidth-36pt\relax
\typeout{MANUSCRIPT-WIDE 2.17}\begin{manuscriptwide}\eqanchor{2.17}\begin{equation*}\tag{2.17}\begin{gathered}\frac{8a}{3}\le\inf\sigma\!\left(K\big|_{\mathscr H_{\mathrm{phys}}\cap\Omega^\perp}\right)\le40a.\end{gathered}\end{equation*}\end{manuscriptwide}
\else \eqanchor{2.17}\begin{equation*}\tag{2.17}\begin{gathered}\frac{8a}{3}\le\inf\sigma\!\left(K\big|_{\mathscr H_{\mathrm{phys}}\cap\Omega^\perp}\right)\le40a.\end{gathered}\end{equation*}\fi

Proof in the Supplement, Appendix \crosspdflink{Entropy_Geometry_and_Lattice_Gaps_Supplement.pdf}{appA.11}{A.11}.\par

\subsubsection*{Spectral form of the two-point function}

The vacuum correlation has the familiar positive spectral representation used in glueball spectroscopy [\hyperlink{ref93}{11}, Eq. (27)]. Here it follows from the actual Hamiltonian already constructed, with no assumed particle masses.\par

\hypertarget{res2.15}{}\noindent \textbf{Proposition \hyperlink{res2.15}{2.7} (connected two-point function and its visible energy).} In Theorem \hyperlink{res2.2}{2.2}, let $f$ be a nonconstant smooth real gauge-invariant multiplication observable. Set $K=H-E_0$, $\phi_f=f-\mu(f)I$ and $v_f=\phi_f\Omega$. Let $\delta_n>0$ be the distinct positive eigenvalues of $K$ on the physical space and $P_n=\mathbf1_{\{\delta_n\}}(K)$. The connected Euclidean two-point function is exactly
\sbox{\equationmeasure}{$\displaystyle\begin{gathered}\begin{aligned}
C_f(t)&:=\langle v_f,e^{-tK}v_f\rangle
       =\sum_n A_n e^{-\delta_n t},\\
A_n&=\|P_nv_f\|^2\ge0,\\
\sum_nA_n&=\|v_f\|^2=\operatorname{Var}_\mu(f)>0,
\qquad t\ge0.
\end{aligned}\end{gathered}$}
\ifdim\wd\equationmeasure>\dimexpr\columnwidth-36pt\relax
\typeout{MANUSCRIPT-WIDE 2.31}\begin{manuscriptwide}\eqanchor{2.31}\begin{equation*}\tag{2.31}\begin{gathered}\begin{aligned}
C_f(t)&:=\langle v_f,e^{-tK}v_f\rangle
       =\sum_n A_n e^{-\delta_n t},\\
A_n&=\|P_nv_f\|^2\ge0,\\
\sum_nA_n&=\|v_f\|^2=\operatorname{Var}_\mu(f)>0,
\qquad t\ge0.
\end{aligned}\end{gathered}\end{equation*}\end{manuscriptwide}
\else \eqanchor{2.31}\begin{equation*}\tag{2.31}\begin{gathered}\begin{aligned}
C_f(t)&:=\langle v_f,e^{-tK}v_f\rangle
       =\sum_n A_n e^{-\delta_n t},\\
A_n&=\|P_nv_f\|^2\ge0,\\
\sum_nA_n&=\|v_f\|^2=\operatorname{Var}_\mu(f)>0,
\qquad t\ge0.
\end{aligned}\end{gathered}\end{equation*}\fi

The series converges absolutely and uniformly for $t\ge0$ at this fixed regulator. If $\delta_f=\min\{\delta_n:A_n>0\}$ and $A_f=\|\mathbf1_{\{\delta_f\}}(K)v_f\|^2$, then $C_f(t)\sim A_fe^{-\delta_ft}$ as $t\to\infty$. One has $\delta_f\ge\Delta(H)$; equality requires overlap with the lowest physical eigenspace.\par

For any supplied nonnegative self-adjoint $K$ with normalized vacuum $\Omega$, $K\Omega=0$, and nonzero physical vector $v\perp\Omega$, including an infinite-volume representation, the general formula is
\sbox{\equationmeasure}{$\displaystyle\begin{gathered}\begin{aligned}
\nu_v(B)&=\|\mathbf1_B(K)v\|^2,\\
C_v(t)&=\int_{[0,\infty)}e^{-Et}\,d\nu_v(E),\\
m_v&=\inf\operatorname{supp}\nu_v.
\end{aligned}\end{gathered}$}
\ifdim\wd\equationmeasure>\dimexpr\columnwidth-36pt\relax
\typeout{MANUSCRIPT-WIDE 2.32}\begin{manuscriptwide}\eqanchor{2.32}\begin{equation*}\tag{2.32}\begin{gathered}\begin{aligned}
\nu_v(B)&=\|\mathbf1_B(K)v\|^2,\\
C_v(t)&=\int_{[0,\infty)}e^{-Et}\,d\nu_v(E),\\
m_v&=\inf\operatorname{supp}\nu_v.
\end{aligned}\end{gathered}\end{equation*}\end{manuscriptwide}
\else \eqanchor{2.32}\begin{equation*}\tag{2.32}\begin{gathered}\begin{aligned}
\nu_v(B)&=\|\mathbf1_B(K)v\|^2,\\
C_v(t)&=\int_{[0,\infty)}e^{-Et}\,d\nu_v(E),\\
m_v&=\inf\operatorname{supp}\nu_v.
\end{aligned}\end{gathered}\end{equation*}\fi

For $t>0$, put $d\nu_{v,t}(E)=e^{-Et}d\nu_v(E)/C_v(t)$. The effective energy obeys
\sbox{\equationmeasure}{$\displaystyle\begin{gathered}\begin{aligned}
E_{\rm eff}(t)&:=-\frac{d}{dt}\log C_v(t)\\
&=\int E\,d\nu_{v,t}(E),\\
E_{\rm eff}'(t)&=-\operatorname{Var}_{\nu_{v,t}}(E)\le0,\\
\lim_{t\to\infty}E_{\rm eff}(t)&=m_v,\\
\lim_{t\to\infty}-\frac1t
 \log\frac{C_v(t)}{C_v(0)}&=m_v,\\
\lim_{t\to\infty}e^{m_vt}C_v(t)&=\nu_v(\{m_v\}).
\end{aligned}\end{gathered}$}
\ifdim\wd\equationmeasure>\dimexpr\columnwidth-36pt\relax
\typeout{MANUSCRIPT-WIDE 2.33}\begin{manuscriptwide}\eqanchor{2.33}\begin{equation*}\tag{2.33}\begin{gathered}\begin{aligned}
E_{\rm eff}(t)&:=-\frac{d}{dt}\log C_v(t)\\
&=\int E\,d\nu_{v,t}(E),\\
E_{\rm eff}'(t)&=-\operatorname{Var}_{\nu_{v,t}}(E)\le0,\\
\lim_{t\to\infty}E_{\rm eff}(t)&=m_v,\\
\lim_{t\to\infty}-\frac1t
 \log\frac{C_v(t)}{C_v(0)}&=m_v,\\
\lim_{t\to\infty}e^{m_vt}C_v(t)&=\nu_v(\{m_v\}).
\end{aligned}\end{gathered}\end{equation*}\end{manuscriptwide}
\else \eqanchor{2.33}\begin{equation*}\tag{2.33}\begin{gathered}\begin{aligned}
E_{\rm eff}(t)&:=-\frac{d}{dt}\log C_v(t)\\
&=\int E\,d\nu_{v,t}(E),\\
E_{\rm eff}'(t)&=-\operatorname{Var}_{\nu_{v,t}}(E)\le0,\\
\lim_{t\to\infty}E_{\rm eff}(t)&=m_v,\\
\lim_{t\to\infty}-\frac1t
 \log\frac{C_v(t)}{C_v(0)}&=m_v,\\
\lim_{t\to\infty}e^{m_vt}C_v(t)&=\nu_v(\{m_v\}).
\end{aligned}\end{gathered}\end{equation*}\fi

Thus a positive leading pure-exponential amplitude requires an energy atom at the visible edge. Proof, spectral truncation bounds and reflection positivity in the Supplement, Appendix \crosspdflink{Entropy_Geometry_and_Lattice_Gaps_Supplement.pdf}{appA.31}{A.13}.\par

\hypertarget{res2.16}{}\noindent \textbf{Proposition \hyperlink{res2.16}{2.8} (an actual interacting plaquette needs multiple energies).} On an isotropic periodic cubic $L^3$ regulator of \hyperlink{eq2.11}{(2.11)}, $L\ge4$, let $f=F_p$ for an elementary plaquette. At $b=0$,
\sbox{\equationmeasure}{$\displaystyle\begin{gathered}C_{F_p}(t)=\frac1{18}e^{-16at/3},\\  t\ge0.\end{gathered}$}
\ifdim\wd\equationmeasure>\dimexpr\columnwidth-36pt\relax
\typeout{MANUSCRIPT-WIDE 2.34}\begin{manuscriptwide}\eqanchor{2.34}\begin{equation*}\tag{2.34}\begin{gathered}C_{F_p}(t)=\frac1{18}e^{-16at/3},\qquad t\ge0.\end{gathered}\end{equation*}\end{manuscriptwide}
\else \eqanchor{2.34}\begin{equation*}\tag{2.34}\begin{gathered}C_{F_p}(t)=\frac1{18}e^{-16at/3},\\  t\ge0.\end{gathered}\end{equation*}\fi

For every $b>0$, at least two distinct $\delta_n$ in \hyperlink{eq2.31}{(2.31)} have $A_n>0$. Consequently $E_{\rm eff}'(t)<0$ for every finite $t>0$. This gives no uniform lower bound on the second weight or on a level separation. Proof in the Supplement, Appendix \crosspdflink{Entropy_Geometry_and_Lattice_Gaps_Supplement.pdf}{appA.32}{A.14}.\par

For $b/a\le1/648$, Proposition \hyperlink{res2.4}{2.4} and Corollary \hyperlink{res2.6}{2.6} give \hyperlink{eq2.16}{(2.16)}, including in infinite volume: the plaquette spectral measure is supported in $[8a/3,\infty)$, has weight greater than $1/60$ and first moment at most $2a/3$. This proves neither an atomic infinite-volume spectrum nor a uniform continuum limit. A single observable may miss lighter states; complete physical coverage is required as in Section \hyperlink{sec6.6}{VI F}.\par

The centered correlation realizes $\langle\phi(0,t)\phi(0,0)\rangle$ at the regulator; the uncentered expression adds $\mu(f)^2$. Here $t$ has inverse-energy units. For Euclidean duration $T$, the factor is $e^{-\delta_nT/\hbar}$; real time gives $e^{-i\delta_nT/\hbar}$. Equation \hyperlink{eq2.33}{(2.33)} describes Euclidean spectral filtering, not evaporation.\par

Continuum fields require renormalized test observables and domain control (Section \hyperlink{sec6.1}{VI A}), including justification of spatial smearing at sharp time. In units $\hbar=c=1$, an isolated positive mass can contribute $t^{-3/2}e^{-Mt}$ at a fixed spatial point in four dimensions because momentum is continuous [\hyperlink{ref99}{12}]. Appendix \crosspdflink{Entropy_Geometry_and_Lattice_Gaps_Supplement.pdf}{appA.31}{A.13} verifies the free-scalar example: a mass atom need not be an energy atom for a localized observable.\par

Thus the strong-coupling interval gives a physical gap and a nonzero finite-energy Wilson excitation at fixed spatial cutoff. In conventional Kogut-Susskind units, $b/a\propto g_0^{-4}$, so weak bare coupling leaves this interval. Spatial cutoff removal, spacetime covariance and a positive physical mass uniform along that trajectory remain unproved; Sections \hyperlink{sec6.2}{VI B}-\hyperlink{sec6.5}{VI E} give intermediate estimates.\par

\hypertarget{sec3}{}\section{Entropy, response and regional certificates}

\hypertarget{sec3.1}{}\subsection{Quantum entropy and physical information}

For a physical Gibbs state, a unitary perturbation injects energy $D(U\rho_\beta U^*\Vert\rho_\beta)/\beta$ [\hyperlink{ref65}{13}, Eqs. (2), (4)-(5)]. Theorem \crosspdflink{Entropy_Geometry_and_Lattice_Gaps_Supplement.pdf}{res3.20}{3.13} proves this identity with explicit operator domains and identifies its normalized vacuum Hessian with the physical energy form. In finite matrices, this Hessian is the BKM metric on the noncommuting tangent $-i[O,\rho_\beta]/\hbar$. The following theorem specializes to gauge-invariant phase pulses, complementing Theorem \crosspdflink{Entropy_Geometry_and_Lattice_Gaps_Supplement.pdf}{res3.15}{3.12}'s inverse-energy susceptibility identity.\par

\hypertarget{res3.1}{}\noindent \textbf{Theorem \hyperlink{res3.1}{3.1} (exact quantum phase cost for the gauge Hamiltonian).} Use Theorem \hyperlink{res2.2}{2.2}'s physical $H$, $K=H-E_0$, vacuum $\Omega$ and law $\mu$. Fix $\hbar>0$ and a smooth real dimensionless gauge-invariant $f$. Multiplication by $e^{-i\alpha f/\hbar}$ is a physical unitary, where $\alpha$ has action units. For every $\beta>0$, define
\sbox{\equationmeasure}{$\displaystyle\begin{gathered}\rho_\beta=\frac{e^{-\beta K}}{\operatorname{Tr}_{\rm phys}e^{-\beta K}},\\  U_{\alpha,f}=e^{-i\alpha f/\hbar},\\  \mathscr D_{\beta,f}(\alpha)=D(U_{\alpha,f}\rho_\beta U_{\alpha,f}^*\Vert\rho_\beta).\end{gathered}$}
\ifdim\wd\equationmeasure>\dimexpr\columnwidth-36pt\relax
\typeout{MANUSCRIPT-WIDE 3.1}\begin{manuscriptwide}\eqanchor{3.1}\begin{equation*}\tag{3.1}\begin{gathered}\rho_\beta=\frac{e^{-\beta K}}{\operatorname{Tr}_{\rm phys}e^{-\beta K}},\\  U_{\alpha,f}=e^{-i\alpha f/\hbar},\\  \mathscr D_{\beta,f}(\alpha)=D(U_{\alpha,f}\rho_\beta U_{\alpha,f}^*\Vert\rho_\beta).\end{gathered}\end{equation*}\end{manuscriptwide}
\else \eqanchor{3.1}\begin{equation*}\tag{3.1}\begin{gathered}\rho_\beta=\frac{e^{-\beta K}}{\operatorname{Tr}_{\rm phys}e^{-\beta K}},\\  U_{\alpha,f}=e^{-i\alpha f/\hbar},\\  \mathscr D_{\beta,f}(\alpha)=D(U_{\alpha,f}\rho_\beta U_{\alpha,f}^*\Vert\rho_\beta).\end{gathered}\end{equation*}\fi

With $M_{\Gamma_a(f,f)}$ denoting multiplication by the electric gradient density, the following identity holds at every pulse amplitude:
\sbox{\equationmeasure}{$\displaystyle\begin{gathered}\beta^{-1}\mathscr D_{\beta,f}(\alpha)=\frac{\alpha^2}{\hbar^2}\operatorname{Tr}\bigl(\rho_\beta M_{\Gamma_a(f,f)}\bigr)\longrightarrow\frac{\alpha^2}{\hbar^2}\mathcal E_\mu(f)\\ (\beta\to\infty).\end{gathered}$}
\ifdim\wd\equationmeasure>\dimexpr\columnwidth-36pt\relax
\typeout{MANUSCRIPT-WIDE 3.2}\begin{manuscriptwide}\eqanchor{3.2}\begin{equation*}\tag{3.2}\begin{gathered}\beta^{-1}\mathscr D_{\beta,f}(\alpha)=\frac{\alpha^2}{\hbar^2}\operatorname{Tr}\bigl(\rho_\beta M_{\Gamma_a(f,f)}\bigr)\longrightarrow\frac{\alpha^2}{\hbar^2}\mathcal E_\mu(f)\\ (\beta\to\infty).\end{gathered}\end{equation*}\end{manuscriptwide}
\else \eqanchor{3.2}\begin{equation*}\tag{3.2}\begin{gathered}\beta^{-1}\mathscr D_{\beta,f}(\alpha)=\frac{\alpha^2}{\hbar^2}\operatorname{Tr}\bigl(\rho_\beta M_{\Gamma_a(f,f)}\bigr)\longrightarrow\frac{\alpha^2}{\hbar^2}\mathcal E_\mu(f)\\ (\beta\to\infty).\end{gathered}\end{equation*}\fi

Consequently, writing $\mathfrak c_H(f)=\lim_{\beta\to\infty}\beta^{-1}\mathscr D_{\beta,f}''(0)$ and $\mathfrak j_\Omega(f)=\left.\partial_\alpha^2(1-|\langle\Omega,U_{\alpha,f}\Omega\rangle|^2)\right|_{\alpha=0}$,
\sbox{\equationmeasure}{$\displaystyle\begin{gathered}\mathfrak c_H(f)=\frac2{\hbar^2}\mathcal E_\mu(f),\\  \mathfrak j_\Omega(f)=\frac2{\hbar^2}\operatorname{Var}_\mu(f),\\  \Delta(H)=\inf_{\operatorname{Var}_\mu(f)>0}\frac{\mathfrak c_H(f)}{\mathfrak j_\Omega(f)}.\end{gathered}$}
\ifdim\wd\equationmeasure>\dimexpr\columnwidth-36pt\relax
\typeout{MANUSCRIPT-WIDE 3.3}\begin{manuscriptwide}\eqanchor{3.3}\begin{equation*}\tag{3.3}\begin{gathered}\mathfrak c_H(f)=\frac2{\hbar^2}\mathcal E_\mu(f),\\  \mathfrak j_\Omega(f)=\frac2{\hbar^2}\operatorname{Var}_\mu(f),\\  \Delta(H)=\inf_{\operatorname{Var}_\mu(f)>0}\frac{\mathfrak c_H(f)}{\mathfrak j_\Omega(f)}.\end{gathered}\end{equation*}\end{manuscriptwide}
\else \eqanchor{3.3}\begin{equation*}\tag{3.3}\begin{gathered}\mathfrak c_H(f)=\frac2{\hbar^2}\mathcal E_\mu(f),\\  \mathfrak j_\Omega(f)=\frac2{\hbar^2}\operatorname{Var}_\mu(f),\\  \Delta(H)=\inf_{\operatorname{Var}_\mu(f)>0}\frac{\mathfrak c_H(f)}{\mathfrak j_\Omega(f)}.\end{gathered}\end{equation*}\fi

The infimum is over all real smooth invariant $f$. Thus the physical gap is precisely uniform coercivity of quantum relative-entropy curvature against vacuum infidelity curvature. Both use the same actual quantum Hamiltonian. The factors of $\hbar$ cancel in this comparison; $\hbar$ already present in physical coefficients of $H$ is not removed. The pulse parameter is neither physical time nor a postulated discrete time step.\par

For every finite $SU(3)$ graph in Theorem \hyperlink{res2.3}{2.3} with $b/a\le1/648$, the established fixed-cutoff estimate gives
\sbox{\equationmeasure}{$\displaystyle\begin{gathered}\mathfrak c_{H_\Lambda}(f)\ge\frac{8a}{3}\mathfrak j_{\Omega_\Lambda}(f).\end{gathered}$}
\ifdim\wd\equationmeasure>\dimexpr\columnwidth-36pt\relax
\typeout{MANUSCRIPT-WIDE 3.4}\begin{manuscriptwide}\eqanchor{3.4}\begin{equation*}\tag{3.4}\begin{gathered}\mathfrak c_{H_\Lambda}(f)\ge\frac{8a}{3}\mathfrak j_{\Omega_\Lambda}(f).\end{gathered}\end{equation*}\end{manuscriptwide}
\else \eqanchor{3.4}\begin{equation*}\tag{3.4}\begin{gathered}\mathfrak c_{H_\Lambda}(f)\ge\frac{8a}{3}\mathfrak j_{\Omega_\Lambda}(f).\end{gathered}\end{equation*}\fi

Its constant is independent of volume in that strong-coupling regime. The continuum quantum theory and an estimate uniform along its physical cutoff-removal trajectory are still required; no infinite-volume Gibbs density operator is assumed here.\par

Proof in the Supplement, Appendix \crosspdflink{Entropy_Geometry_and_Lattice_Gaps_Supplement.pdf}{appA.26}{A.27}.\par

Corollary 12 of [\hyperlink{ref2}{1}] identifies the positive metric loss $G^\Lambda=g^{\rm BKM}-\Lambda^*g^{\rm BKM}$ as the Hessian of a channel's relative-entropy deficit. This comparison assigns no sign to a difference of Riemann curvatures. For a finite-matrix physical model, Corollary \crosspdflink{Entropy_Geometry_and_Lattice_Gaps_Supplement.pdf}{res3.22}{3.15} shows that a uniform lower bound on this loss, normalized by the physical vacuum-coherence weight of Proposition \crosspdflink{Entropy_Geometry_and_Lattice_Gaps_Supplement.pdf}{res3.21}{3.14}, certifies the energy gap. When each unitary path branch preserves the physical Gibbs state, the loss equals their averaged squared BKM tangent separation. A quantitative separation bound therefore supplies a precise conditional path-information criterion. If all branches are ordinary gauge transformations, their restrictions to the physical sector \hyperlink{eq2.3}{(2.3)} are the identity and the loss vanishes. A positive certificate therefore requires branches that act nontrivially on physical states.\par

The equal-weight branches $I$ and $2P_\Omega-I$ recover the exact gap quotient using the true physical vacuum projector. Their extension to heat-trace Hamiltonians uses a finite spectral form core, including for compact-link gauge operators; no extension of arbitrary channels is claimed. Source Hessians are taken at fixed temperature before the zero-temperature limit; these limits cannot generally be reversed. Discarded-record entropy does not measure injected energy. Physical calibration and positive coercivity uniform in volume and cutoff remain unproved for continuum Yang-Mills.\par

For a sequence of finite regulators, the target is a common lower bound $\mathfrak c_{H_n}(f)\ge m\mathfrak j_{\Omega_n}(f)$ with $m>0$ in fixed physical energy units, for every smooth real invariant $f$ at each regulator. Take the zero-temperature limit at each regulator first; a joint low-temperature and cutoff limit would need separate uniform thermal control. The resulting physical gaps give the contractions used by Theorem \hyperlink{res6.12}{6.27}, subject to its independent reconstruction, pairing-convergence and survival assumptions. After relativistic reconstruction, an energy gap $m$ corresponds to mass $m/c^2$; in units $c=1$ these coincide. This formulation uses quantum coherences without presuming that an action unit fixes $m$.\par

\subsubsection*{Configuration entropy and electric Fisher information}

Use Theorem \hyperlink{res2.2}{2.2}'s actual ground law $\mu$ and electric form $\mathcal E_\mu$. For a smooth real invariant source $f$, the exponential tilt $d\nu_s=e^{sf}d\mu/\int e^{sf}d\mu$ has relative entropy $D_f(s)=D(\nu_s\Vert\mu)$ and Hessian $D_f''(0)=\operatorname{Var}_\mu(f)$. Its electric Fisher information $I_\mu(r)=\int r\,\Gamma_a(\log r,\log r)d\mu$ satisfies $\partial_s^2 I_\mu(d\nu_s/d\mu)|_{s=0}=2\mathcal E_\mu(f)$. Thus the physical gap inequality is exactly the corresponding Hessian inequality on every smooth physical direction (Theorem \crosspdflink{Entropy_Geometry_and_Lattice_Gaps_Supplement.pdf}{res3.7}{3.6}). The gradients are configuration-space electric derivatives, not source-parameter derivatives; the tilted laws need not be ground laws of $H-sf$.\par

Finite source spaces approximate the gap from above when they exhaust the centered form domain in form norm (Proposition \crosspdflink{Entropy_Geometry_and_Lattice_Gaps_Supplement.pdf}{res3.9}{3.8}). A lower bound therefore requires control of omitted directions. The global Fisher-affinity alternative is given in Theorem \crosspdflink{Entropy_Geometry_and_Lattice_Gaps_Supplement.pdf}{res3.8}{3.7}; neither criterion asserts a logarithmic Sobolev inequality for arbitrary densities.\par

Shen, Zhu and Zhu establish log-Sobolev and Poincare inequalities for the infinite-volume Wilson lattice Gibbs measure, with exponential decay of correlations between separated smooth cylinder observables, under explicit strong-coupling assumptions [\hyperlink{ref105}{14}, Theorem 1.4 and Corollary 1.6]. Their Langevin dynamics uses an auxiliary time. The present law is $d\mu=\Omega^2dm$, whose diffusion semigroup is unitarily equivalent to $e^{-t(H-E_0)}$ in physical Euclidean time (Theorem \hyperlink{res2.2}{2.2}). Applying their estimates to this ground law requires a separate comparison.\par

\subsubsection*{Balanced records in physical Euclidean time}

We measure record loss under the physical Euclidean evolution of Theorem \hyperlink{res2.2}{2.2}. Its ground-state transform defines the reversible Markov semigroup
\sbox{\equationmeasure}{$\displaystyle\begin{gathered}P_t=W^{-1}e^{-t(H-E_0)}W=e^{tL},\\  Lf=\sum_ea_e\left(\Delta_ef+2\langle\nabla_e\log\Omega,\nabla_ef\rangle\right).\end{gathered}$}
\ifdim\wd\equationmeasure>\dimexpr\columnwidth-36pt\relax
\typeout{MANUSCRIPT-WIDE 3.5}\begin{manuscriptwide}\eqanchor{3.5}\begin{equation*}\tag{3.5}\begin{gathered}P_t=W^{-1}e^{-t(H-E_0)}W=e^{tL},\qquad Lf=\sum_ea_e\left(\Delta_ef+2\langle\nabla_e\log\Omega,\nabla_ef\rangle\right).\end{gathered}\end{equation*}\end{manuscriptwide}
\else \eqanchor{3.5}\begin{equation*}\tag{3.5}\begin{gathered}P_t=W^{-1}e^{-t(H-E_0)}W=e^{tL},\\  Lf=\sum_ea_e\left(\Delta_ef+2\langle\nabla_e\log\Omega,\nabla_ef\rangle\right).\end{gathered}\end{equation*}\fi

It preserves $\mu$, constants, and gauge invariance. Its time is the physical Euclidean time of $H$, in the units of Section \hyperlink{sec2.2}{II B}. For a real invariant $u$ with $\int u\,d\mu=0$ and $\|u\|_\infty\le1$, choose a balanced binary record $R\in\{-1,1\}$ with conditional configuration laws $d\mu_\pm=(1\pm u)d\mu$. Both are probability laws and their mixture is the stationary carrier $\mu$. Keeping $R$ fixed and evolving both laws by $P_t$ gives densities $1\pm P_tu$. With natural logarithms and $0\log0=0$, their mutual information is
\sbox{\equationmeasure}{$\displaystyle\begin{gathered}\mathscr I_\mu(u)=\int\beta(u)d\mu,\\  \beta(z)=\frac{(1+z)\log(1+z)+(1-z)\log(1-z)}2,\\  I(R:X_t)=\mathscr I_\mu(P_tu).\end{gathered}$}
\ifdim\wd\equationmeasure>\dimexpr\columnwidth-36pt\relax
\typeout{MANUSCRIPT-WIDE 3.6}\begin{manuscriptwide}\eqanchor{3.6}\begin{equation*}\tag{3.6}\begin{gathered}\mathscr I_\mu(u)=\int\beta(u)d\mu,\qquad \beta(z)=\frac{(1+z)\log(1+z)+(1-z)\log(1-z)}2,\qquad I(R:X_t)=\mathscr I_\mu(P_tu).\end{gathered}\end{equation*}\end{manuscriptwide}
\else \eqanchor{3.6}\begin{equation*}\tag{3.6}\begin{gathered}\mathscr I_\mu(u)=\int\beta(u)d\mu,\\  \beta(z)=\frac{(1+z)\log(1+z)+(1-z)\log(1-z)}2,\\  I(R:X_t)=\mathscr I_\mu(P_tu).\end{gathered}\end{equation*}\fi

\hypertarget{res3.2}{}\noindent \textbf{Theorem \hyperlink{res3.2}{3.2} (binary information, electric energy, and the physical gap).} For smooth real invariant $u$ with zero mean and $\|u\|_\infty<1$, set $u_t=P_tu$ and
\sbox{\equationmeasure}{$\displaystyle\begin{gathered}\Phi_{\pm,t}=\Omega\sqrt{1\pm u_t},\\  \overline E_t=\frac12\sum_{\pm}\big\|(H-E_0)^{1/2}\Phi_{\pm,t}\big\|^2.\end{gathered}$}
\ifdim\wd\equationmeasure>\dimexpr\columnwidth-36pt\relax
\typeout{MANUSCRIPT-WIDE 3.7}\begin{manuscriptwide}\eqanchor{3.7}\begin{equation*}\tag{3.7}\begin{gathered}\Phi_{\pm,t}=\Omega\sqrt{1\pm u_t},\qquad \overline E_t=\frac12\sum_{\pm}\big\|(H-E_0)^{1/2}\Phi_{\pm,t}\big\|^2.\end{gathered}\end{equation*}\end{manuscriptwide}
\else \eqanchor{3.7}\begin{equation*}\tag{3.7}\begin{gathered}\Phi_{\pm,t}=\Omega\sqrt{1\pm u_t},\\  \overline E_t=\frac12\sum_{\pm}\big\|(H-E_0)^{1/2}\Phi_{\pm,t}\big\|^2.\end{gathered}\end{equation*}\fi

These physical vectors are normalized, and
\sbox{\equationmeasure}{$\displaystyle\begin{gathered}-\frac{d}{dt}\mathscr I_\mu(u_t)=\int\frac{\Gamma_a(u_t,u_t)}{1-u_t^2}d\mu=4\overline E_t.\end{gathered}$}
\ifdim\wd\equationmeasure>\dimexpr\columnwidth-36pt\relax
\typeout{MANUSCRIPT-WIDE 3.8}\begin{manuscriptwide}\eqanchor{3.8}\begin{equation*}\tag{3.8}\begin{gathered}-\frac{d}{dt}\mathscr I_\mu(u_t)=\int\frac{\Gamma_a(u_t,u_t)}{1-u_t^2}d\mu=4\overline E_t.\end{gathered}\end{equation*}\end{manuscriptwide}
\else \eqanchor{3.8}\begin{equation*}\tag{3.8}\begin{gathered}-\frac{d}{dt}\mathscr I_\mu(u_t)=\int\frac{\Gamma_a(u_t,u_t)}{1-u_t^2}d\mu=4\overline E_t.\end{gathered}\end{equation*}\fi

For any $\gamma\ge0$, the following are equivalent: (i) $\Delta(H)\ge\gamma$; (ii) $\overline E_0\ge(\gamma/2)\mathscr I_\mu(u)$ for every such smooth $u$; (iii)
\sbox{\equationmeasure}{$\displaystyle\begin{gathered}\mathscr I_\mu(P_tu)\le e^{-2\gamma t}\mathscr I_\mu(u)\\ (t\ge0)\end{gathered}$}
\ifdim\wd\equationmeasure>\dimexpr\columnwidth-36pt\relax
\typeout{MANUSCRIPT-WIDE 3.9}\begin{manuscriptwide}\eqanchor{3.9}\begin{equation*}\tag{3.9}\begin{gathered}\mathscr I_\mu(P_tu)\le e^{-2\gamma t}\mathscr I_\mu(u)\qquad(t\ge0)\end{gathered}\end{equation*}\end{manuscriptwide}
\else \eqanchor{3.9}\begin{equation*}\tag{3.9}\begin{gathered}\mathscr I_\mu(P_tu)\le e^{-2\gamma t}\mathscr I_\mu(u)\\ (t\ge0)\end{gathered}\end{equation*}\fi

for every bounded centered real invariant $u$ with $\|u\|_\infty\le1$; (iv) the same inequality holds for all such smooth $u$ at one prescribed time $t_*>0$. In fact, for every $t>0$,
\sbox{\equationmeasure}{$\displaystyle\begin{gathered}\sup_{\substack{u\ \text{real, invariant},\ \mu(u)=0\\\|u\|_\infty\le1,\ \mathscr I_\mu(u)>0}}\frac{\mathscr I_\mu(P_tu)}{\mathscr I_\mu(u)}=e^{-2t\Delta(H)}.\end{gathered}$}
\ifdim\wd\equationmeasure>\dimexpr\columnwidth-36pt\relax
\typeout{MANUSCRIPT-WIDE 3.10}\begin{manuscriptwide}\eqanchor{3.10}\begin{equation*}\tag{3.10}\begin{gathered}\sup_{\substack{u\ \text{real, invariant},\ \mu(u)=0\\\|u\|_\infty\le1,\ \mathscr I_\mu(u)>0}}\frac{\mathscr I_\mu(P_tu)}{\mathscr I_\mu(u)}=e^{-2t\Delta(H)}.\end{gathered}\end{equation*}\end{manuscriptwide}
\else \eqanchor{3.10}\begin{equation*}\tag{3.10}\begin{gathered}\sup_{\substack{u\ \text{real, invariant},\ \mu(u)=0\\\|u\|_\infty\le1,\ \mathscr I_\mu(u)>0}}\frac{\mathscr I_\mu(P_tu)}{\mathscr I_\mu(u)}=e^{-2t\Delta(H)}.\end{gathered}\end{equation*}\fi

The supremum is unchanged if restricted to smooth $u$ with $\|u\|_\infty<1$. For any smooth centered real invariant $f$,
\sbox{\equationmeasure}{$\displaystyle\begin{gathered}\begin{aligned}
\left.\frac{d^2}{d\epsilon^2}\mathscr I_\mu(\epsilon P_tf)\right|_{\epsilon=0}
&=\|P_tf\|_{L^2(\mu)}^2\\
&=\langle Wf,e^{-2t(H-E_0)}Wf\rangle.
\end{aligned}\end{gathered}$}
\ifdim\wd\equationmeasure>\dimexpr\columnwidth-36pt\relax
\typeout{MANUSCRIPT-WIDE 3.11}\begin{manuscriptwide}[28]\eqanchor{3.11}\begin{equation*}\tag{3.11}\begin{gathered}\begin{aligned}
\left.\frac{d^2}{d\epsilon^2}\mathscr I_\mu(\epsilon P_tf)\right|_{\epsilon=0}
&=\|P_tf\|_{L^2(\mu)}^2\\
&=\langle Wf,e^{-2t(H-E_0)}Wf\rangle.
\end{aligned}\end{gathered}\end{equation*}\end{manuscriptwide}
\else \eqanchor{3.11}\begin{equation*}\tag{3.11}\begin{gathered}\begin{aligned}
\left.\frac{d^2}{d\epsilon^2}\mathscr I_\mu(\epsilon P_tf)\right|_{\epsilon=0}
&=\|P_tf\|_{L^2(\mu)}^2\\
&=\langle Wf,e^{-2t(H-E_0)}Wf\rangle.
\end{aligned}\end{gathered}\end{equation*}\fi

Proof in the Supplement, Appendix \crosspdflink{Entropy_Geometry_and_Lattice_Gaps_Supplement.pdf}{appA.15}{A.18}.\par

For every finite box covered by Theorem \hyperlink{res2.3}{2.3}, the proved bound $\Delta(H)\ge8a/3$ gives $\mathscr I_\mu(P_tu)\le e^{-16at/3}\mathscr I_\mu(u)$ and $\overline E_0\ge(4a/3)\mathscr I_\mu(u)$ when $b/a\le1/648$. These constants are independent of the box volume. This consequence uses each finite-box ground-state law; no thermodynamic configuration path measure is asserted.\par

The stationary-mixture criterion has optimal constants $\gamma=\Delta(H)$ in \hyperlink{eq3.9}{(3.9)} and $c=\Delta(H)/2$ in $\overline E_0\ge c\mathscr I_\mu(u)$. It asserts no logarithmic Sobolev inequality for arbitrary densities. Its dynamics and electric form use the interacting ground law; an unrelated observation channel cannot replace $P_t$.\par

The square-root lifts \hyperlink{eq3.7}{(3.7)} are generally not linear Euclidean or real-time evolutions of $\Phi_{\pm,0}$, and their excitation energy is not apparatus work. Vacuum components let energy and information vanish with the bias; the positive floor applies to the vacuum complement. Equation \hyperlink{eq3.11}{(3.11)} pairs binary time $t$ with reflected separation $2t$. A common small-bias contraction on a linear $L^2$-dense physical family suffices for the converse gap implication; a finite archive leaves omitted modes uncontrolled. No Yang-Mills continuum contraction is established.\par

\hypertarget{sec3.2}{}\subsection{Retained response and regional currents}

A finite observable family controls only its retained span. The sharp two-block and many-block comparisons (Theorems \crosspdflink{Entropy_Geometry_and_Lattice_Gaps_Supplement.pdf}{res3.11}{3.9} and \crosspdflink{Entropy_Geometry_and_Lattice_Gaps_Supplement.pdf}{res3.12}{3.10}; see also [\hyperlink{ref8}{15}, \hyperlink{ref9}{16}]) offer alternatives whose coupling bounds also require proof. Energy normalization is independent: Hamiltonian rescaling preserves the ground law and static entropy geometry but changes the gap.\par

\subsubsection*{Physical susceptibility and omitted modes}

A physical-source response weights soft excitations more strongly than ordinary covariance. Let $K\ge0$ be self-adjoint, $K\Omega=0$, $\|\Omega\|=1$, and $\mathcal H_0=\Omega^\perp$. No uniqueness of the vacuum is assumed. For $v\in\mathcal H_0$ define the extended inverse form
\sbox{\equationmeasure}{$\displaystyle\begin{gathered}\begin{aligned}
\chi_K(v)&:=2\lim_{\varepsilon\downarrow0}
\langle v,(K+\varepsilon)^{-1}v\rangle\\
&=2\int_0^\infty\langle v,e^{-tK}v\rangle\,dt\\
&\in[0,\infty].
\end{aligned}\end{gathered}$}
\ifdim\wd\equationmeasure>\dimexpr\columnwidth-36pt\relax
\typeout{MANUSCRIPT-WIDE 3.12}\begin{manuscriptwide}\eqanchor{3.12}\begin{equation*}\tag{3.12}\begin{gathered}\begin{aligned}
\chi_K(v)&:=2\lim_{\varepsilon\downarrow0}
\langle v,(K+\varepsilon)^{-1}v\rangle\\
&=2\int_0^\infty\langle v,e^{-tK}v\rangle\,dt\\
&\in[0,\infty].
\end{aligned}\end{gathered}\end{equation*}\end{manuscriptwide}
\else \eqanchor{3.12}\begin{equation*}\tag{3.12}\begin{gathered}\begin{aligned}
\chi_K(v)&:=2\lim_{\varepsilon\downarrow0}
\langle v,(K+\varepsilon)^{-1}v\rangle\\
&=2\int_0^\infty\langle v,e^{-tK}v\rangle\,dt\\
&\in[0,\infty].
\end{aligned}\end{gathered}\end{equation*}\fi

Thus a nonzero zero-mode component gives infinite response. If the Hilbert space is finite dimensional, $K$ has a unique vacuum and $O$ is Hermitian, define $e_O(\lambda)=\min\operatorname{spec}(K-\lambda O)$ near zero. For $v=(I-P_\Omega)O\Omega$, the zero-source perturbation identity proved in Appendix \crosspdflink{Entropy_Geometry_and_Lattice_Gaps_Supplement.pdf}{appA.20}{A.21} gives $\chi_K(v)=-e_O''(0)$, with $\|v\|^2=\operatorname{Var}_\Omega(O)$, without a response-bound hypothesis. This compares the physical response Hessian with covariance, rather than replacing response by covariance curvature.\par

\hypertarget{res3.3}{}\noindent \textbf{Theorem \hyperlink{res3.3}{3.3} (normalized response criterion).} For $C>0$, the following are equivalent: (i) $K\ge C^{-1}(I-P_\Omega)$ as forms; (ii) $\chi_K(v)\le2C\|v\|^2$ for every $v\in\mathcal H_0$; (iii) the same inequality holds on a complex linear dense subspace of $\mathcal H_0$. Each condition implies $\ker K=\mathbb C\Omega$. When $\mathcal H_0\ne\{0\}$, write $\Delta=\inf\operatorname{spec}(K|_{\mathcal H_0})$. The optimal response constant is $1/\Delta$, interpreted as infinity when $\Delta=0$.\par

Proof in the Supplement, Appendix \crosspdflink{Entropy_Geometry_and_Lattice_Gaps_Supplement.pdf}{appA.20}{A.21}. This zero-source criterion requires neither a tilted-interval response bound nor retained flux. On a complete family it is equivalent to the spectral gap.\par

\hypertarget{res3.4}{}\noindent \textbf{Theorem \hyperlink{res3.4}{3.4} (retained response with an omitted-mode bound).} Let $S\subset\mathcal H_0$ be a closed subspace, with orthogonal projection $P_S$. Suppose $C>0$, $a\ge0$ and $0\le b<1$ satisfy
\sbox{\equationmeasure}{$\displaystyle\begin{gathered}\chi_K(v)\le2C\|v\|^2\ (v\in S),\\  \|(I-P_S)\psi\|^2\\ \le a\|K^{1/2}\psi\|^2+b\|\psi\|^2\\ (\psi\in D(K^{1/2})\cap\mathcal H_0).\end{gathered}$}
\ifdim\wd\equationmeasure>\dimexpr\columnwidth-36pt\relax
\typeout{MANUSCRIPT-WIDE 3.13}\begin{manuscriptwide}\eqanchor{3.13}\begin{equation*}\tag{3.13}\begin{gathered}\chi_K(v)\le2C\|v\|^2\ (v\in S),\qquad \|(I-P_S)\psi\|^2\le a\|K^{1/2}\psi\|^2+b\|\psi\|^2\ (\psi\in D(K^{1/2})\cap\mathcal H_0).\end{gathered}\end{equation*}\end{manuscriptwide}
\else \eqanchor{3.13}\begin{equation*}\tag{3.13}\begin{gathered}\chi_K(v)\le2C\|v\|^2\ (v\in S),\\  \|(I-P_S)\psi\|^2\\ \le a\|K^{1/2}\psi\|^2+b\|\psi\|^2\\ (\psi\in D(K^{1/2})\cap\mathcal H_0).\end{gathered}\end{equation*}\fi

Then the vacuum is unique and
\sbox{\equationmeasure}{$\displaystyle\begin{gathered}K\ge\frac{1-b}{C+a}(I-P_\Omega).\end{gathered}$}
\ifdim\wd\equationmeasure>\dimexpr\columnwidth-36pt\relax
\typeout{MANUSCRIPT-WIDE 3.14}\begin{manuscriptwide}\eqanchor{3.14}\begin{equation*}\tag{3.14}\begin{gathered}K\ge\frac{1-b}{C+a}(I-P_\Omega).\end{gathered}\end{equation*}\end{manuscriptwide}
\else \eqanchor{3.14}\begin{equation*}\tag{3.14}\begin{gathered}K\ge\frac{1-b}{C+a}(I-P_\Omega).\end{gathered}\end{equation*}\fi

The projection need not preserve the form domain. For a finite family $v_1,\ldots,v_N$ spanning $S$, finite individual responses define the inverse-form matrix $X$, and the first assumption is exactly
\sbox{\equationmeasure}{$\displaystyle\begin{gathered}G_{ij}=\langle v_i,v_j\rangle,\\  X_{ij}=2\langle K^{-1/2}v_i,K^{-1/2}v_j\rangle,\\  X\preceq2CG.\end{gathered}$}
\ifdim\wd\equationmeasure>\dimexpr\columnwidth-36pt\relax
\typeout{MANUSCRIPT-WIDE 3.15}\begin{manuscriptwide}\eqanchor{3.15}\begin{equation*}\tag{3.15}\begin{gathered}G_{ij}=\langle v_i,v_j\rangle,\quad X_{ij}=2\langle K^{-1/2}v_i,K^{-1/2}v_j\rangle,\qquad X\preceq2CG.\end{gathered}\end{equation*}\end{manuscriptwide}
\else \eqanchor{3.15}\begin{equation*}\tag{3.15}\begin{gathered}G_{ij}=\langle v_i,v_j\rangle,\\  X_{ij}=2\langle K^{-1/2}v_i,K^{-1/2}v_j\rangle,\\  X\preceq2CG.\end{gathered}\end{equation*}\fi

Here the inverse acts on $(\ker K)^\perp$; finiteness in \hyperlink{eq3.12}{(3.12)} ensures that each $v_i$ lies there and in its inverse-form domain. Matrix order includes all complex combinations and handles dependent probes through the Gram null space.\par

Proof in the Supplement, Appendix \crosspdflink{Entropy_Geometry_and_Lattice_Gaps_Supplement.pdf}{appA.21}{A.22}. The omitted-mode estimate is a separate physical assumption, not a consequence of a finite response matrix. In regulator applications the retained family may grow with volume and resolution. Constants $C,a$ have inverse-energy units and must be uniform in physical units.\par

Even a bounded source cyclic for the excited spectral algebra can have finite susceptibility in a gapless system (Proposition \crosspdflink{Entropy_Geometry_and_Lattice_Gaps_Supplement.pdf}{res3.14}{3.11}). Density obtained by spectral filtering supplies no uniform response bound; the regional criterion makes the additional estimates explicit.\par

\subsubsection*{A finite regional certificate}

The omitted-mode assumption can be derived from local conditional inequalities and an energy estimate on an exceptional set. Retained response can in turn be certified by finitely many vector fields. Use Theorem \hyperlink{res2.2}{2.2}'s actual ground law on $M=G^{\mathsf E}$, its physical space $\mathcal H_\mu=L^2(\mu)^K$, and its smooth invariant form core $\mathscr C$. Write $Df=(\sqrt{a_e}\nabla_e f)_e$, with its closed physical realization, and let $\mathscr Z$ be the Hilbert space of square-integrable edge-tangent fields. Then $\mathcal E_\mu(f)=\|Df\|_{\mathscr Z}^2$ and $A_\mu=D^*D=W^{-1}(H-E_0)W$, where $Wf=\Omega f$.\par

The result also holds in any specified probability-space ground-state realization: $W$ is unitary from a closed physical subspace of $L^2(\mu)$, $W1=\Omega$, and $W^{-1}(H-E_0)W=D^*D$, where $D$ is closed into an $L^2$ space $\mathscr Z$ of pointwise gradient fields, $\Gamma_a(f,f)=|Df|^2$, and its form core $\mathscr C$ contains constants. Cell indicators must lie in that physical subspace; every local integral and weak identity must use this same realization. These structural assumptions, rather than unitary equivalence alone, preserve the regional argument.\par

\hypertarget{res3.5}{}\noindent \textbf{Theorem \hyperlink{res3.5}{3.5} (regional current certificate).} Let $D_1,\ldots,D_N$ be pairwise disjoint measurable gauge-invariant configuration cells with $p_i=\mu(D_i)>0$. Put $U=\bigcup_iD_i$, $u_i=1_{D_i}-p_i$, $S=\operatorname{span}_{\mathbb C}\{u_i\}$, and let $P_S$ be its orthogonal projection. Set $\bar f_i=p_i^{-1}\int_{D_i}f\,d\mu$. Suppose $C_{\rm loc},C_{\rm resp}>0$, $a_0\ge0$ and $0\le b_0<1$ satisfy
\sbox{\equationmeasure}{$\displaystyle\begin{gathered}\begin{gathered}
\int_{D_i}|f-\bar f_i|^2d\mu\le C_{\rm loc}\int_{D_i}\Gamma_a(f,f)d\mu\\
(f\in\mathscr C),\\
\int_{U^c}|f|^2d\mu\le a_0\mathcal E_\mu(f)+b_0\|f\|_2^2\\
(f\in\mathscr C,\ \mu(f)=0).
\end{gathered}\end{gathered}$}
\ifdim\wd\equationmeasure>\dimexpr\columnwidth-36pt\relax
\typeout{MANUSCRIPT-WIDE 3.16}\begin{manuscriptwide}\eqanchor{3.16}\begin{equation*}\tag{3.16}\begin{gathered}\begin{gathered}
\int_{D_i}|f-\bar f_i|^2d\mu\le C_{\rm loc}\int_{D_i}\Gamma_a(f,f)d\mu\\
(f\in\mathscr C),\\
\int_{U^c}|f|^2d\mu\le a_0\mathcal E_\mu(f)+b_0\|f\|_2^2\\
(f\in\mathscr C,\ \mu(f)=0).
\end{gathered}\end{gathered}\end{equation*}\end{manuscriptwide}
\else \eqanchor{3.16}\begin{equation*}\tag{3.16}\begin{gathered}\begin{gathered}
\int_{D_i}|f-\bar f_i|^2d\mu\le C_{\rm loc}\int_{D_i}\Gamma_a(f,f)d\mu\\
(f\in\mathscr C),\\
\int_{U^c}|f|^2d\mu\le a_0\mathcal E_\mu(f)+b_0\|f\|_2^2\\
(f\in\mathscr C,\ \mu(f)=0).
\end{gathered}\end{gathered}\end{equation*}\fi

Supply $J_1,\ldots,J_N\in\mathscr Z$ satisfying the global weak identities and finite Hermitian matrix inequality
\sbox{\equationmeasure}{$\displaystyle\begin{gathered}\begin{gathered}
\langle u_i,f\rangle_\mu=\langle J_i,Df\rangle_{\mathscr Z}\quad(f\in\mathscr C),\\
R_{ij}=\langle J_i,J_j\rangle_{\mathscr Z},\\
\mathsf G_{ij}=\langle u_i,u_j\rangle_\mu=p_i\delta_{ij}-p_ip_j,\\
R\preceq C_{\rm resp}\mathsf G.
\end{gathered}\end{gathered}$}
\ifdim\wd\equationmeasure>\dimexpr\columnwidth-36pt\relax
\typeout{MANUSCRIPT-WIDE 3.17}\begin{manuscriptwide}\eqanchor{3.17}\begin{equation*}\tag{3.17}\begin{gathered}\begin{gathered}
\langle u_i,f\rangle_\mu=\langle J_i,Df\rangle_{\mathscr Z}\quad(f\in\mathscr C),\\
R_{ij}=\langle J_i,J_j\rangle_{\mathscr Z},\\
\mathsf G_{ij}=\langle u_i,u_j\rangle_\mu=p_i\delta_{ij}-p_ip_j,\\
R\preceq C_{\rm resp}\mathsf G.
\end{gathered}\end{gathered}\end{equation*}\end{manuscriptwide}
\else \eqanchor{3.17}\begin{equation*}\tag{3.17}\begin{gathered}\begin{gathered}
\langle u_i,f\rangle_\mu=\langle J_i,Df\rangle_{\mathscr Z}\quad(f\in\mathscr C),\\
R_{ij}=\langle J_i,J_j\rangle_{\mathscr Z},\\
\mathsf G_{ij}=\langle u_i,u_j\rangle_\mu=p_i\delta_{ij}-p_ip_j,\\
R\preceq C_{\rm resp}\mathsf G.
\end{gathered}\end{gathered}\end{equation*}\fi

Then every centered physical form-domain vector satisfies the first bound below, and every $v\in S$ satisfies the second:
\sbox{\equationmeasure}{$\displaystyle\begin{gathered}\|(I-P_S)f\|_2^2\le(a_0+C_{\rm loc})\mathcal E_\mu(f)+b_0\|f\|_2^2,\\  \chi_{A_\mu}(v)\le2C_{\rm resp}\|v\|_2^2.\end{gathered}$}
\ifdim\wd\equationmeasure>\dimexpr\columnwidth-36pt\relax
\typeout{MANUSCRIPT-WIDE 3.18}\begin{manuscriptwide}\eqanchor{3.18}\begin{equation*}\tag{3.18}\begin{gathered}\|(I-P_S)f\|_2^2\le(a_0+C_{\rm loc})\mathcal E_\mu(f)+b_0\|f\|_2^2,\qquad \chi_{A_\mu}(v)\le2C_{\rm resp}\|v\|_2^2.\end{gathered}\end{equation*}\end{manuscriptwide}
\else \eqanchor{3.18}\begin{equation*}\tag{3.18}\begin{gathered}\|(I-P_S)f\|_2^2\le(a_0+C_{\rm loc})\mathcal E_\mu(f)+b_0\|f\|_2^2,\\  \chi_{A_\mu}(v)\le2C_{\rm resp}\|v\|_2^2.\end{gathered}\end{equation*}\fi

Consequently, on the physical invariant space,
\sbox{\equationmeasure}{$\displaystyle\begin{gathered}H-E_0\ge\frac{1-b_0}{a_0+C_{\rm loc}+C_{\rm resp}}(I-P_\Omega).\end{gathered}$}
\ifdim\wd\equationmeasure>\dimexpr\columnwidth-36pt\relax
\typeout{MANUSCRIPT-WIDE 3.19}\begin{manuscriptwide}\eqanchor{3.19}\begin{equation*}\tag{3.19}\begin{gathered}H-E_0\ge\frac{1-b_0}{a_0+C_{\rm loc}+C_{\rm resp}}(I-P_\Omega).\end{gathered}\end{equation*}\end{manuscriptwide}
\else \eqanchor{3.19}\begin{equation*}\tag{3.19}\begin{gathered}H-E_0\ge\frac{1-b_0}{a_0+C_{\rm loc}+C_{\rm resp}}(I-P_\Omega).\end{gathered}\end{equation*}\fi

Proof in the Supplement, Appendix \crosspdflink{Entropy_Geometry_and_Lattice_Gaps_Supplement.pdf}{appA.23}{A.24}. The matrix inequality includes every complex combination, including dependent indicators; the indicator projection need not preserve the form domain. Local-cell control, exceptional-set energy and finitely many global weak identities give both inputs to Theorem \hyperlink{res3.4}{3.4}. Proposition \hyperlink{res4.5}{4.6} supplies a test for the exceptional estimate.\par

A reference conditional law $\nu_i$ on $D_i$ suffices if its Poincare constant $\lambda_i>0$ uses the same restricted electric gradient and $d\mu(\cdot\mid D_i)=r_i\,d\nu_i$, with $0<\operatorname*{ess\,inf}r_i\le\operatorname*{ess\,sup}r_i<\infty$. The bounded-density argument in Supplement equation \crosspdflink{Entropy_Geometry_and_Lattice_Gaps_Supplement.pdf}{eqA.7}{(A.7)} gives $C_{\rm loc}=\max_i[(\operatorname*{ess\,sup}r_i/\operatorname*{ess\,inf}r_i)/\lambda_i]$. These inequalities must hold on restrictions of the physical core. Disconnected cells need a valid conditional inequality or further labels; diameter alone is insufficient.\par

The $J_i$ are duals of the configuration-space electric gradient, not spacetime chromoelectric tubes. Their identities hold on the global invariant core, including transport across artificial cell interfaces. Separate no-flux cell equations or independent singlet projections on spatial boundaries do not establish \hyperlink{eq3.17}{(3.17)}. Detectors built from continuous spatial regions may select these configuration cells; the two kinds of region remain distinct.\par

In the fixed compact setting, the minimum squared current norm for a centered source $v$ is $\langle v,A_\mu^{-1}v\rangle=\chi_{A_\mu}(v)/2$, attained by $DA_\mu^{-1}v$. Every other weak solution differs by a vector orthogonal to the range of $D$. Uniform certification still requires an independent current construction and norm bound.\par

\subsubsection*{Gibbs source response}

For the actual physical Gibbs family $\rho_{\beta,h}=e^{-\beta(K-hO)}/\operatorname{Tr}e^{-\beta(K-hO)}$, Theorem \crosspdflink{Entropy_Geometry_and_Lattice_Gaps_Supplement.pdf}{res3.15}{3.12} identifies a normalized log-Euclidean Renyi derivative, for $h\ne0$, with $\mathcal R_\beta(h)=\partial_h\operatorname{Tr}(\rho_{\beta,h}O)=\beta\operatorname{Cov}^{\rm KM}_{\rho_{\beta,h}}(O,O)$ [\hyperlink{ref60}{17}, \hyperlink{ref62}{18}]. Its hypotheses are $K\ge0$ with unique normalized vacuum $\Omega$, a Hilbert space of dimension greater than one, $\operatorname{Tr}e^{-\beta_0K}<\infty$ for some $\beta_0>0$, bounded Hermitian $O$, and $\beta\ge\beta_0$. The ordered limit is $\lim_{\beta\to\infty}\lim_{h\to0}\mathcal R_\beta(h)=\chi_K(v)$, where $v=(O-\langle\Omega,O\Omega\rangle I)\Omega$.\par

A common bound by $2C\|v\|^2$ supplies the response criterion only for the vectors represented, including every complex combination in the required dense physical family. For the real gauge-configuration Hamiltonians, real multiplication sources suffice when their centered vectors are dense in the real physical space, because the response and norm split under complex conjugation. The Gibbs identity supplies no regulator-uniform thermal estimate, infinite-volume Gibbs state, or continuum construction; physical normalization, pairing convergence and reflected survival remain separate requirements.\par

\hypertarget{sec4}{}\section{Screening and control across scales}

The regional criteria require bounds on the actual ground law. Central-slice approximation accesses that law; screening and exceptional-set estimates control its fluctuations, and the multiscale criterion tracks iteration losses. All comparisons use the specified electric form.\par

\hypertarget{sec4.1}{}\subsection{Accessing the actual spatial ground law}

\hypertarget{res4.1}{}\noindent \textbf{Proposition \hyperlink{res4.1}{4.1} (central-slice approximation).} At a fixed finite regulator of Theorem \hyperlink{res2.2}{2.2}, put
\sbox{\equationmeasure}{$\displaystyle\begin{gathered}v_T=\frac{e^{-TH}1}{\|e^{-TH}1\|_{L^2(m)}},\\  d\mu_T=v_T^2dm,\\  T>0.\end{gathered}$}
\ifdim\wd\equationmeasure>\dimexpr\columnwidth-36pt\relax
\typeout{MANUSCRIPT-WIDE 4.1}\begin{manuscriptwide}\eqanchor{4.1}\begin{equation*}\tag{4.1}\begin{gathered}v_T=\frac{e^{-TH}1}{\|e^{-TH}1\|_{L^2(m)}},\qquad d\mu_T=v_T^2dm,\qquad T>0.\end{gathered}\end{equation*}\end{manuscriptwide}
\else \eqanchor{4.1}\begin{equation*}\tag{4.1}\begin{gathered}v_T=\frac{e^{-TH}1}{\|e^{-TH}1\|_{L^2(m)}},\\  d\mu_T=v_T^2dm,\\  T>0.\end{gathered}\end{equation*}\fi

Then $v_T$ is positive, smooth and invariant, and $v_T\to\Omega$ in every $C^k$ norm as $T\to\infty$. The law $\mu_T$ is the central-slice marginal of a symmetric Euclidean slab of half-height $T$ with constant endpoint states. If, for some sequence $T_j\to\infty$ and one $\gamma>0$,
\sbox{\equationmeasure}{$\displaystyle\begin{gathered}\operatorname{Var}_{\mu_{T_j}}(f)\le\gamma^{-1}\int\Gamma_a(f,f)d\mu_{T_j}\end{gathered}$}
\ifdim\wd\equationmeasure>\dimexpr\columnwidth-36pt\relax
\typeout{MANUSCRIPT-WIDE 4.2}\begin{manuscriptwide}\eqanchor{4.2}\begin{equation*}\tag{4.2}\begin{gathered}\operatorname{Var}_{\mu_{T_j}}(f)\le\gamma^{-1}\int\Gamma_a(f,f)d\mu_{T_j}\end{gathered}\end{equation*}\end{manuscriptwide}
\else \eqanchor{4.2}\begin{equation*}\tag{4.2}\begin{gathered}\operatorname{Var}_{\mu_{T_j}}(f)\le\gamma^{-1}\int\Gamma_a(f,f)d\mu_{T_j}\end{gathered}\end{equation*}\fi

holds for every smooth invariant $f$, then $\Delta(H)\ge\gamma$.\par

Proof in the Supplement, Appendix \crosspdflink{Entropy_Geometry_and_Lattice_Gaps_Supplement.pdf}{appB.1}{B.1}.\par

Proposition \hyperlink{res4.1}{4.1} makes the actual transfer dynamics' central-slice laws a screening target. A common bound along $T_j\to\infty$ passes to each regulator's ground law without a cutoff-uniform convergence rate. Continuum transfer additionally requires positive $\gamma$ in common physical units, complete-sector coverage, reconstruction and observable survival (Sections \hyperlink{sec6.1}{VI A}, \hyperlink{sec6.5}{VI E}, \hyperlink{sec6.6}{VI F}). Neither \hyperlink{eq4.16}{(4.16)} nor \hyperlink{eq4.2}{(4.2)} is proved along the Yang-Mills continuum trajectory.\par

\hypertarget{sec4.2}{}\subsection{Screening and conditional Fisher information}

Theorem \hyperlink{res3.2}{3.2} needs a Poincare inequality, without requiring positive weighted curvature or a full logarithmic Sobolev inequality. Spatial mixing [\hyperlink{ref50}{19}, Theorem 2.3] and coupling-based curvature [\hyperlink{ref100}{20}, Theorem 2] provide established entropy-factorization criteria. Applying them requires bounds on the actual conditional laws or Markov kernels; BKM state-space curvature alone supplies neither. The following projection argument states the buffer-screening bound and electric-form comparison needed here.\par

Let $\mu$ be a probability law and $\mathcal F_A,\mathcal F_B$ two sub-$\sigma$-algebras, interpreted as the variables outside regions $A,B$. Write
\sbox{\equationmeasure}{$\displaystyle\begin{gathered}P_Af=\mathbb E_\mu[f\mid\mathcal F_A],\\  P_Bf=\mathbb E_\mu[f\mid\mathcal F_B],\\  \Pi f=\mu(f)1,\\  V_D(f)=\|f-P_Df\|_2^2.\end{gathered}$}
\ifdim\wd\equationmeasure>\dimexpr\columnwidth-36pt\relax
\typeout{MANUSCRIPT-WIDE 4.3}\begin{manuscriptwide}\eqanchor{4.3}\begin{equation*}\tag{4.3}\begin{gathered}P_Af=\mathbb E_\mu[f\mid\mathcal F_A],\quad P_Bf=\mathbb E_\mu[f\mid\mathcal F_B],\quad \Pi f=\mu(f)1,\quad V_D(f)=\|f-P_Df\|_2^2.\end{gathered}\end{equation*}\end{manuscriptwide}
\else \eqanchor{4.3}\begin{equation*}\tag{4.3}\begin{gathered}P_Af=\mathbb E_\mu[f\mid\mathcal F_A],\\  P_Bf=\mathbb E_\mu[f\mid\mathcal F_B],\\  \Pi f=\mu(f)1,\\  V_D(f)=\|f-P_Df\|_2^2.\end{gathered}\end{equation*}\fi

\hypertarget{res4.2}{}\noindent \textbf{Theorem \hyperlink{res4.2}{4.2} (projection screening).} On $L^2(\mu)$, or a closed subspace containing constants and preserved by both projections, suppose $q=\|P_AP_B-\Pi\|<1$, with the norm on the specified space. Then
\sbox{\equationmeasure}{$\displaystyle\begin{gathered}\operatorname{Var}_\mu(f)\le\frac{V_A(f)+V_B(f)}{1-q}.\end{gathered}$}
\ifdim\wd\equationmeasure>\dimexpr\columnwidth-36pt\relax
\typeout{MANUSCRIPT-WIDE 4.4}\begin{manuscriptwide}\eqanchor{4.4}\begin{equation*}\tag{4.4}\begin{gathered}\operatorname{Var}_\mu(f)\le\frac{V_A(f)+V_B(f)}{1-q}.\end{gathered}\end{equation*}\end{manuscriptwide}
\else \eqanchor{4.4}\begin{equation*}\tag{4.4}\begin{gathered}\operatorname{Var}_\mu(f)\le\frac{V_A(f)+V_B(f)}{1-q}.\end{gathered}\end{equation*}\fi

Proof in the Supplement, Appendix \crosspdflink{Entropy_Geometry_and_Lattice_Gaps_Supplement.pdf}{appB.2}{B.2}.\par

The exterior variables must not contain an uncontrolled common variable: any nonconstant $f$ fixed by both projections forces $q=1$. If a shared algebra $\mathcal C$ is retained, assuming $\|P_AP_B-P_{\mathcal C}\|\le q<1$ and $\mathcal C\subset\mathcal F_A\cap\mathcal F_B$ yields only a bound for $\|f-P_{\mathcal C}f\|_2^2$, provided $P_{\mathcal C}$ also preserves the specified space. Fluctuations in $\mathcal C$ still require control. In a gauge application, conditional projections must preserve the chosen physical space. Independent singlet projection on artificial boundaries can delete crossing Wilson observables; it is not conditional screening.\par

\hypertarget{res4.3}{}\noindent \textbf{Corollary \hyperlink{res4.3}{4.3} (electric comparison).} Use Theorem \hyperlink{res2.2}{2.2}'s ground law and physical space. Suppose \hyperlink{eq4.4}{(4.4)} holds there and, on its smooth invariant core,
\sbox{\equationmeasure}{$\displaystyle\begin{gathered}V_D(f)\\ \le\lambda_D^{-1}\mathcal E_D(f)\ (D=A,B),\\  \mathcal E_A(f)+\mathcal E_B(f)\\ \le M\mathcal E_\mu(f),\\  \lambda_A,\lambda_B,M>0.\end{gathered}$}
\ifdim\wd\equationmeasure>\dimexpr\columnwidth-36pt\relax
\typeout{MANUSCRIPT-WIDE 4.5}\begin{manuscriptwide}\eqanchor{4.5}\begin{equation*}\tag{4.5}\begin{gathered}V_D(f)\le\lambda_D^{-1}\mathcal E_D(f)\ (D=A,B),\qquad \mathcal E_A(f)+\mathcal E_B(f)\le M\mathcal E_\mu(f),\quad \lambda_A,\lambda_B,M>0.\end{gathered}\end{equation*}\end{manuscriptwide}
\else \eqanchor{4.5}\begin{equation*}\tag{4.5}\begin{gathered}V_D(f)\\ \le\lambda_D^{-1}\mathcal E_D(f)\ (D=A,B),\\  \mathcal E_A(f)+\mathcal E_B(f)\\ \le M\mathcal E_\mu(f),\\  \lambda_A,\lambda_B,M>0.\end{gathered}\end{equation*}\fi

Then $\Delta(H)\ge(1-q)\min(\lambda_A,\lambda_B)/M$.\par

Proof in the Supplement, Appendix \crosspdflink{Entropy_Geometry_and_Lattice_Gaps_Supplement.pdf}{appB.3}{B.3}.\par

When $\mathcal E_D$ contains exactly the electric derivatives inside $D$, $M$ is their overlap multiplicity. A resampling gap becomes physical only through this form comparison.\par

On the full $L^2(\mu)$ space, for $\mathcal F_A=\sigma(X)$ and $\mathcal F_B=\sigma(Y)$ with standard Borel-valued variables, a concrete sufficient bound uses their joint law $\nu_{XY}$ and marginals. If $k=d\nu_{XY}/d(\nu_X\otimes\nu_Y)$ exists, then
\sbox{\equationmeasure}{$\displaystyle\begin{gathered}q\le\|k-1\|_{L^2(\nu_X\otimes\nu_Y)}.\end{gathered}$}
\ifdim\wd\equationmeasure>\dimexpr\columnwidth-36pt\relax
\typeout{MANUSCRIPT-WIDE 4.6}\begin{manuscriptwide}\eqanchor{4.6}\begin{equation*}\tag{4.6}\begin{gathered}q\le\|k-1\|_{L^2(\nu_X\otimes\nu_Y)}.\end{gathered}\end{equation*}\end{manuscriptwide}
\else \eqanchor{4.6}\begin{equation*}\tag{4.6}\begin{gathered}q\le\|k-1\|_{L^2(\nu_X\otimes\nu_Y)}.\end{gathered}\end{equation*}\fi

For centered $g(X),h(Y)$, covariance is $\int\overline g h(k-1)d\nu_Xd\nu_Y$; Cauchy-Schwarz and maximal correlation give \hyperlink{eq4.6}{(4.6)}. An $L^\infty$ bound also suffices. Small mutual information does not: $X=Y\sim\operatorname{Bernoulli}(\epsilon)$, $0<\epsilon<1$, has $I(X:Y)\to0$ as $\epsilon\downarrow0$, but its normalized centered indicator gives $q=1$. Rare-event probabilities and finite detector covariances therefore cannot replace the operator bound. Optimizing over finite cap-observable families gives lower bounds on $q$, whereas \hyperlink{eq4.4}{(4.4)} needs an upper bound. Such tests can falsify screening constants; certification requires omitted-direction control, as in Proposition \crosspdflink{Entropy_Geometry_and_Lattice_Gaps_Supplement.pdf}{res3.9}{3.8}.\par

The projection norm in Theorem \hyperlink{res4.2}{4.2} can be bounded through the Fisher information of an actual conditional law. This requires its complete parameter space and an independently controlled marginal Poincare constant. Ordinary sectional curvature or a finite detector family's Fisher matrix does not supply either hypothesis.\par

Let $(M_X,g_X)$ and $(M_Y,g_Y)$ be smooth compact connected Riemannian manifolds without boundary. Fix the product metric, write $dx,dy$ for their volume measures, and set
\sbox{\equationmeasure}{$\displaystyle\begin{gathered}d\mu=Z^{-1}e^{-W(x,y)}dx\,dy,\\  W\in C^\infty(M_X\times M_Y;\mathbb R),\\  Z_y=\int_{M_X}e^{-W(x,y)}dx,\\  d\nu_y=Z_y^{-1}e^{-W(x,y)}dx.\end{gathered}$}
\ifdim\wd\equationmeasure>\dimexpr\columnwidth-36pt\relax
\typeout{MANUSCRIPT-WIDE 4.7}\begin{manuscriptwide}\eqanchor{4.7}\begin{equation*}\tag{4.7}\begin{gathered}d\mu=Z^{-1}e^{-W(x,y)}dx\,dy,\quad W\in C^\infty(M_X\times M_Y;\mathbb R),\quad Z_y=\int_{M_X}e^{-W(x,y)}dx,\quad d\nu_y=Z_y^{-1}e^{-W(x,y)}dx.\end{gathered}\end{equation*}\end{manuscriptwide}
\else \eqanchor{4.7}\begin{equation*}\tag{4.7}\begin{gathered}d\mu=Z^{-1}e^{-W(x,y)}dx\,dy,\\  W\in C^\infty(M_X\times M_Y;\mathbb R),\\  Z_y=\int_{M_X}e^{-W(x,y)}dx,\\  d\nu_y=Z_y^{-1}e^{-W(x,y)}dx.\end{gathered}\end{equation*}\fi

The $Y$-marginal is $d\bar\mu=Z^{-1}Z_y\,dy$. Let $P_X,P_Y$ be conditional expectation onto $X,Y$, respectively, on the full complex $L^2(\mu)$, and put $\Pi f=\mu(f)1$. For real tangent vectors $v,w\in T_yM_Y$, define the conditional Fisher tensor
\sbox{\equationmeasure}{$\displaystyle\begin{gathered}J_y(v,w)=\operatorname{Cov}_{\nu_y}\big(d_yW[v],d_yW[w]\big),\\  q=\|P_XP_Y-\Pi\|.\end{gathered}$}
\ifdim\wd\equationmeasure>\dimexpr\columnwidth-36pt\relax
\typeout{MANUSCRIPT-WIDE 4.8}\begin{manuscriptwide}\eqanchor{4.8}\begin{equation*}\tag{4.8}\begin{gathered}J_y(v,w)=\operatorname{Cov}_{\nu_y}\big(d_yW[v],d_yW[w]\big),\qquad q=\|P_XP_Y-\Pi\|.\end{gathered}\end{equation*}\end{manuscriptwide}
\else \eqanchor{4.8}\begin{equation*}\tag{4.8}\begin{gathered}J_y(v,w)=\operatorname{Cov}_{\nu_y}\big(d_yW[v],d_yW[w]\big),\\  q=\|P_XP_Y-\Pi\|.\end{gathered}\end{equation*}\fi

Here real covariance has its usual bilinear meaning; all variances of complex functions below use squared absolute values.\par

\hypertarget{res4.4}{}\noindent \textbf{Theorem \hyperlink{res4.4}{4.4} (conditional Fisher screening).} Suppose that, for some $\beta>0$ and $C\ge0$,
\sbox{\equationmeasure}{$\displaystyle\begin{gathered}\operatorname{Var}_{\bar\mu}(h)\le\beta^{-1}\int|\nabla_Yh|^2d\bar\mu\\ (h\in H^1(M_Y)),\\  J_y\le Cg_Y\\ (y\in M_Y).\end{gathered}$}
\ifdim\wd\equationmeasure>\dimexpr\columnwidth-36pt\relax
\typeout{MANUSCRIPT-WIDE 4.9}\begin{manuscriptwide}\eqanchor{4.9}\begin{equation*}\tag{4.9}\begin{gathered}\operatorname{Var}_{\bar\mu}(h)\le\beta^{-1}\int|\nabla_Yh|^2d\bar\mu\quad(h\in H^1(M_Y)),\qquad J_y\le Cg_Y\quad(y\in M_Y).\end{gathered}\end{equation*}\end{manuscriptwide}
\else \eqanchor{4.9}\begin{equation*}\tag{4.9}\begin{gathered}\operatorname{Var}_{\bar\mu}(h)\le\beta^{-1}\int|\nabla_Yh|^2d\bar\mu\\ (h\in H^1(M_Y)),\\  J_y\le Cg_Y\\ (y\in M_Y).\end{gathered}\end{equation*}\fi

Then
\sbox{\equationmeasure}{$\displaystyle\begin{gathered}q^2\le\frac{C}{\beta+C}<1.\end{gathered}$}
\ifdim\wd\equationmeasure>\dimexpr\columnwidth-36pt\relax
\typeout{MANUSCRIPT-WIDE 4.10}\begin{manuscriptwide}\eqanchor{4.10}\begin{equation*}\tag{4.10}\begin{gathered}q^2\le\frac{C}{\beta+C}<1.\end{gathered}\end{equation*}\end{manuscriptwide}
\else \eqanchor{4.10}\begin{equation*}\tag{4.10}\begin{gathered}q^2\le\frac{C}{\beta+C}<1.\end{gathered}\end{equation*}\fi

A sufficient condition for the second hypothesis is a uniform $X$-conditional Poincare inequality with constant $\alpha>0$, together with the conditional mixed-Hessian bound
\sbox{\equationmeasure}{$\displaystyle\begin{gathered}\operatorname{Var}_{\nu_y}(u)\le\alpha^{-1}\int|\nabla_Xu|^2d\nu_y,\\  \mathbb E_{\nu_y}[B_y^*B_y]\le R^2I_{T_yM_Y},\\  B_yv=\nabla_X(d_yW[v]),\end{gathered}$}
\ifdim\wd\equationmeasure>\dimexpr\columnwidth-36pt\relax
\typeout{MANUSCRIPT-WIDE 4.11}\begin{manuscriptwide}\eqanchor{4.11}\begin{equation*}\tag{4.11}\begin{gathered}\operatorname{Var}_{\nu_y}(u)\le\alpha^{-1}\int|\nabla_Xu|^2d\nu_y,\qquad \mathbb E_{\nu_y}[B_y^*B_y]\le R^2I_{T_yM_Y},\qquad B_yv=\nabla_X(d_yW[v]),\end{gathered}\end{equation*}\end{manuscriptwide}
\else \eqanchor{4.11}\begin{equation*}\tag{4.11}\begin{gathered}\operatorname{Var}_{\nu_y}(u)\le\alpha^{-1}\int|\nabla_Xu|^2d\nu_y,\\  \mathbb E_{\nu_y}[B_y^*B_y]\le R^2I_{T_yM_Y},\\  B_yv=\nabla_X(d_yW[v]),\end{gathered}\end{equation*}\fi

for every $y$, every $u\in H^1(M_X)$, and one $R\ge0$. The adjoint uses the fixed factor metrics. In that case $C=R^2/\alpha$ is admissible and $q^2\le R^2/(\alpha\beta+R^2)$.\par

Moreover, with $W_{\mathrm{eff}}(y)=-\log Z_y$, one has the exact tensor identity
\sbox{\equationmeasure}{$\displaystyle\begin{gathered}\operatorname{Hess}_YW_{\mathrm{eff}}=\mathbb E_{\nu_y}[\operatorname{Hess}_YW]-J_y.\end{gathered}$}
\ifdim\wd\equationmeasure>\dimexpr\columnwidth-36pt\relax
\typeout{MANUSCRIPT-WIDE 4.12}\begin{manuscriptwide}\eqanchor{4.12}\begin{equation*}\tag{4.12}\begin{gathered}\operatorname{Hess}_YW_{\mathrm{eff}}=\mathbb E_{\nu_y}[\operatorname{Hess}_YW]-J_y.\end{gathered}\end{equation*}\end{manuscriptwide}
\else \eqanchor{4.12}\begin{equation*}\tag{4.12}\begin{gathered}\operatorname{Hess}_YW_{\mathrm{eff}}=\mathbb E_{\nu_y}[\operatorname{Hess}_YW]-J_y.\end{gathered}\end{equation*}\fi

Proof in the Supplement, Appendix \crosspdflink{Entropy_Geometry_and_Lattice_Gaps_Supplement.pdf}{appB.4}{B.4}.\par

The tensor $J$ is the Fisher information of the full conditional family $y\mapsto\nu_y$, whose score is $-d_yW+\nu_y(d_yW)$. Equation \hyperlink{eq4.12}{(4.12)} gives the exact correction to the effective-potential Hessian after integrating out $X$; it does not identify a detector manifold's Riemann curvature tensor.\par

Condition \hyperlink{eq4.11}{(4.11)} allows large mixed derivatives on sets of small conditional probability, but must hold after every required conditioning; an unconditional moment bound is insufficient. Likewise, finitely many cap tests do not certify \hyperlink{eq4.8}{(4.8)}'s full norm. Theorem \hyperlink{res4.4}{4.4} supplies sufficient inputs, not their verification for Yang-Mills.\par

For physical use, constant rescalings of the factor metrics must encode the electric coefficients, the law must be an identified ground law or Proposition \hyperlink{res4.1}{4.1}'s central-slice law, and both conditional projections must preserve the physical space. Buffer integration uses the actual marginal; $W$ is generally not the local four-dimensional Wilson action. Uniform marginal and conditional bounds, gauge-compatible boundary data, physical normalization, electric comparison and continuum transfer remain to be established along the continuum trajectory.\par

\subsubsection*{Actual finite-block starting bounds}

\hypertarget{res4.7}{}\noindent \textbf{Proposition \hyperlink{res4.7}{4.5} (explicit conditional and marginal screening bounds).} Use Theorem \hyperlink{res6.7}{6.2}'s compact connected nontrivial group, metric, actual full ground state and electric weights. Let $D_G=\operatorname{diam}(G)$ and define, for every nonempty finite free-link set $F$,
\sbox{\equationmeasure}{$\displaystyle\begin{gathered}\begin{aligned}
D_e&=\frac{B_e}{a_e r_G},\\
d_F&=2D_G\sum_{e\in F}D_e,\\
\lambda_F&=c_G\min_{e\in F}a_e\,e^{-d_F}.
\end{aligned}\end{gathered}$}
\ifdim\wd\equationmeasure>\dimexpr\columnwidth-36pt\relax
\typeout{MANUSCRIPT-WIDE 4.17}\begin{manuscriptwide}\eqanchor{4.17}\begin{equation*}\tag{4.17}\begin{gathered}\begin{aligned}
D_e&=\frac{B_e}{a_e r_G},\\
d_F&=2D_G\sum_{e\in F}D_e,\\
\lambda_F&=c_G\min_{e\in F}a_e\,e^{-d_F}.
\end{aligned}\end{gathered}\end{equation*}\end{manuscriptwide}
\else \eqanchor{4.17}\begin{equation*}\tag{4.17}\begin{gathered}\begin{aligned}
D_e&=\frac{B_e}{a_e r_G},\\
d_F&=2D_G\sum_{e\in F}D_e,\\
\lambda_F&=c_G\min_{e\in F}a_e\,e^{-d_F}.
\end{aligned}\end{gathered}\end{equation*}\fi

Starting from $\mu=\Omega^2m$, fix any disjoint exterior links at any value $\xi$, retain $F$, and integrate all remaining links. Denote the resulting probability law on $G^F$ by $\mu_{F,\xi}$. Its smooth positive Haar density $p_{F,\xi}$ obeys
\sbox{\equationmeasure}{$\displaystyle\begin{gathered}\begin{aligned}
\operatorname{osc}_{G^F}\log p_{F,\xi}&\le d_F,\\
\operatorname{Var}_{\mu_{F,\xi}}(f)
&\le\lambda_F^{-1}
\int\sum_{e\in F}a_e|\nabla_e f|^2\,d\mu_{F,\xi}.
\end{aligned}\end{gathered}$}
\ifdim\wd\equationmeasure>\dimexpr\columnwidth-36pt\relax
\typeout{MANUSCRIPT-WIDE 4.18}\begin{manuscriptwide}\eqanchor{4.18}\begin{equation*}\tag{4.18}\begin{gathered}\begin{aligned}
\operatorname{osc}_{G^F}\log p_{F,\xi}&\le d_F,\\
\operatorname{Var}_{\mu_{F,\xi}}(f)
&\le\lambda_F^{-1}
\int\sum_{e\in F}a_e|\nabla_e f|^2\,d\mu_{F,\xi}.
\end{aligned}\end{gathered}\end{equation*}\end{manuscriptwide}
\else \eqanchor{4.18}\begin{equation*}\tag{4.18}\begin{gathered}\begin{aligned}
\operatorname{osc}_{G^F}\log p_{F,\xi}&\le d_F,\\
\operatorname{Var}_{\mu_{F,\xi}}(f)
&\le\lambda_F^{-1}
\int\sum_{e\in F}a_e|\nabla_e f|^2\,d\mu_{F,\xi}.
\end{aligned}\end{gathered}\end{equation*}\fi

The inequality holds on the complete weighted $H^1(G^F)$ form domain, including complex functions, uniformly in $\xi$ and in the links integrated out.\par

For two disjoint nonempty retained link sets $X,Y$, form their actual joint marginal conditional on $\xi$ and let $P_X,P_Y$ be conditional expectation onto their respective coordinate algebras. Let $\Pi$ project onto constants. Then, with the electric factor metric $g_Y^{\rm el}=\bigoplus_{e\in Y}a_e^{-1}g$,
\sbox{\equationmeasure}{$\displaystyle\begin{gathered}\begin{aligned}
C_Y&=4\sum_{e\in Y}a_eD_e^2,\\
J_y&\preceq C_Yg_Y^{\rm el},\\
\|P_XP_Y-\Pi\|^2
&\le\frac{C_Y}{\lambda_Y+C_Y}<1.
\end{aligned}\end{gathered}$}
\ifdim\wd\equationmeasure>\dimexpr\columnwidth-36pt\relax
\typeout{MANUSCRIPT-WIDE 4.19}\begin{manuscriptwide}\eqanchor{4.19}\begin{equation*}\tag{4.19}\begin{gathered}\begin{aligned}
C_Y&=4\sum_{e\in Y}a_eD_e^2,\\
J_y&\preceq C_Yg_Y^{\rm el},\\
\|P_XP_Y-\Pi\|^2
&\le\frac{C_Y}{\lambda_Y+C_Y}<1.
\end{aligned}\end{gathered}\end{equation*}\end{manuscriptwide}
\else \eqanchor{4.19}\begin{equation*}\tag{4.19}\begin{gathered}\begin{aligned}
C_Y&=4\sum_{e\in Y}a_eD_e^2,\\
J_y&\preceq C_Yg_Y^{\rm el},\\
\|P_XP_Y-\Pi\|^2
&\le\frac{C_Y}{\lambda_Y+C_Y}<1.
\end{aligned}\end{gathered}\end{equation*}\fi

Here $J$ is the conditional Fisher tensor of the full $X$-coordinate family given $Y=y$, as in Theorem \hyperlink{res4.4}{4.4}. Interchanging $X,Y$ gives another valid upper bound; the smaller may be used. Every constant uses derivatives of the actual full potential. Thus bounds on these local derivatives give one estimate independent of total volume and exterior values. Conditioning does not assert any new physical boundary condition or identify the conditional law with a frozen bare block vacuum.\par

Proof in the Supplement, Appendix \crosspdflink{Entropy_Geometry_and_Lattice_Gaps_Supplement.pdf}{appB.7}{B.7}.\par

These bounds supply finite-block marginal constants without an assumed global gap. Their Wilson specialization and scaling are given in Appendix \crosspdflink{Entropy_Geometry_and_Lattice_Gaps_Supplement.pdf}{appB.7}{B.7}. The stated screening bound has no decay with buffer distance and deteriorates toward one at weak bare coupling.\par

\hypertarget{sec4.3}{}\subsection{Exceptional energy and scale losses}

Theorem \hyperlink{res3.5}{3.5} needs an energy-weighted exceptional-set estimate; small probability alone does not control a normalized function supported there. The Lyapunov method [\hyperlink{ref57}{21}, Lemma 1.1] supplies a sufficient estimate. Keeping several cells leaves intercell transitions to the retained-current certificate.\par

\hypertarget{res4.5}{}\noindent \textbf{Proposition \hyperlink{res4.5}{4.6} (a physical Lyapunov certificate).} In the compact-link setting of Theorem \hyperlink{res2.2}{2.2}, let $L$ be the actual ground-state diffusion generator in \hyperlink{eq3.5}{(3.5)}. Let $U$ be a measurable invariant subset and let $h>0$ be smooth and invariant. Suppose $\kappa>0$ and $d\ge0$ satisfy, almost everywhere,
\sbox{\equationmeasure}{$\displaystyle\begin{gathered}-\frac{Lh}{h}\ge\kappa\mathbf1_{U^c}-d\mathbf1_U.\end{gathered}$}
\ifdim\wd\equationmeasure>\dimexpr\columnwidth-36pt\relax
\typeout{MANUSCRIPT-WIDE 4.13}\begin{manuscriptwide}\eqanchor{4.13}\begin{equation*}\tag{4.13}\begin{gathered}-\frac{Lh}{h}\ge\kappa\mathbf1_{U^c}-d\mathbf1_U.\end{gathered}\end{equation*}\end{manuscriptwide}
\else \eqanchor{4.13}\begin{equation*}\tag{4.13}\begin{gathered}-\frac{Lh}{h}\ge\kappa\mathbf1_{U^c}-d\mathbf1_U.\end{gathered}\end{equation*}\fi

Then every physical form-domain function satisfies
\sbox{\equationmeasure}{$\displaystyle\begin{gathered}\int_{U^c}|f|^2d\mu\le\frac{\mathcal E_\mu(f)+d\|f\|_2^2}{\kappa+d}.\end{gathered}$}
\ifdim\wd\equationmeasure>\dimexpr\columnwidth-36pt\relax
\typeout{MANUSCRIPT-WIDE 4.14}\begin{manuscriptwide}\eqanchor{4.14}\begin{equation*}\tag{4.14}\begin{gathered}\int_{U^c}|f|^2d\mu\le\frac{\mathcal E_\mu(f)+d\|f\|_2^2}{\kappa+d}.\end{gathered}\end{equation*}\end{manuscriptwide}
\else \eqanchor{4.14}\begin{equation*}\tag{4.14}\begin{gathered}\int_{U^c}|f|^2d\mu\le\frac{\mathcal E_\mu(f)+d\|f\|_2^2}{\kappa+d}.\end{gathered}\end{equation*}\fi

When $U=\bigcup_iD_i$ is the union of Theorem \hyperlink{res3.5}{3.5}'s cells, its exceptional-set hypothesis holds with $a_0=(\kappa+d)^{-1}$ and $b_0=d/(\kappa+d)<1$. Under that theorem's remaining local-cell and current assumptions, it gives $\Delta(H)\ge\kappa/[1+(\kappa+d)(C_{\rm loc}+C_{\rm resp})]$.\par

One may verify \hyperlink{eq4.13}{(4.13)} from the physical Hamiltonian. If $\phi>0$ is smooth and invariant and $\overline E\ge E_0$, then
\sbox{\equationmeasure}{$\displaystyle\begin{gathered}\frac{H\phi}{\phi}-\overline E\ge\kappa\mathbf1_{U^c}-d\mathbf1_U\end{gathered}$}
\ifdim\wd\equationmeasure>\dimexpr\columnwidth-36pt\relax
\typeout{MANUSCRIPT-WIDE 4.15}\begin{manuscriptwide}\eqanchor{4.15}\begin{equation*}\tag{4.15}\begin{gathered}\frac{H\phi}{\phi}-\overline E\ge\kappa\mathbf1_{U^c}-d\mathbf1_U\end{gathered}\end{equation*}\end{manuscriptwide}
\else \eqanchor{4.15}\begin{equation*}\tag{4.15}\begin{gathered}\frac{H\phi}{\phi}-\overline E\ge\kappa\mathbf1_{U^c}-d\mathbf1_U\end{gathered}\end{equation*}\fi

is sufficient, taking $h=\phi/\Omega$. Here $H\phi$ is its differential expression on a smooth function. A normalized trial state's Rayleigh quotient supplies an upper bound $\overline E$; useful uniform constants still require proof.\par

Proof in the Supplement, Appendix \crosspdflink{Entropy_Geometry_and_Lattice_Gaps_Supplement.pdf}{appB.5}{B.5}.\par

The inequality tests escape energy in the actual configuration law. Neither an auxiliary curvature-tail bound nor a Lyapunov function for an unrelated sampler establishes it. Its use in Yang-Mills remains conditional on local-cell control, globally valid current identities, and constants surviving the continuum trajectory.\par

When screening uses Theorem \hyperlink{res4.4}{4.4}, its product-law assumptions must hold on every conditional cover, uniformly in exterior data. Together with local electric Poincare bounds and summable losses, these give the following sufficient multiscale criterion for finite coordinate regulators; their indexing regions may come from covers of a continuous spatial manifold.\par

\hypertarget{res4.6}{}\noindent \textbf{Theorem \hyperlink{res4.6}{4.7} (multiscale conditional coercivity).} Fix a regulated product configuration space with additive electric gradient form. Assume every kernel below has a closable local gradient form with the stated differentiable core, and use its closure. Supply a hierarchy $\mathfrak M_s$ of conditional probability kernels on regions, indexed also by their exterior data. All conditioning uses the same supplied law. Assume:\par

\noindent \textbf{(i)} Every kernel in $\mathfrak M_0$ satisfies the Poincare inequality with constant $\lambda_0>0$ against its local electric form.
\textbf{(ii)} For each kernel $\nu\in\mathfrak M_s$, $s\ge1$, there are finitely many covers $A_j\cup B_j=\Lambda_\nu$ and weights $w_j\ge0$, $\sum_jw_j=1$. Every conditional kernel obtained by fixing the exterior of $A_j$ or $B_j$ belongs to $\bigcup_{r<s}\mathfrak M_r$, almost surely, with the same retained coordinates and electric weights.
\textbf{(iii)} Each cover satisfies $\|P_{A_j}P_{B_j}-\Pi_\nu\|\le q_s<1$, where $q_s\ge0$ on the full conditional $L^2$ space. For every electric derivative, its weighted multiplicity in the covers is at most $1+\varepsilon_s$, where $\varepsilon_s\ge0$.\par

Assume these statements for every kernel and almost every exterior datum needed by the recursion, and on a common differentiable core containing constants and closed under the restrictions used in the conditional estimates. Then every kernel in $\mathfrak M_s$ obeys
\sbox{\equationmeasure}{$\displaystyle\begin{gathered}\operatorname{Var}_\nu(f)\le\lambda_s^{-1}\mathcal E_\nu(f),\\  \lambda_s=\lambda_0\prod_{r=1}^s\frac{1-q_r}{1+\varepsilon_r}.\end{gathered}$}
\ifdim\wd\equationmeasure>\dimexpr\columnwidth-36pt\relax
\typeout{MANUSCRIPT-WIDE 4.16}\begin{manuscriptwide}\eqanchor{4.16}\begin{equation*}\tag{4.16}\begin{gathered}\operatorname{Var}_\nu(f)\le\lambda_s^{-1}\mathcal E_\nu(f),\qquad \lambda_s=\lambda_0\prod_{r=1}^s\frac{1-q_r}{1+\varepsilon_r}.\end{gathered}\end{equation*}\end{manuscriptwide}
\else \eqanchor{4.16}\begin{equation*}\tag{4.16}\begin{gathered}\operatorname{Var}_\nu(f)\le\lambda_s^{-1}\mathcal E_\nu(f),\\  \lambda_s=\lambda_0\prod_{r=1}^s\frac{1-q_r}{1+\varepsilon_r}.\end{gathered}\end{equation*}\fi

If $\sum_r(q_r+\varepsilon_r)<\infty$, then $\inf_s\lambda_s>0$. Closure extends the inequalities to their electric form domains. For a full ground law among these kernels, restriction to invariant functions gives the physical gap bound.\par

Proof in the Supplement, Appendix \crosspdflink{Entropy_Geometry_and_Lattice_Gaps_Supplement.pdf}{appB.6}{B.6}.\par

Full conditional-space control is stronger than physical-sector control and retains exterior gauge data. The kernels are not separately chosen regional ground states; a boundary-sector version needs compatible invariant domains and projections throughout. Fixed losses can destroy the certificate: $q_s=q\in(0,1)$ and $\varepsilon_s=0$ give $\lambda_s=\lambda_0(1-q)^s$. Using Theorem \hyperlink{res4.4}{4.4}'s sufficient bound in \hyperlink{eq4.16}{(4.16)} requires summability of $\sqrt{C_s/(\beta_s+C_s)}$ and overlap losses. Displaced overlaps may reduce average multiplicity, but require construction and screening estimates. Bounded cover degree and local coercivity do not exclude collective soft modes. Related multiscale strategies [\hyperlink{ref51}{22}] supply none of these Yang-Mills inputs. Across regulators, require a common positive base constant in physical units and one positive lower bound on the complete products in \hyperlink{eq4.16}{(4.16)}. Sufficient hypotheses are regulator-independent envelopes $q_{r,h}\le\bar q_r<1$, $\varepsilon_{r,h}\le\bar\varepsilon_r$ with $\sum_r(\bar q_r+\bar\varepsilon_r)<\infty$. Cutoffwise summability is insufficient; Appendix \crosspdflink{Entropy_Geometry_and_Lattice_Gaps_Supplement.pdf}{appB.7}{B.7} gives explicit failures of that inference.\par

\hypertarget{sec5}{}\section{Geometric inputs and physical identification}

Entropy geometry can organize candidate states, connections and regions. Its relevance to a mass gap is determined by a comparison with the actual physical energy form. Theorems \hyperlink{res3.1}{3.1} and \crosspdflink{Entropy_Geometry_and_Lattice_Gaps_Supplement.pdf}{res3.8}{3.7} make that comparison precise for quantum relative entropy and configuration Fisher information; Theorems \hyperlink{res3.5}{3.5} and \hyperlink{res4.6}{4.7} require physical current and scale estimates. A geometric metric or an auxiliary generator does not supply those estimates.\par

\noindent \textbf{Pinching and conditional tube energy.} Pinching motivates a possible confinement mechanism: transverse flux concentration maintained by a cost for spreading. This differs from sectional-curvature bounds, a narrow geometric neck, concentration of a probability law, and quantum dephasing. None alone identifies a physical energy penalty.\par

The inverse-area and occupied-volume competition below is the familiar fixed-flux bag balance, explicitly written in [\hyperlink{ref103}{23}, Section 2.3, Eq. (2.3)]. We use it as a conditional physical-energy certificate, whose vacuum subtraction, relaxation, complete-sector coverage and uniform constants remain to be proved.\par

For a conditional area balance, let a spatial tube have finite physical length $L>0$, measurable areas $0<A(s)<\infty$, and volume disintegration $dV=ds\,dA_s$. Supply constants $q>0$ and $\mathcal B>0$ and an independently established comparison $E_{\rm rel}\ge\int_0^L[q/A(s)+\mathcal B A(s)]\,ds$ for every member of the identified tube class, after vacuum subtraction and relaxation. The inverse-area coefficient represents a concentration constraint: in a scalar quadratic normalization, a real profile $u\in L^2(S)$ on a section $S$ of area $A$, with fixed nonzero flux $\Phi=\int_Su\,dA$ satisfies $\frac12\int_Su^2dA\ge\Phi^2/(2A)$ by Cauchy-Schwarz, giving $q=\Phi^2/2$. A non-Abelian use requires a specified gauge-covariant flux and energy identification. The residual density $\mathcal B$ must not count the same field energy twice.\par

The arithmetic-geometric mean inequality gives $q/A+\mathcal B A\ge2\sqrt{q\mathcal B}$, and hence $E_{\rm rel}\ge2\sqrt{q\mathcal B}\,L$. Positive tension is not a neutral vacuum gap: this bound supplies no positive floor if $L$ can approach zero. To contribute to the latter, the comparison must establish Theorem \hyperlink{res6.42}{6.11}'s bounds on the complete centered physical form core after hidden variables relax, including local low modes and localization errors, uniformly across regulators and surrounding configurations.\par

Color neutrality constrains admissible operators, not particle number or energy. In pure SU(3), a quadratic curvature singlet can overlap a single composite excitation; regulated numerical glueball calculations study such matrix elements [\hyperlink{ref93}{11}]. A two-particle threshold already presupposes physical asymptotic masses. A confinement-based lower bound must use the actual physical energy, the same vacuum subtraction and the complete physical sector.\par

Under the geometric input assumptions stated in the Introduction, the qutrit foundation [\hyperlink{ref2}{1}] gives $\mathfrak{so}(8)=\mathfrak h_8\oplus\mathfrak m_{20}$, with $\mathfrak h_8=\operatorname{ad}\mathfrak{su}(3)$ and $(\mathfrak m_{20})_{\mathbb C}=10\oplus\overline{10}$. These conjugate components form one real irreducible module. Their zero center charge permits adjoint screening [\hyperlink{ref94}{24}]. The quadratic singlet norm and adjoint curvature feedback are different contractions. Proposition \crosspdflink{Entropy_Geometry_and_Lattice_Gaps_Supplement.pdf}{res6.44}{6.30}, in the Supplement, Appendix \crosspdflink{Entropy_Geometry_and_Lattice_Gaps_Supplement.pdf}{appC.43}{C.15}, bounds the feedback and constructs a flat SO(8) connection with nonzero projected color curvature. Consequently the completed curvature energy need not dominate the projected color energy. The reduction has global group $\mathrm{PSU}(3)$. Identifying it with an $SU(3)$ spacetime gauge field carrying fundamental Wilson observables requires a specified $SU(3)$ lift [\hyperlink{ref2}{1}, Remark A3]. A smooth bundle over contractible $\mathbb R^4$ admits such a lift; the $SU(3)$ regulator of Section \hyperlink{sec2}{II} is supplied independently.\par

A separately supplied spacetime reduction field with positive kinetic coefficient can oppose that feedback. If $\|F_{A_c}\|_\infty\le K$, its supplied restoring coefficient $\mu^2>\sqrt3K$ gives complementary classical quadratic coercivity. At a Yang-Mills critical background the estimate also applies after removing infinitesimal gauge redundancy. The residual color Hessian remains uncontrolled. The entropy tensor identity $\|D^\Psi C^J\|^2=27\|\Phi_{20}\|^2$ fixes a geometric normalization [\hyperlink{ref2}{1}, Theorem 18], but supplies neither this physical coefficient nor a quantum gap.\par

Section \hyperlink{sec6}{VI} places these required comparisons with the relaxed physical energy within the hierarchy and reconstruction framework.\par

\hypertarget{sec6}{}\section{Conditional continuum transfer}

Our Hamiltonians have fixed physical energy units. For a dimensionless transfer operator $T_n=e^{-\widehat H_n}=e^{-a_{t,n}H_n^{\rm phys}}$,
\sbox{\equationmeasure}{$\displaystyle\begin{gathered}H_n^{\rm phys}=a_{t,n}^{-1}\widehat H_n,\\ \Delta_n^{\rm phys}=a_{t,n}^{-1}\widehat\Delta_n.\end{gathered}$}
\ifdim\wd\equationmeasure>\dimexpr\columnwidth-36pt\relax
\typeout{MANUSCRIPT-WIDE 6.1}\begin{manuscriptwide}\eqanchor{6.1}\begin{equation*}\tag{6.1}\begin{gathered}H_n^{\rm phys}=a_{t,n}^{-1}\widehat H_n,\qquad\Delta_n^{\rm phys}=a_{t,n}^{-1}\widehat\Delta_n.\end{gathered}\end{equation*}\end{manuscriptwide}
\else \eqanchor{6.1}\begin{equation*}\tag{6.1}\begin{gathered}H_n^{\rm phys}=a_{t,n}^{-1}\widehat H_n,\\ \Delta_n^{\rm phys}=a_{t,n}^{-1}\widehat\Delta_n.\end{gathered}\end{equation*}\fi

A closing dimensionless gap can coexist with a positive physical gap. The transfer route below uses reflected observable pairings and requires a surviving physical excitation.\par

Two constructions connect regulated energies to limiting dynamics. In the \textbf{relaxation route}, the retained form minimizes the full energy over eliminated variables. For their uniform gap consequences, Theorems \hyperlink{res6.38}{6.7} and \hyperlink{res6.45}{6.12} use an independently established infrared lower bound and full-energy estimates for discarded components, with a uniform bound on the total inverse-shell loss; a summable sequence of inverse shell bounds is sufficient. Comparing regulators by Theorem \hyperlink{res6.46}{6.23} also requires spectral tightness to retain a densely defined limiting generator.\par

In the \textbf{energy-preserving embedding route}, Theorem \hyperlink{res6.8}{6.21} and Proposition \hyperlink{res6.32}{6.22} use exact compression and summable dynamical defects to construct limiting dynamics. A common physical decay estimate is a separate input to spectral transfer. Relaxation need not preserve energy under the natural retained injection, so Theorem \hyperlink{res6.45}{6.12} does not automatically furnish these embeddings. For the Yang-Mills application, both routes require surviving renormalized physical observables and the field-theoretic identification specified below.\par

Continuum construction requires controlled dynamics at a fixed physical scale and nonzero renormalized physical observables. Changing units cannot extend Section \hyperlink{sec2}{II}: for $h,g_h>0$ and fixed $C_E,C_B>0$, the coefficients $a_h=C_Eg_h^2/h$ and $b_h=C_B/(g_h^2h)$ give $b_h/a_h=(C_B/C_E)g_h^{-4}\to\infty$ as $g_h\to0$. The following blocking, positive-time, smoothing, and interaction estimates do not assume the four-dimensional interacting continuum contraction.\par

\hypertarget{sec6.1}{}\subsection{Reconstructed target and physical time}

Let $\mathcal A$ be a unital complex algebra of Euclidean test observables, with involution $*$ and a linear time-reflection involution $\theta$ preserving $1$ and commuting with $*$. Let $\mathcal A_+\subset\mathcal A$ be a unital linear subspace preserved by unital linear forward physical-time translations $\tau_t$, $t\ge0$, with $\tau_0=I$ and $\tau_t1=1$. In the intended application, $\mathcal A_+$ is the polynomial algebra of specified renormalized local gauge-invariant fields supported at strictly positive times, with the constant included. Reflection acts on field components according to their Euclidean transformation law.\par

Supply a normalized linear functional $S$ on $\mathcal A$, a Hilbert space $\mathcal H$, a linear map $q:\mathcal A_+\to\mathcal H$ with dense range, a normalized vector $\Omega=q(1)$, and a nonnegative self-adjoint $H$. Use inner products antilinear in the first argument. Assume $S(F)=\langle\Omega,q(F)\rangle$ for every $F\in\mathcal A_+$. For every $F,G\in\mathcal A_+$ and $t\ge0$, impose the following reconstruction identity.
\sbox{\equationmeasure}{$\displaystyle\begin{gathered}\langle q(F),e^{-tH}q(G)\rangle=S\big((\theta F)^*\tau_tG\big),\\  H\Omega=0.\end{gathered}$}
\ifdim\wd\equationmeasure>\dimexpr\columnwidth-36pt\relax
\typeout{MANUSCRIPT-WIDE 6.2}\begin{manuscriptwide}\eqanchor{6.2}\begin{equation*}\tag{6.2}\begin{gathered}\langle q(F),e^{-tH}q(G)\rangle=S\big((\theta F)^*\tau_tG\big),\qquad H\Omega=0.\end{gathered}\end{equation*}\end{manuscriptwide}
\else \eqanchor{6.2}\begin{equation*}\tag{6.2}\begin{gathered}\langle q(F),e^{-tH}q(G)\rangle=S\big((\theta F)^*\tau_tG\big),\\  H\Omega=0.\end{gathered}\end{equation*}\fi

At zero time, $q$ identifies the null vectors of the reflected form. Vacuum uniqueness and quantum-field locality are not assumed here. Ordinary Euclidean variance cannot replace a nonzero centered reflected norm: white noise gives positive ordinary variance but a vacuum-only reflection quotient (Remark \crosspdflink{Entropy_Geometry_and_Lattice_Gaps_Supplement.pdf}{res6.27}{6.42}).\par

A quantum-field application must additionally supply locality, covariance, the spectral condition and distributional field domains. The full Osterwalder-Schrader theorem [\hyperlink{ref14}{25}, Section IV.1] requires its stated distribution spaces, regularity and axioms $E1$-$E4$; its clustering axiom $E4$ [\hyperlink{ref13}{26}] already includes the corresponding vacuum-uniqueness conclusion. It cannot derive that conclusion here from assumptions omitting clustering. Reflection positivity and fixed-order moment convergence alone do not replace reconstruction.\par

\hypertarget{sec6.2}{}\subsection{Actual vacua and compatible laws}

We first record what follows from actual finite-regulator vacua: compatible static marginals and local score estimates. These inputs do not by themselves identify a compatible physical dynamics or supply a uniform infrared gap.\par

\subsubsection*{Constructing the static hierarchy}

Actual finite-regulator vacua need not agree under coarse holonomy maps. Their subsequential marginals nevertheless supply a compatible static hierarchy. The underlying graphs may consist of paths on a continuous spatial manifold; the following construction concerns their compact holonomy carriers.\par

\hypertarget{res6.1}{}\noindent \textbf{Proposition \hyperlink{res6.1}{6.1} (static hierarchy from regulated vacua).} Let $X_n=G^{\mathsf E_n}$ be a countable family with continuous surjections $p_{n,m}:X_m\to X_n$ for $m\ge n$, satisfying $p_{n,n}=I$ and $p_{n,m}p_{m,N}=p_{n,N}$. Assume equivariance under compact gauge groups $K_n$, with compatible surjective group maps $K_m\to K_n$. Let $\mu_N^{\rm act}$ be the actual invariant ground law of a smooth finite-regulator Hamiltonian with positive electric coefficients. There are a single subsequence $N_k\to\infty$ and invariant probabilities $\mu_n$ such that, for every fixed $n\le m$,
\sbox{\equationmeasure}{$\displaystyle\begin{gathered}\begin{aligned}
(p_{n,N_k})_*\mu_{N_k}^{\rm act}&\rightharpoonup\mu_n,\\
(p_{n,m})_*\mu_m&=\mu_n.
\end{aligned}\end{gathered}$}
\ifdim\wd\equationmeasure>\dimexpr\columnwidth-36pt\relax
\typeout{MANUSCRIPT-WIDE 6.3}\begin{manuscriptwide}\eqanchor{6.3}\begin{equation*}\tag{6.3}\begin{gathered}\begin{aligned}
(p_{n,N_k})_*\mu_{N_k}^{\rm act}&\rightharpoonup\mu_n,\\
(p_{n,m})_*\mu_m&=\mu_n.
\end{aligned}\end{gathered}\end{equation*}\end{manuscriptwide}
\else \eqanchor{6.3}\begin{equation*}\tag{6.3}\begin{gathered}\begin{aligned}
(p_{n,N_k})_*\mu_{N_k}^{\rm act}&\rightharpoonup\mu_n,\\
(p_{n,m})_*\mu_m&=\mu_n.
\end{aligned}\end{gathered}\end{equation*}\fi

There is a unique Radon probability on $X_\infty=\varprojlim X_n$ with these marginals. All finite products of continuous physical cylinder observables have convergent expectations along that subsequence. Pullback gives compatible physical $L^2$ isometries preserving the constant vacuum vector.\par

Proof in the Supplement, Appendix \crosspdflink{Entropy_Geometry_and_Lattice_Gaps_Supplement.pdf}{appC.1}{C.1}, using compact projective measure extension [\hyperlink{ref84}{27}, Theorem 1]. The static hierarchy comes from actual vacua without requiring native coarse vacua to agree. Its marginals may be singular; uniqueness, compatible energies, time correlations and a nonzero centered observable remain unproved.\par

An exact dynamical benchmark is serial edge subdivision with only new degree-two vertices. For the product map $p$, choose $a_e=\sum_{r\subset e}a'_r$ and $V'=V\circ p$. Bi-invariance and internal gauge reduction make pullback unitary onto the entire fine physical space, with $H'J=JH$ and zero defects; Appendix \crosspdflink{Entropy_Geometry_and_Lattice_Gaps_Supplement.pdf}{appC.1}{C.1} supplies the argument, extending the additive-Laplacian construction [\hyperlink{ref84}{27}, Theorem 5]. This adds no physical modes. With homogeneous $a_h=C_Eg_h^2/h$, dyadic compatibility requires $g_{h/2}^2=g_h^2/4$, rather than a logarithmically running ratio tending to one.\par

\subsubsection*{Actual vacuum gradients and holonomy bounds}

A maximum-principle estimate controls the actual vacuum on arbitrarily large finite graphs, without a small-coupling assumption.\par

\hypertarget{res6.7}{}\noindent \textbf{Theorem \hyperlink{res6.7}{6.2} (local vacuum gradients and uniform holonomy errors).} Let $G$ be a compact connected Lie group with a bi-invariant metric satisfying $\operatorname{Ric}_G\ge r_Gg$, $r_G>0$. On a finite product of links let $H=-\sum_ea_e\Delta_e+V$, with $a_e>0$ and smooth real gauge-invariant $V$, and let $\Omega>0$ be its normalized ground state. With $B_e=\|\nabla_eV\|_\infty$,
\sbox{\equationmeasure}{$\displaystyle\begin{gathered}\|\nabla_e\log\Omega\|_\infty
\le\frac{B_e}{a_er_G}.\end{gathered}$}
\ifdim\wd\equationmeasure>\dimexpr\columnwidth-36pt\relax
\typeout{MANUSCRIPT-WIDE 6.20}\begin{manuscriptwide}\eqanchor{6.20}\begin{equation*}\tag{6.20}\begin{gathered}\|\nabla_e\log\Omega\|_\infty
\le\frac{B_e}{a_er_G}.\end{gathered}\end{equation*}\end{manuscriptwide}
\else \eqanchor{6.20}\begin{equation*}\tag{6.20}\begin{gathered}\|\nabla_e\log\Omega\|_\infty
\le\frac{B_e}{a_er_G}.\end{gathered}\end{equation*}\fi

For a simple cycle $C$, retain its full based holonomy $W$. Use the actual ground law, the electric lift \crosspdflink{Entropy_Geometry_and_Lattice_Gaps_Supplement.pdf}{eqC.34}{(C.34)}, and put
\sbox{\equationmeasure}{$\displaystyle\begin{gathered}\begin{aligned}
\kappa_C&=\sum_{e\in C}a_e,&
s_C&=\frac{2}{r_G\kappa_C}\sum_{e\in C}B_e,\\
c_C&=\kappa_Cs_C^2,&
\operatorname{tr}\mathcal I(w)&\le s_C^2.
\end{aligned}\end{gathered}$}
\ifdim\wd\equationmeasure>\dimexpr\columnwidth-36pt\relax
\typeout{MANUSCRIPT-WIDE 6.21}\begin{manuscriptwide}\eqanchor{6.21}\begin{equation*}\tag{6.21}\begin{gathered}\begin{aligned}
\kappa_C&=\sum_{e\in C}a_e,&
s_C&=\frac{2}{r_G\kappa_C}\sum_{e\in C}B_e,\\
c_C&=\kappa_Cs_C^2,&
\operatorname{tr}\mathcal I(w)&\le s_C^2.
\end{aligned}\end{gathered}\end{equation*}\end{manuscriptwide}
\else \eqanchor{6.21}\begin{equation*}\tag{6.21}\begin{gathered}\begin{aligned}
\kappa_C&=\sum_{e\in C}a_e,&
s_C&=\frac{2}{r_G\kappa_C}\sum_{e\in C}B_e,\\
c_C&=\kappa_Cs_C^2,&
\operatorname{tr}\mathcal I(w)&\le s_C^2.
\end{aligned}\end{gathered}\end{equation*}\fi

The same Fisher bound holds after fixing any exterior links disjoint from $C$, uniformly in their values. For the full stationary ground diffusion, let $K=-L$, let $A$ be the exact marginal form operator on $W$, and let $Jf=f(W)$. Then
\sbox{\equationmeasure}{$\displaystyle\begin{gathered}\begin{aligned}
\|J^*e^{-tK}J-e^{-tA}\|&\le tc_C/2,\\
D(\mathbb P_W^T\Vert\overline{\mathbb P}^{\,T})
&\le Tc_C/4.
\end{aligned}\end{gathered}$}
\ifdim\wd\equationmeasure>\dimexpr\columnwidth-36pt\relax
\typeout{MANUSCRIPT-WIDE 6.22}\begin{manuscriptwide}\eqanchor{6.22}\begin{equation*}\tag{6.22}\begin{gathered}\begin{aligned}
\|J^*e^{-tK}J-e^{-tA}\|&\le tc_C/2,\\
D(\mathbb P_W^T\Vert\overline{\mathbb P}^{\,T})
&\le Tc_C/4.
\end{aligned}\end{gathered}\end{equation*}\end{manuscriptwide}
\else \eqanchor{6.22}\begin{equation*}\tag{6.22}\begin{gathered}\begin{aligned}
\|J^*e^{-tK}J-e^{-tA}\|&\le tc_C/2,\\
D(\mathbb P_W^T\Vert\overline{\mathbb P}^{\,T})
&\le Tc_C/4.
\end{aligned}\end{gathered}\end{equation*}\fi

Here the second law is the stationary marginal Markov law. All constants depend on the indicated cycle weights and local derivatives, not on total volume. In the $SU(3)$ normalization of Section \hyperlink{sec2}{II}, if $a_e=a$, each link meets at most $\nu$ Wilson plaquettes, and each plaquette coefficient is $b\ge0$, then
\sbox{\equationmeasure}{$\displaystyle\begin{gathered}r_G=\frac34,\\ 
s_C^2\le\frac{32\nu^2}{27}\left(\frac ba\right)^2.\end{gathered}$}
\ifdim\wd\equationmeasure>\dimexpr\columnwidth-36pt\relax
\typeout{MANUSCRIPT-WIDE 6.23}\begin{manuscriptwide}\eqanchor{6.23}\begin{equation*}\tag{6.23}\begin{gathered}r_G=\frac34,\qquad
s_C^2\le\frac{32\nu^2}{27}\left(\frac ba\right)^2.\end{gathered}\end{equation*}\end{manuscriptwide}
\else \eqanchor{6.23}\begin{equation*}\tag{6.23}\begin{gathered}r_G=\frac34,\\ 
s_C^2\le\frac{32\nu^2}{27}\left(\frac ba\right)^2.\end{gathered}\end{equation*}\fi

Proof in the Supplement, Appendix \crosspdflink{Entropy_Geometry_and_Lattice_Gaps_Supplement.pdf}{appC.10}{C.4}. That proof also gives an explicit volume-independent positive-time Wilson fluctuation at every fixed finite $b/a$. Conditioning uses the actual full vacuum; a frozen bare block Hamiltonian generally has a different vacuum. The conditional Fisher assertion alone does not identify dynamics with an evolving exterior.\par

The formula controls surrounding configurations across graphs at each regulator, but supplies no summable physical error budget. For a cycle of fixed length in links, its upper bound scales as $1/(hg_h^6)$ under the coefficients at the start of Section \hyperlink{sec6}{VI}; this deterioration does not bound the actual error below. Generated interactions enter $B_e$, and changing the electric symbol requires a new comparison. Joint-symbol and local-energy estimates must still combine multiple loops or overlapping blocks into a single defect bound.\par

\subsubsection*{Improved scores and conditional curvature}

\hypertarget{res6.48}{}\noindent \textbf{Proposition \hyperlink{res6.48}{6.3} (uniform square-root vacuum scores).} Use the actual full finite Hamiltonian $H=-\sum_ea_e\Delta_e+V$ with $a_e>0$ and smooth real $V$, and its normalized positive ground state $\Omega$. Let $G$ be compact and connected with a bi-invariant metric, diameter $D_G$ and nonnegative Ricci curvature. With all other links denoted by $\xi$, put
\sbox{\equationmeasure}{$\displaystyle\begin{gathered}\begin{aligned}
B_i&=\|\nabla_iV\|_\infty,\\
M_i&=\sup_\xi\bigl(\max_xV(x,\xi)-\min_xV(x,\xi)\bigr),\\
\sigma_i&=\frac32\sqrt{\frac{M_i+\frac32D_GB_i}{a_i}}.
\end{aligned}\end{gathered}$}
\ifdim\wd\equationmeasure>\dimexpr\columnwidth-36pt\relax
\typeout{MANUSCRIPT-WIDE 6.104}\begin{manuscriptwide}\eqanchor{6.104}\begin{equation*}\tag{6.104}\begin{gathered}\begin{aligned}
B_i&=\|\nabla_iV\|_\infty,\\
M_i&=\sup_\xi\bigl(\max_xV(x,\xi)-\min_xV(x,\xi)\bigr),\\
\sigma_i&=\frac32\sqrt{\frac{M_i+\frac32D_GB_i}{a_i}}.
\end{aligned}\end{gathered}\end{equation*}\end{manuscriptwide}
\else \eqanchor{6.104}\begin{equation*}\tag{6.104}\begin{gathered}\begin{aligned}
B_i&=\|\nabla_iV\|_\infty,\\
M_i&=\sup_\xi\bigl(\max_xV(x,\xi)-\min_xV(x,\xi)\bigr),\\
\sigma_i&=\frac32\sqrt{\frac{M_i+\frac32D_GB_i}{a_i}}.
\end{aligned}\end{gathered}\end{equation*}\fi

Then $|\nabla_i\log\Omega|\le\sigma_i$ at every configuration. If $\operatorname{Ric}_G\ge r_Gg>0$, replace $\sigma_i$ by its minimum with $B_i/(a_ir_G)$. Any verified uniform bound $\omega_i$ on $\operatorname{osc}_i\log\Omega$, including Proposition \hyperlink{res6.41}{6.9}'s bound, also gives
\sbox{\equationmeasure}{$\displaystyle\begin{gathered}\|\nabla_i\log\Omega\|_\infty
\le\inf_{t>0}\left\{
\sqrt{\frac{\omega_i+M_it}{a_it}}+\frac{B_it}{2}
\right\}.\end{gathered}$}
\ifdim\wd\equationmeasure>\dimexpr\columnwidth-36pt\relax
\typeout{MANUSCRIPT-WIDE 6.105}\begin{manuscriptwide}\eqanchor{6.105}\begin{equation*}\tag{6.105}\begin{gathered}\|\nabla_i\log\Omega\|_\infty
\le\inf_{t>0}\left\{
\sqrt{\frac{\omega_i+M_it}{a_it}}+\frac{B_it}{2}
\right\}.\end{gathered}\end{equation*}\end{manuscriptwide}
\else \eqanchor{6.105}\begin{equation*}\tag{6.105}\begin{gathered}\|\nabla_i\log\Omega\|_\infty
\le\inf_{t>0}\left\{
\sqrt{\frac{\omega_i+M_it}{a_it}}+\frac{B_it}{2}
\right\}.\end{gathered}\end{equation*}\fi

These are actual pointwise bounds, independent of the number of other links. Their local data must still be controlled along a physical cutoff trajectory. The doubled bounds for $\log\rho$, $\rho=\Omega^2$, survive marginalizing other coordinates and conditioning coordinates other than $i$.\par

For a simple cycle $C$, Theorem \hyperlink{res6.7}{6.2}'s Fisher, semigroup and path-entropy estimates remain valid with
\sbox{\equationmeasure}{$\displaystyle\begin{gathered}\begin{aligned}
\kappa_C&=\sum_{e\in C}a_e,\\
s_C^{\rm new}&=\frac2{\kappa_C}\sum_{e\in C}a_e\sigma_e,\\
c_C^{\rm new}&=\kappa_C(s_C^{\rm new})^2 .
\end{aligned}\end{gathered}$}
\ifdim\wd\equationmeasure>\dimexpr\columnwidth-36pt\relax
\typeout{MANUSCRIPT-WIDE 6.106}\begin{manuscriptwide}\eqanchor{6.106}\begin{equation*}\tag{6.106}\begin{gathered}\begin{aligned}
\kappa_C&=\sum_{e\in C}a_e,\\
s_C^{\rm new}&=\frac2{\kappa_C}\sum_{e\in C}a_e\sigma_e,\\
c_C^{\rm new}&=\kappa_C(s_C^{\rm new})^2 .
\end{aligned}\end{gathered}\end{equation*}\end{manuscriptwide}
\else \eqanchor{6.106}\begin{equation*}\tag{6.106}\begin{gathered}\begin{aligned}
\kappa_C&=\sum_{e\in C}a_e,\\
s_C^{\rm new}&=\frac2{\kappa_C}\sum_{e\in C}a_e\sigma_e,\\
c_C^{\rm new}&=\kappa_C(s_C^{\rm new})^2 .
\end{aligned}\end{gathered}\end{equation*}\fi

For homogeneous $SU(3)$ Wilson weights, plaquette coefficient $b$ and at most $\nu$ incident plaquettes per link,
\sbox{\equationmeasure}{$\displaystyle\begin{gathered}\begin{aligned}
\sigma_i&\le\frac32
\sqrt{\frac{3\nu}{2}\left(1+\frac{D_G}{\sqrt6}\right)\frac ba},\\
c_C^{\rm new}&\le\frac{27}{2}\nu
\left(1+\frac{D_G}{\sqrt6}\right)|C|\,b .
\end{aligned}\end{gathered}$}
\ifdim\wd\equationmeasure>\dimexpr\columnwidth-36pt\relax
\typeout{MANUSCRIPT-WIDE 6.107}\begin{manuscriptwide}\eqanchor{6.107}\begin{equation*}\tag{6.107}\begin{gathered}\begin{aligned}
\sigma_i&\le\frac32
\sqrt{\frac{3\nu}{2}\left(1+\frac{D_G}{\sqrt6}\right)\frac ba},\\
c_C^{\rm new}&\le\frac{27}{2}\nu
\left(1+\frac{D_G}{\sqrt6}\right)|C|\,b .
\end{aligned}\end{gathered}\end{equation*}\end{manuscriptwide}
\else \eqanchor{6.107}\begin{equation*}\tag{6.107}\begin{gathered}\begin{aligned}
\sigma_i&\le\frac32
\sqrt{\frac{3\nu}{2}\left(1+\frac{D_G}{\sqrt6}\right)\frac ba},\\
c_C^{\rm new}&\le\frac{27}{2}\nu
\left(1+\frac{D_G}{\sqrt6}\right)|C|\,b .
\end{aligned}\end{gathered}\end{equation*}\fi

The old linear score estimate can be better near $b/a=0$. The new pointwise estimate does not tensorize eliminated interiors, remove the varied-link sum in finite-fiber comparisons, or make the physical refinement errors summable. The stochastic derivative formula is from [\hyperlink{ref96}{5}, Theorem 2.2]; its use here selects one coordinate of the full interacting vacuum, averages over the actual exterior paths, and closes the explicit pointwise estimate. Proof in the Supplement, Appendix \crosspdflink{Entropy_Geometry_and_Lattice_Gaps_Supplement.pdf}{appC.47}{C.5}.\par

For the coefficients $a_\ell=C_Eg_\ell^2/\ell$ and $b_\ell=C_B/(g_\ell^2\ell)$, the new fixed-link-count cycle certificate is $O(1/(\ell g_\ell^2))$, improving the earlier $O(1/(\ell g_\ell^6))$ certificate. It still grows on weak-bare-coupling refinement. Neither upper estimate is a lower bound on the actual error.\par

Positive pointwise curvature can fail for an actual Wilson vacuum. Proposition \crosspdflink{Entropy_Geometry_and_Lattice_Gaps_Supplement.pdf}{res6.50}{6.29} (Supplement, Appendix \crosspdflink{Entropy_Geometry_and_Lattice_Gaps_Supplement.pdf}{appC.49}{C.6}) conditions the product holonomy of two edge-disjoint SU(3) plaquettes and finds negative conditional weighted Ricci curvature on open sets sampled by physical invariant gradients at large magnetic/electric ratio. It identifies rare retained configurations, but good-set estimates still require control of eliminated variables and exceptional-set energy. This does not prove deterioration of the conditional Poincare constant; direct energy estimates may remain useful.\par

\hypertarget{sec6.3}{}\subsection{Exact reduction and the retained infrared form}

Exact elimination retains discarded response and temporal memory. We pass from the abstract relaxed form to actual-vacuum comparisons, then separate the retained infrared estimate from the shell bounds needed for a hierarchy.\par

\subsubsection*{Exact blocking and omitted modes}

\hypertarget{res6.3}{}\noindent \textbf{Theorem \hyperlink{res6.3}{6.4} (exact blocking and omitted sectors).} Let $T$ be a positive self-adjoint contraction, let $T\Omega=\Omega$ for a normalized vector $\Omega$, and let $a_t>0$ be its physical time step. Put $r=\|T|_{\Omega^\perp}\|$ and suppose $0<r\le1$. For every positive integer $k$,
\sbox{\equationmeasure}{$\displaystyle\begin{gathered}\|T^k|_{\Omega^\perp}\|=r^k,\\  -\frac{\log(r^k)}{ka_t}=-\frac{\log r}{a_t}.\end{gathered}$}
\ifdim\wd\equationmeasure>\dimexpr\columnwidth-36pt\relax
\typeout{MANUSCRIPT-WIDE 6.6}\begin{manuscriptwide}\eqanchor{6.6}\begin{equation*}\tag{6.6}\begin{gathered}\|T^k|_{\Omega^\perp}\|=r^k,\qquad -\frac{\log(r^k)}{ka_t}=-\frac{\log r}{a_t}.\end{gathered}\end{equation*}\end{manuscriptwide}
\else \eqanchor{6.6}\begin{equation*}\tag{6.6}\begin{gathered}\|T^k|_{\Omega^\perp}\|=r^k,\\  -\frac{\log(r^k)}{ka_t}=-\frac{\log r}{a_t}.\end{gathered}\end{equation*}\fi

More generally, let $S$ be a positive self-adjoint contraction on a centered physical Hilbert space, let $J:\mathscr K_c\to\mathscr K$ be an isometry, and set $P=JJ^*$, $Q=I-P$, and $A_n=J^*S^nJ$. Then
\sbox{\equationmeasure}{$\displaystyle\begin{gathered}A_2-A_1^2=J^*SQSJ=(QSJ)^*(QSJ)\ge0.\end{gathered}$}
\ifdim\wd\equationmeasure>\dimexpr\columnwidth-36pt\relax
\typeout{MANUSCRIPT-WIDE 6.7}\begin{manuscriptwide}\eqanchor{6.7}\begin{equation*}\tag{6.7}\begin{gathered}A_2-A_1^2=J^*SQSJ=(QSJ)^*(QSJ)\ge0.\end{gathered}\end{equation*}\end{manuscriptwide}
\else \eqanchor{6.7}\begin{equation*}\tag{6.7}\begin{gathered}A_2-A_1^2=J^*SQSJ=(QSJ)^*(QSJ)\ge0.\end{gathered}\end{equation*}\fi

If $\|PSP\|\le\rho<1$, $\|QSQ\|\le d<1$, and $\|QSP\|\le\eta$, where $\rho,d,\eta\ge0$, then
\sbox{\equationmeasure}{$\displaystyle\begin{gathered}\|S\|\le\frac{\rho+d+\sqrt{(\rho-d)^2+4\eta^2}}2<1\\ \text{provided}\\ \eta^2<(1-\rho)(1-d).\end{gathered}$}
\ifdim\wd\equationmeasure>\dimexpr\columnwidth-36pt\relax
\typeout{MANUSCRIPT-WIDE 6.8}\begin{manuscriptwide}\eqanchor{6.8}\begin{equation*}\tag{6.8}\begin{gathered}\|S\|\le\frac{\rho+d+\sqrt{(\rho-d)^2+4\eta^2}}2<1\quad\text{provided}\quad\eta^2<(1-\rho)(1-d).\end{gathered}\end{equation*}\end{manuscriptwide}
\else \eqanchor{6.8}\begin{equation*}\tag{6.8}\begin{gathered}\|S\|\le\frac{\rho+d+\sqrt{(\rho-d)^2+4\eta^2}}2<1\\ \text{provided}\\ \eta^2<(1-\rho)(1-d).\end{gathered}\end{equation*}\fi

Proof in the Supplement, Appendix \crosspdflink{Entropy_Geometry_and_Lattice_Gaps_Supplement.pdf}{appC.3}{C.7}.\par

Exact temporal blocking preserves the physical spectral edge. Compression can lose excursions through omitted modes, so iterating $A_1$ generally changes the dynamics. A retained contraction also leaves the complement uncontrolled: $S=\operatorname{diag}(\rho,1-\epsilon)$, with $P$ the first-coordinate projection, has zero cross term and retained norm $\rho$, while its full norm approaches one as $\epsilon\downarrow0$.\par

Theorem \hyperlink{res3.2}{3.2}'s physical information experiment also measures the memory lost in \hyperlink{eq6.7}{(6.7)}. The relevant loss is information about the initial perturbation that remains in the full configuration but is absent from the current coarse observation.\par

\hypertarget{res6.34}{}\noindent \textbf{Proposition \hyperlink{res6.34}{6.5} (discarded information and physical memory).} Use the actual stationary ground-state process of \hyperlink{eq3.5}{(3.5)}, with semigroup $T_t=P_t$. Let $\pi:M\to Y$ be a measurable gauge-equivariant observation into a standard Borel space with its specified induced gauge action, and put $\nu=\pi_*\mu$. On the physical invariant spaces let $Jf=f\circ\pi$, $P=JJ^*$, $Q=I-P$, $M_t=J^*T_tJ$ and $\delta_t=\|QT_tJ\|$. For real bounded centered dimensionless invariant $f$, prepare a balanced record $R=\pm1$ with conditional initial laws $(1\pm\epsilon Jf)d\mu$, where $|\epsilon|\|f\|_\infty<1$. Keep $R$ fixed while the configuration evolves, and write $Y_t=\pi(X_t)$ and $\mathcal L_t(\epsilon,f)=I(R:X_t\mid Y_t)$. Then
\sbox{\equationmeasure}{$\displaystyle\begin{gathered}\begin{aligned}
\left.\partial_\epsilon^2\mathcal L_t(\epsilon,f)\right|_{\epsilon=0}
&=\|QT_tJf\|^2\\
&=\langle f,(M_{2t}-M_t^2)f\rangle_\nu,\\
\|M_{2t}-M_t^2\|&=\delta_t^2.
\end{aligned}\end{gathered}$}
\ifdim\wd\equationmeasure>\dimexpr\columnwidth-36pt\relax
\typeout{MANUSCRIPT-WIDE 6.68}\begin{manuscriptwide}\eqanchor{6.68}\begin{equation*}\tag{6.68}\begin{gathered}\begin{aligned}
\left.\partial_\epsilon^2\mathcal L_t(\epsilon,f)\right|_{\epsilon=0}
&=\|QT_tJf\|^2\\
&=\langle f,(M_{2t}-M_t^2)f\rangle_\nu,\\
\|M_{2t}-M_t^2\|&=\delta_t^2.
\end{aligned}\end{gathered}\end{equation*}\end{manuscriptwide}
\else \eqanchor{6.68}\begin{equation*}\tag{6.68}\begin{gathered}\begin{aligned}
\left.\partial_\epsilon^2\mathcal L_t(\epsilon,f)\right|_{\epsilon=0}
&=\|QT_tJf\|^2\\
&=\langle f,(M_{2t}-M_t^2)f\rangle_\nu,\\
\|M_{2t}-M_t^2\|&=\delta_t^2.
\end{aligned}\end{gathered}\end{equation*}\fi

The supremum of the Hessian divided by $\|f\|_{L^2(\nu)}^2$ over all such nonzero $f$ is $\delta_t^2$, with value zero if the centered coarse space is trivial. For every integer $n\ge1$,
\sbox{\equationmeasure}{$\displaystyle\begin{gathered}\|M_{nt}-M_t^n\|\le\frac{n(n-1)}2\delta_t^2.\end{gathered}$}
\ifdim\wd\equationmeasure>\dimexpr\columnwidth-36pt\relax
\typeout{MANUSCRIPT-WIDE 6.69}\begin{manuscriptwide}\eqanchor{6.69}\begin{equation*}\tag{6.69}\begin{gathered}\|M_{nt}-M_t^n\|\le\frac{n(n-1)}2\delta_t^2.\end{gathered}\end{equation*}\end{manuscriptwide}
\else \eqanchor{6.69}\begin{equation*}\tag{6.69}\begin{gathered}\|M_{nt}-M_t^n\|\le\frac{n(n-1)}2\delta_t^2.\end{gathered}\end{equation*}\fi

Proof and finite-amplitude bounds in the Supplement, Appendix \crosspdflink{Entropy_Geometry_and_Lattice_Gaps_Supplement.pdf}{appC.33}{C.8}. The time $t$ has inverse-energy units; the information, bias and operator errors are dimensionless. These identities use the actual physical Euclidean dynamics and require no instantaneous generator-defect bound. A uniform estimate must control the normalized complete retained family. The loss is a measure of omitted prediction, not injected energy or particle mass: $J=I$ gives zero loss even for gapless dynamics. It is also distinct from $M_t-e^{-tA}$ for the instantaneous marginal-form generator.\par

Stable coarse observations can nevertheless conceal arbitrarily slow physical modes. On $\mathbb T^2$, the generator $K_\varepsilon=E_*(-\partial_y^2-\varepsilon\partial_z^2)$ has full gap $\varepsilon E_*$ for $E_*>0$ and $0<\varepsilon\le1$, while observation of $y$ alone has an exact semigroup with gap $E_*$ and zero memory. A record encoded as $1\pm a\cos z$, $0<a<1$, is invisible even to the complete observed history. Appendix \crosspdflink{Entropy_Geometry_and_Lattice_Gaps_Supplement.pdf}{appC.33}{C.8} proves its surviving full-system information bound. Such a record is initially hidden, unlike the retained preparations in Proposition \hyperlink{res6.34}{6.5}. Stable information loss can therefore produce apparent stationarity for an observer; physical stability additionally requires hidden-sector and coupling control.\par

Let $S_f$ be reflection positive with bounded positive transfer $T_f$. Assume $B^*$ preserves products, conjugation and constants, intertwines reflection, maps the coarse positive-time algebra into the fine positive-time algebra, and intertwines one coarse shift with $k$ fine shifts. Set $S_c(F)=S_f(B^*F)$. Equality of reflected pairings makes $J[F]=[B^*F]$ well defined and extends it isometrically. The coarse shift descends to bounded $T_c$ with $JT_c=T_f^kJ$. The closed range of $J$ is invariant and hence reducing for the self-adjoint $T_f^k$, proving positivity and exact dynamics there. The orthogonal physical sector remains uncontrolled. Reflection equivariance alone does not ensure the required half-space support.\par

Boundary gauge representations require similar care. A crossing Wilson loop is a global singlet formed from open path segments transforming at their endpoints. Independent averaging at a boundary endpoint kills a single fundamental matrix factor because its Haar mean is zero. Thus imposing separate singlet constraints on each segment removes a globally physical observable. A reduction must retain the matching boundary representations or quantify the effect of their exclusion. Complete history retention and complete boundary-sector control are different requirements.\par

\subsubsection*{Energy remaining after relaxation}

The compressed resolvent records memory without assuming an autonomous coarse semigroup. Its relaxed form is the minimum physical form energy over allowed discarded components at fixed retained vector. Static marginal compatibility alone does not identify this form.\par

\hypertarget{res6.35}{}\noindent \textbf{Theorem \hyperlink{res6.35}{6.6} (exact relaxed form and physical gap).} Let $\mathcal H_R,\mathcal H_U$ be Hilbert spaces, let $s$ be a densely defined closed nonnegative form on $\mathcal H_R$ with associated self-adjoint operator $S\ge0$, and let $D\ge\delta I$, $\delta>0$, be self-adjoint on $\mathcal H_U$. Let $L:\mathcal H_R\to\mathcal H_U$ be bounded with $\|L\|\le\kappa$. Define $W(x,y)=(x,y+Lx)$ and
\sbox{\equationmeasure}{$\displaystyle\begin{gathered}\begin{aligned}
D(q)&=W^{-1}\bigl(D(S^{1/2})\oplus D(D^{1/2})\bigr),\\
q[(x,y)]&=s[x]+\|D^{1/2}(y+Lx)\|^2.
\end{aligned}\end{gathered}$}
\ifdim\wd\equationmeasure>\dimexpr\columnwidth-36pt\relax
\typeout{MANUSCRIPT-WIDE 6.70}\begin{manuscriptwide}\eqanchor{6.70}\begin{equation*}\tag{6.70}\begin{gathered}\begin{aligned}
D(q)&=W^{-1}\bigl(D(S^{1/2})\oplus D(D^{1/2})\bigr),\\
q[(x,y)]&=s[x]+\|D^{1/2}(y+Lx)\|^2.
\end{aligned}\end{gathered}\end{equation*}\end{manuscriptwide}
\else \eqanchor{6.70}\begin{equation*}\tag{6.70}\begin{gathered}\begin{aligned}
D(q)&=W^{-1}\bigl(D(S^{1/2})\oplus D(D^{1/2})\bigr),\\
q[(x,y)]&=s[x]+\|D^{1/2}(y+Lx)\|^2.
\end{aligned}\end{gathered}\end{equation*}\fi

Then $q$ is closed and densely defined and determines a self-adjoint $K\ge0$. For $x\in D(S^{1/2})$, its minimum over all $y$ with $(x,y)\in D(q)$ is $s[x]$, attained at $y=-Lx$. For $z>0$, put
\sbox{\equationmeasure}{$\displaystyle\begin{gathered}\begin{aligned}
L_z&=D(D+z)^{-1}L,\\
F_z&=S+zI+zL^*D(D+z)^{-1}L.
\end{aligned}\end{gathered}$}
\ifdim\wd\equationmeasure>\dimexpr\columnwidth-36pt\relax
\typeout{MANUSCRIPT-WIDE 6.71}\begin{manuscriptwide}\eqanchor{6.71}\begin{equation*}\tag{6.71}\begin{gathered}\begin{aligned}
L_z&=D(D+z)^{-1}L,\\
F_z&=S+zI+zL^*D(D+z)^{-1}L.
\end{aligned}\end{gathered}\end{equation*}\end{manuscriptwide}
\else \eqanchor{6.71}\begin{equation*}\tag{6.71}\begin{gathered}\begin{aligned}
L_z&=D(D+z)^{-1}L,\\
F_z&=S+zI+zL^*D(D+z)^{-1}L.
\end{aligned}\end{gathered}\end{equation*}\fi

The operator $F_z$ is self-adjoint on $D(S)$, with $F_z\ge zI$. If $\iota_Rx=(x,0)$, then the exact retained resolvent is
\sbox{\equationmeasure}{$\displaystyle\begin{gathered}\iota_R^*(K+z)^{-1}\iota_R=F_z^{-1}.\end{gathered}$}
\ifdim\wd\equationmeasure>\dimexpr\columnwidth-36pt\relax
\typeout{MANUSCRIPT-WIDE 6.72}\begin{manuscriptwide}\eqanchor{6.72}\begin{equation*}\tag{6.72}\begin{gathered}\iota_R^*(K+z)^{-1}\iota_R=F_z^{-1}.\end{gathered}\end{equation*}\end{manuscriptwide}
\else \eqanchor{6.72}\begin{equation*}\tag{6.72}\begin{gathered}\iota_R^*(K+z)^{-1}\iota_R=F_z^{-1}.\end{gathered}\end{equation*}\fi

If additionally $S\ge mI$, $m>0$, then
\sbox{\equationmeasure}{$\displaystyle\begin{gathered}\begin{aligned}
K&\ge\lambda_*(m,\delta,\kappa)I,\\
\tau&=m+\delta(1+\kappa^2),\\
\lambda_*&=\frac{\tau-\sqrt{\tau^2-4m\delta}}2>0.
\end{aligned}\end{gathered}$}
\ifdim\wd\equationmeasure>\dimexpr\columnwidth-36pt\relax
\typeout{MANUSCRIPT-WIDE 6.73}\begin{manuscriptwide}\eqanchor{6.73}\begin{equation*}\tag{6.73}\begin{gathered}\begin{aligned}
K&\ge\lambda_*(m,\delta,\kappa)I,\\
\tau&=m+\delta(1+\kappa^2),\\
\lambda_*&=\frac{\tau-\sqrt{\tau^2-4m\delta}}2>0.
\end{aligned}\end{gathered}\end{equation*}\end{manuscriptwide}
\else \eqanchor{6.73}\begin{equation*}\tag{6.73}\begin{gathered}\begin{aligned}
K&\ge\lambda_*(m,\delta,\kappa)I,\\
\tau&=m+\delta(1+\kappa^2),\\
\lambda_*&=\frac{\tau-\sqrt{\tau^2-4m\delta}}2>0.
\end{aligned}\end{gathered}\end{equation*}\fi

This bound is sharp for these three constants. It also gives $\iota_R^*K^{-1}\iota_R=S^{-1}$. When $K$ is the actual vacuum-subtracted Hamiltonian on its entire vacuum complement, common $m,\delta,\kappa$ imply common physical decay $\|e^{-tK}\|\le e^{-t\lambda_*}$. No gap is asserted when a positive lower bound on $S$ is unavailable.\par

Proof in the Supplement, Appendix \crosspdflink{Entropy_Geometry_and_Lattice_Gaps_Supplement.pdf}{appC.34}{C.9}.\par

\noindent \textbf{Physical and domain interpretation.} The map $x\mapsto(x,-Lx)$ is an energy-minimizing lift, not a conditional expectation. The bare injection $\iota_R$ need not preserve the form domain; the bounded resolvent compression in \hyperlink{eq6.72}{(6.72)} remains well defined. The theorem therefore does not assume that $D^{1/2}L$ exists on all retained vectors. In a regular block representation $K=\left(\begin{smallmatrix}A&B^*\\B&D\end{smallmatrix}\right)$, the identification is $L=D^{-1}B$ and $S=A-B^*D^{-1}B$. These products are literal only when their domains have been justified; in a singular limit the normal form, rather than a difference of divergent operators, is the definition. Shorting for bounded operators [\hyperlink{ref101}{7}] and for nonnegative selfadjoint relations [\hyperlink{ref104}{8}, Section 3], together with Feshbach-Schur theory [\hyperlink{ref90}{6}], provides the operator background; this formulation specifies the minimizing lift, physical domains and comparison constants.\par

\subsubsection*{Covariance-free response}

A vertical energy bound alone controls the error in replacing the exact retained response by its relaxed static resolvent. Covariance enters the separate comparison with instantaneous static energy in Theorem \hyperlink{res6.36}{6.8} below.\par

\hypertarget{res6.38}{}\noindent \textbf{Theorem \hyperlink{res6.38}{6.7} (covariance-free relaxed response).} Let $q$ be a densely defined closed nonnegative form on $\mathcal H$, with operator $K$, and let $J:\mathcal H_R\to\mathcal H$ be an isometry. Set $P=JJ^*$, $Q=I-P$, and suppose
\sbox{\equationmeasure}{$\displaystyle\begin{gathered}q[g]\ge\delta\|Qg\|^2
\\ (g\in D(q)),\\  \delta>0.\end{gathered}$}
\ifdim\wd\equationmeasure>\dimexpr\columnwidth-36pt\relax
\typeout{MANUSCRIPT-WIDE 6.79}\begin{manuscriptwide}\eqanchor{6.79}\begin{equation*}\tag{6.79}\begin{gathered}q[g]\ge\delta\|Qg\|^2
\quad(g\in D(q)),\qquad \delta>0.\end{gathered}\end{equation*}\end{manuscriptwide}
\else \eqanchor{6.79}\begin{equation*}\tag{6.79}\begin{gathered}q[g]\ge\delta\|Qg\|^2
\\ (g\in D(q)),\\  \delta>0.\end{gathered}\end{equation*}\fi

On $V_R=J^*D(q)$, define the relaxed form and its positive-frequency counterpart by
\sbox{\equationmeasure}{$\displaystyle\begin{gathered}\begin{aligned}
s[x]&=\inf_{J^*g=x}q[g],\\
f_z[x]&=\inf_{J^*g=x}\bigl(q[g]+z\|g\|^2\bigr),
\qquad z>0,
\end{aligned}\end{gathered}$}
\ifdim\wd\equationmeasure>\dimexpr\columnwidth-36pt\relax
\typeout{MANUSCRIPT-WIDE 6.80}\begin{manuscriptwide}\eqanchor{6.80}\begin{equation*}\tag{6.80}\begin{gathered}\begin{aligned}
s[x]&=\inf_{J^*g=x}q[g],\\
f_z[x]&=\inf_{J^*g=x}\bigl(q[g]+z\|g\|^2\bigr),
\qquad z>0,
\end{aligned}\end{gathered}\end{equation*}\end{manuscriptwide}
\else \eqanchor{6.80}\begin{equation*}\tag{6.80}\begin{gathered}\begin{aligned}
s[x]&=\inf_{J^*g=x}q[g],\\
f_z[x]&=\inf_{J^*g=x}\bigl(q[g]+z\|g\|^2\bigr),
\qquad z>0,
\end{aligned}\end{gathered}\end{equation*}\fi

where the infima run over $g\in D(q)$. Both forms are closed and their minima are attained uniquely. Their operators $S$ and $F_z$ satisfy
\sbox{\equationmeasure}{$\displaystyle\begin{gathered}\begin{aligned}
s[x]+z\|x\|^2
&\le f_z[x]\\
&\le(1+z/\delta)s[x]+z\|x\|^2,\\
F_z^{-1}&=J^*(K+z)^{-1}J.
\end{aligned}\end{gathered}$}
\ifdim\wd\equationmeasure>\dimexpr\columnwidth-36pt\relax
\typeout{MANUSCRIPT-WIDE 6.81}\begin{manuscriptwide}\eqanchor{6.81}\begin{equation*}\tag{6.81}\begin{gathered}\begin{aligned}
s[x]+z\|x\|^2
&\le f_z[x]\\
&\le(1+z/\delta)s[x]+z\|x\|^2,\\
F_z^{-1}&=J^*(K+z)^{-1}J.
\end{aligned}\end{gathered}\end{equation*}\end{manuscriptwide}
\else \eqanchor{6.81}\begin{equation*}\tag{6.81}\begin{gathered}\begin{aligned}
s[x]+z\|x\|^2
&\le f_z[x]\\
&\le(1+z/\delta)s[x]+z\|x\|^2,\\
F_z^{-1}&=J^*(K+z)^{-1}J.
\end{aligned}\end{gathered}\end{equation*}\fi

In particular, $E_z=(S+z)^{-1}-F_z^{-1}$ is nonnegative and
\sbox{\equationmeasure}{$\displaystyle\begin{gathered}\|E_z\|\le
\frac{1}{\delta(1+\sqrt{1+z/\delta})^2}
\le\frac1{4\delta}.\end{gathered}$}
\ifdim\wd\equationmeasure>\dimexpr\columnwidth-36pt\relax
\typeout{MANUSCRIPT-WIDE 6.82}\begin{manuscriptwide}\eqanchor{6.82}\begin{equation*}\tag{6.82}\begin{gathered}\|E_z\|\le
\frac{1}{\delta(1+\sqrt{1+z/\delta})^2}
\le\frac1{4\delta}.\end{gathered}\end{equation*}\end{manuscriptwide}
\else \eqanchor{6.82}\begin{equation*}\tag{6.82}\begin{gathered}\|E_z\|\le
\frac{1}{\delta(1+\sqrt{1+z/\delta})^2}
\le\frac1{4\delta}.\end{gathered}\end{equation*}\fi

Neither preservation of $D(q)$ by $P$, a covariance bound, nor a positive infrared gap is assumed. If one separately proves $s[x]\ge m\|x\|^2$, $m>0$, then the sharp uniform consequence is
\sbox{\equationmeasure}{$\displaystyle\begin{gathered}K\ge\frac{m\delta}{m+\delta}I.\end{gathered}$}
\ifdim\wd\equationmeasure>\dimexpr\columnwidth-36pt\relax
\typeout{MANUSCRIPT-WIDE 6.83}\begin{manuscriptwide}\eqanchor{6.83}\begin{equation*}\tag{6.83}\begin{gathered}K\ge\frac{m\delta}{m+\delta}I.\end{gathered}\end{equation*}\end{manuscriptwide}
\else \eqanchor{6.83}\begin{equation*}\tag{6.83}\begin{gathered}K\ge\frac{m\delta}{m+\delta}I.\end{gathered}\end{equation*}\fi

For a physical mass gap, these hypotheses must hold on the entire physical vacuum complement.\par

The exact horizontal-vertical splitting of Theorem \hyperlink{res6.36}{6.8} gives \hyperlink{eq6.79}{(6.79)} at each of its finite regulators, so \hyperlink{eq6.82}{(6.82)} applies independently of the conditional covariance. A stronger conditional conclusion holds along an exactly compatible tower. Let $q_j$ be densely defined closed nonnegative forms and let $J_j:\mathcal H_{j-1}\to\mathcal H_j$ be isometries and suppose $q_j$ relaxes exactly to $q_{j-1}$, including domains, under $J_j^*$. If every $q_j$ satisfies \hyperlink{eq6.79}{(6.79)} with $Q_j=I-J_jJ_j^*$ and $\delta_j>0$, a single independently established base gap $q_0[x]\ge m_0\|x\|^2$, $m_0>0$, yields
\sbox{\equationmeasure}{$\displaystyle\begin{gathered}K_N\ge
\left(m_0^{-1}+\sum_{j=1}^N\delta_j^{-1}\right)^{-1}I.\end{gathered}$}
\ifdim\wd\equationmeasure>\dimexpr\columnwidth-36pt\relax
\typeout{MANUSCRIPT-WIDE 6.84}\begin{manuscriptwide}\eqanchor{6.84}\begin{equation*}\tag{6.84}\begin{gathered}K_N\ge
\left(m_0^{-1}+\sum_{j=1}^N\delta_j^{-1}\right)^{-1}I.\end{gathered}\end{equation*}\end{manuscriptwide}
\else \eqanchor{6.84}\begin{equation*}\tag{6.84}\begin{gathered}K_N\ge
\left(m_0^{-1}+\sum_{j=1}^N\delta_j^{-1}\right)^{-1}I.\end{gathered}\end{equation*}\fi

Thus summable inverse shell energies propagate that one infrared gap uniformly through refinement. For example, $\delta_j\ge\delta_*b^j$, $\delta_*>0$, $b>1$, gives a common lower bound $[m_0^{-1}+1/(\delta_*(b-1))]^{-1}$, while \hyperlink{eq6.82}{(6.82)} gives summable response errors. This extension assumes actual compatibility and vertical estimates for the generated forms; it does not derive them or establish cross-cutoff operator convergence. For the physical target, the base gap and shell estimates must also be uniform in spatial volume. These remain independent Yang-Mills requirements. Proof, including sharpness of \hyperlink{eq6.83}{(6.83)}, in the Supplement, Appendix \crosspdflink{Entropy_Geometry_and_Lattice_Gaps_Supplement.pdf}{appC.37}{C.10}.\par

\subsubsection*{Relaxation under the actual Yang-Mills vacuum}

The relaxed-energy comparison can be proved for a specified map under the actual vacuum. Use $H=-\sum_ea_e\Delta_e+V$ with $a_e>0$ and smooth real gauge-invariant $V$, its ground law $\mu=\rho\,dm$ with $\rho=\Omega^2$ for the normalized positive ground state of energy $E_0$, and the physical form $q[f]=\int\sum_ea_e|\nabla_e f|^2\,d\mu$ of \hyperlink{eq2.7}{(2.7)}. The group is nontrivial, compact and connected, with the fixed bi-invariant metric. Choose two distinct composable links and retain $\pi(u,v,\xi)=(uv,\xi)$, where $\xi$ contains every other fine link. Let $J$ be pullback, $P=JJ^*$, $Q=I-P$ and $q_A[x]=q[Jx]$ under the actual marginal $\nu$; let $A$ be its associated operator.\par

Put $s_e=\|\nabla_e\log\rho\|_\infty$. The group diameter $D_G$ and Haar gap $c_G$ give
\sbox{\equationmeasure}{$\displaystyle\begin{gathered}\delta=\frac{a_1a_2}{a_1+a_2}\,c_G
 e^{-D_G(s_1+s_2)}>0.\end{gathered}$}
\ifdim\wd\equationmeasure>\dimexpr\columnwidth-36pt\relax
\typeout{MANUSCRIPT-WIDE 6.74}\begin{manuscriptwide}\eqanchor{6.74}\begin{equation*}\tag{6.74}\begin{gathered}\delta=\frac{a_1a_2}{a_1+a_2}\,c_G
 e^{-D_G(s_1+s_2)}>0.\end{gathered}\end{equation*}\end{manuscriptwide}
\else \eqanchor{6.74}\begin{equation*}\tag{6.74}\begin{gathered}\delta=\frac{a_1a_2}{a_1+a_2}\,c_G
 e^{-D_G(s_1+s_2)}>0.\end{gathered}\end{equation*}\fi

Let $\mathcal G_h$ be the electric-orthogonal horizontal gradient for this product map, including every $\xi$ derivative, weighted by $\sqrt{a_1+a_2}$ on the product and $\sqrt{a_e}$ on the others. Appendix \crosspdflink{Entropy_Geometry_and_Lattice_Gaps_Supplement.pdf}{appC.35}{C.11} verifies the exact lift and energy splitting. Define
\sbox{\equationmeasure}{$\displaystyle\begin{gathered}\begin{aligned}
h&=\mathcal G_h\log\rho,\\
C&=\operatorname*{ess\,sup}_{y}
\|\operatorname{Cov}_\mu(h\mid\pi=y)\|_{\rm op},\\
C&\le\frac{(a_1s_1+a_2s_2)^2}{a_1+a_2}
 +\sum_{e\ne1,2}a_es_e^2.
\end{aligned}\end{gathered}$}
\ifdim\wd\equationmeasure>\dimexpr\columnwidth-36pt\relax
\typeout{MANUSCRIPT-WIDE 6.75}\begin{manuscriptwide}\eqanchor{6.75}\begin{equation*}\tag{6.75}\begin{gathered}\begin{aligned}
h&=\mathcal G_h\log\rho,\\
C&=\operatorname*{ess\,sup}_{y}
\|\operatorname{Cov}_\mu(h\mid\pi=y)\|_{\rm op},\\
C&\le\frac{(a_1s_1+a_2s_2)^2}{a_1+a_2}
 +\sum_{e\ne1,2}a_es_e^2.
\end{aligned}\end{gathered}\end{equation*}\end{manuscriptwide}
\else \eqanchor{6.75}\begin{equation*}\tag{6.75}\begin{gathered}\begin{aligned}
h&=\mathcal G_h\log\rho,\\
C&=\operatorname*{ess\,sup}_{y}
\|\operatorname{Cov}_\mu(h\mid\pi=y)\|_{\rm op},\\
C&\le\frac{(a_1s_1+a_2s_2)^2}{a_1+a_2}
 +\sum_{e\ne1,2}a_es_e^2.
\end{aligned}\end{gathered}\end{equation*}\fi

This finite constant retains covariance between components; conditional independence is not assumed.\par

\hypertarget{res6.36}{}\noindent \textbf{Theorem \hyperlink{res6.36}{6.8} (actual relaxed energy and exact response).} All infima below are over $g\in D(q)$. In this setting, $P,Q$ preserve the form domain and the restricted discarded operator satisfies $D\ge\delta I$. For $x\in D(A^{1/2})$, the infimum $s[x]=\inf_{J^*g=x}q[g]$ is attained uniquely, defines a closed form on $D(A^{1/2})$, and obeys
\sbox{\equationmeasure}{$\displaystyle\begin{gathered}\frac{\delta}{\delta+C}q_A[x]\le s[x]\le q_A[x].\end{gathered}$}
\ifdim\wd\equationmeasure>\dimexpr\columnwidth-36pt\relax
\typeout{MANUSCRIPT-WIDE 6.76}\begin{manuscriptwide}\eqanchor{6.76}\begin{equation*}\tag{6.76}\begin{gathered}\frac{\delta}{\delta+C}q_A[x]\le s[x]\le q_A[x].\end{gathered}\end{equation*}\end{manuscriptwide}
\else \eqanchor{6.76}\begin{equation*}\tag{6.76}\begin{gathered}\frac{\delta}{\delta+C}q_A[x]\le s[x]\le q_A[x].\end{gathered}\end{equation*}\fi

For $z>0$, the form $f_z[x]=\inf_{J^*g=x}\{q[g]+z\|g\|^2\}$ is closed on the same domain. If $F_z$ is its operator and $K=\Omega^{-1}(H-E_0)\Omega$, then
\sbox{\equationmeasure}{$\displaystyle\begin{gathered}\begin{aligned}
\theta_z&=\frac{\delta+z}{\delta+z+C},\\
z\|x\|^2+\theta_zq_A[x]
&\le f_z[x]\le z\|x\|^2+q_A[x],\\
J^*(K+z)^{-1}J&=F_z^{-1},\\
(A+z)^{-1}&\le F_z^{-1}\le(\theta_zA+z)^{-1}.
\end{aligned}\end{gathered}$}
\ifdim\wd\equationmeasure>\dimexpr\columnwidth-36pt\relax
\typeout{MANUSCRIPT-WIDE 6.77}\begin{manuscriptwide}\eqanchor{6.77}\begin{equation*}\tag{6.77}\begin{gathered}\begin{aligned}
\theta_z&=\frac{\delta+z}{\delta+z+C},\\
z\|x\|^2+\theta_zq_A[x]
&\le f_z[x]\le z\|x\|^2+q_A[x],\\
J^*(K+z)^{-1}J&=F_z^{-1},\\
(A+z)^{-1}&\le F_z^{-1}\le(\theta_zA+z)^{-1}.
\end{aligned}\end{gathered}\end{equation*}\end{manuscriptwide}
\else \eqanchor{6.77}\begin{equation*}\tag{6.77}\begin{gathered}\begin{aligned}
\theta_z&=\frac{\delta+z}{\delta+z+C},\\
z\|x\|^2+\theta_zq_A[x]
&\le f_z[x]\le z\|x\|^2+q_A[x],\\
J^*(K+z)^{-1}J&=F_z^{-1},\\
(A+z)^{-1}&\le F_z^{-1}\le(\theta_zA+z)^{-1}.
\end{aligned}\end{gathered}\end{equation*}\fi

No smallness condition $C<\delta$ is required. These are form and resolvent orders, not orders between the corresponding semigroups. The conclusions hold on the full spaces and their gauge-invariant restrictions. If the actual retained physical form independently satisfies $q_A[x]\ge a_R\|x\|^2$ on its centered subspace, $a_R>0$, the complete physical gap obeys
\sbox{\equationmeasure}{$\displaystyle\begin{gathered}\begin{aligned}
\tau_R&=a_R+\delta+C,\\
\Delta(K)&\ge
\frac{\tau_R-\sqrt{\tau_R^2-4a_R\delta}}2>0.
\end{aligned}\end{gathered}$}
\ifdim\wd\equationmeasure>\dimexpr\columnwidth-36pt\relax
\typeout{MANUSCRIPT-WIDE 6.78}\begin{manuscriptwide}\eqanchor{6.78}\begin{equation*}\tag{6.78}\begin{gathered}\begin{aligned}
\tau_R&=a_R+\delta+C,\\
\Delta(K)&\ge
\frac{\tau_R-\sqrt{\tau_R^2-4a_R\delta}}2>0.
\end{aligned}\end{gathered}\end{equation*}\end{manuscriptwide}
\else \eqanchor{6.78}\begin{equation*}\tag{6.78}\begin{gathered}\begin{aligned}
\tau_R&=a_R+\delta+C,\\
\Delta(K)&\ge
\frac{\tau_R-\sqrt{\tau_R^2-4a_R\delta}}2>0.
\end{aligned}\end{gathered}\end{equation*}\fi

Proof, explicit score bounds and energy-relative dressing estimates in the Supplement, Appendix \crosspdflink{Entropy_Geometry_and_Lattice_Gaps_Supplement.pdf}{appC.35}{C.11}. Theorem \hyperlink{res6.7}{6.2} supplies $s_e\le2\|\nabla_eV\|_\infty/(a_er_G)$ when $\operatorname{Ric}_G\ge r_Gg>0$. The covariance norm in \hyperlink{eq6.75}{(6.75)} can be much smaller than its displayed trace bound, which can grow with volume. A pure edge subdivision may discard no physical degree of freedom. This is an actual finite-regulator comparison, not a uniform continuum gap. It connects conditional information bounds to relaxed energy and exact response; applying this sufficient comparison uniformly requires control of the tensor, discarded coercivity and retained infrared form.\par

\subsubsection*{Actual fiber coercivity}

\hypertarget{res6.41}{}\noindent \textbf{Proposition \hyperlink{res6.41}{6.9} (local vacuum oscillation and fiber coercivity).} Consider the actual positive normalized ground state $\Omega$ of the finite compact-link Hamiltonian in Theorem \hyperlink{res6.36}{6.8}. Let $G$ be compact and connected, $n=\dim G$, and let $p_\tau$ be the heat kernel of $e^{\tau\Delta_G}$ relative to normalized Haar measure. Define
\sbox{\equationmeasure}{$\displaystyle\begin{gathered}\begin{aligned}
M_i&=\sup_\xi\left[\max_vV(v,\xi)-\min_vV(v,\xi)\right],\\
H_G(\tau)&=\frac{\max_{x,z}p_\tau(x,z)}{\min_{x,z}p_\tau(x,z)},\\
\Phi_G(m)&=\inf_{\tau>0}\{m\tau+\log H_G(\tau)\},\\
\omega_i&=\Phi_G(M_i/a_i).
\end{aligned}\end{gathered}$}
\ifdim\wd\equationmeasure>\dimexpr\columnwidth-36pt\relax
\typeout{MANUSCRIPT-WIDE 6.89}\begin{manuscriptwide}\eqanchor{6.89}\begin{equation*}\tag{6.89}\begin{gathered}\begin{aligned}
M_i&=\sup_\xi\left[\max_vV(v,\xi)-\min_vV(v,\xi)\right],\\
H_G(\tau)&=\frac{\max_{x,z}p_\tau(x,z)}{\min_{x,z}p_\tau(x,z)},\\
\Phi_G(m)&=\inf_{\tau>0}\{m\tau+\log H_G(\tau)\},\\
\omega_i&=\Phi_G(M_i/a_i).
\end{aligned}\end{gathered}\end{equation*}\end{manuscriptwide}
\else \eqanchor{6.89}\begin{equation*}\tag{6.89}\begin{gathered}\begin{aligned}
M_i&=\sup_\xi\left[\max_vV(v,\xi)-\min_vV(v,\xi)\right],\\
H_G(\tau)&=\frac{\max_{x,z}p_\tau(x,z)}{\min_{x,z}p_\tau(x,z)},\\
\Phi_G(m)&=\inf_{\tau>0}\{m\tau+\log H_G(\tau)\},\\
\omega_i&=\Phi_G(M_i/a_i).
\end{aligned}\end{gathered}\end{equation*}\fi

Here $v$ is link $i$ and $\xi$ contains every other link. For every surrounding configuration, the actual vacuum and the two-link fiber of Theorem \hyperlink{res6.36}{6.8} satisfy
\sbox{\equationmeasure}{$\displaystyle\begin{gathered}\begin{aligned}
\operatorname{osc}_v\log\Omega(v,\xi)&\le\omega_i,\\
q_v[g]&\ge\delta_{\rm FK}\|Qg\|^2,\\
\delta_{\rm FK}&=a_vc_Ge^{-2(\omega_1+\omega_2)},\\
a_v&=\frac{a_1a_2}{a_1+a_2}.
\end{aligned}\end{gathered}$}
\ifdim\wd\equationmeasure>\dimexpr\columnwidth-36pt\relax
\typeout{MANUSCRIPT-WIDE 6.90}\begin{manuscriptwide}\eqanchor{6.90}\begin{equation*}\tag{6.90}\begin{gathered}\begin{aligned}
\operatorname{osc}_v\log\Omega(v,\xi)&\le\omega_i,\\
q_v[g]&\ge\delta_{\rm FK}\|Qg\|^2,\\
\delta_{\rm FK}&=a_vc_Ge^{-2(\omega_1+\omega_2)},\\
a_v&=\frac{a_1a_2}{a_1+a_2}.
\end{aligned}\end{gathered}\end{equation*}\end{manuscriptwide}
\else \eqanchor{6.90}\begin{equation*}\tag{6.90}\begin{gathered}\begin{aligned}
\operatorname{osc}_v\log\Omega(v,\xi)&\le\omega_i,\\
q_v[g]&\ge\delta_{\rm FK}\|Qg\|^2,\\
\delta_{\rm FK}&=a_vc_Ge^{-2(\omega_1+\omega_2)},\\
a_v&=\frac{a_1a_2}{a_1+a_2}.
\end{aligned}\end{gathered}\end{equation*}\fi

No gap on the complete physical sector is assumed. Since a bi-invariant group metric has nonnegative Ricci curvature, the Li-Yau inequality [\hyperlink{ref91}{28}] gives the explicit bound
\sbox{\equationmeasure}{$\displaystyle\begin{gathered}\begin{aligned}
H_G(\tau)&\le2^n\exp\!\left(\frac{3D_G^2}{4\tau}\right),\\
\Phi_G(m)&\le n\log2+D_G\sqrt{3m}\quad(m>0),\\
\Phi_G(0)&=0.
\end{aligned}\end{gathered}$}
\ifdim\wd\equationmeasure>\dimexpr\columnwidth-36pt\relax
\typeout{MANUSCRIPT-WIDE 6.91}\begin{manuscriptwide}\eqanchor{6.91}\begin{equation*}\tag{6.91}\begin{gathered}\begin{aligned}
H_G(\tau)&\le2^n\exp\!\left(\frac{3D_G^2}{4\tau}\right),\\
\Phi_G(m)&\le n\log2+D_G\sqrt{3m}\quad(m>0),\\
\Phi_G(0)&=0.
\end{aligned}\end{gathered}\end{equation*}\end{manuscriptwide}
\else \eqanchor{6.91}\begin{equation*}\tag{6.91}\begin{gathered}\begin{aligned}
H_G(\tau)&\le2^n\exp\!\left(\frac{3D_G^2}{4\tau}\right),\\
\Phi_G(m)&\le n\log2+D_G\sqrt{3m}\quad(m>0),\\
\Phi_G(0)&=0.
\end{aligned}\end{gathered}\end{equation*}\fi

When Theorem \hyperlink{res6.7}{6.2} applies, each $\omega_i$ in \hyperlink{eq6.90}{(6.90)} may be replaced by $\min\{\Phi_G(M_i/a_i),D_GB_i/(a_ir_G)\}$. The exact heat-kernel optimization can improve the explicit square-root bound. These are oscillation estimates; they do not replace the pointwise score bounds used elsewhere.\par

A complementary variational argument gives the local electric-energy bound $a_i\int|\nabla_i\Omega|^2dm\le M_i$. Its proof and resulting improvements to the averaged constants of Proposition \hyperlink{res6.39}{6.10} appear in Appendix \crosspdflink{Entropy_Geometry_and_Lattice_Gaps_Supplement.pdf}{appC.40}{C.12}. Coordinatewise density comparison likewise improves the conditional finite-block certificates of Proposition \hyperlink{res4.7}{4.5}.\par

Bounded plaquette incidence makes these Wilson constants volume independent. At large $b/a$, the explicit fiber exponent grows as $\sqrt{b/a}$ rather than $b/a$, improving discarded-sector coercivity uniformly in surrounding configurations. A retained infrared gap, uniform physical refinement and the full-sector essential-supremum covariance bound remain unproved.\par

Proof in the Supplement, Appendix \crosspdflink{Entropy_Geometry_and_Lattice_Gaps_Supplement.pdf}{appC.40}{C.12}.\par

\subsubsection*{Averaged covariance and smooth probes}

\hypertarget{res6.39}{}\noindent \textbf{Proposition \hyperlink{res6.39}{6.10} (averaged covariance and smooth-probe response).} In Theorem \hyperlink{res6.36}{6.8} suppose $\operatorname{Ric}_G\ge r_Gg$, $r_G>0$, and set $B_i=\|\nabla_iV\|_\infty$ for the two merged links. Retain its actual horizontal score $h$, marginal $\nu$, fiber bound $\delta$, and $s_i=\|\nabla_i\log\rho\|_\infty$. Put
\sbox{\equationmeasure}{$\displaystyle\begin{gathered}\begin{aligned}
a_v&=\frac{a_1a_2}{a_1+a_2},\\
\Sigma(y)&=\operatorname{Cov}_\mu(h\mid\pi=y),\\
C_{\rm pair}&=\frac{(a_1s_1+a_2s_2)^2}{a_1+a_2},\\
\overline C&=C_{\rm pair}+\frac{a_v}{\delta}
\left(\frac{B_1}{\sqrt{a_1r_G}}+
\frac{B_2}{\sqrt{a_2r_G}}\right)^2.
\end{aligned}\end{gathered}$}
\ifdim\wd\equationmeasure>\dimexpr\columnwidth-36pt\relax
\typeout{MANUSCRIPT-WIDE 6.85}\begin{manuscriptwide}\eqanchor{6.85}\begin{equation*}\tag{6.85}\begin{gathered}\begin{aligned}
a_v&=\frac{a_1a_2}{a_1+a_2},\\
\Sigma(y)&=\operatorname{Cov}_\mu(h\mid\pi=y),\\
C_{\rm pair}&=\frac{(a_1s_1+a_2s_2)^2}{a_1+a_2},\\
\overline C&=C_{\rm pair}+\frac{a_v}{\delta}
\left(\frac{B_1}{\sqrt{a_1r_G}}+
\frac{B_2}{\sqrt{a_2r_G}}\right)^2.
\end{aligned}\end{gathered}\end{equation*}\end{manuscriptwide}
\else \eqanchor{6.85}\begin{equation*}\tag{6.85}\begin{gathered}\begin{aligned}
a_v&=\frac{a_1a_2}{a_1+a_2},\\
\Sigma(y)&=\operatorname{Cov}_\mu(h\mid\pi=y),\\
C_{\rm pair}&=\frac{(a_1s_1+a_2s_2)^2}{a_1+a_2},\\
\overline C&=C_{\rm pair}+\frac{a_v}{\delta}
\left(\frac{B_1}{\sqrt{a_1r_G}}+
\frac{B_2}{\sqrt{a_2r_G}}\right)^2.
\end{aligned}\end{gathered}\end{equation*}\fi

Then
\sbox{\equationmeasure}{$\displaystyle\begin{gathered}\begin{aligned}
\int\operatorname{tr}\Sigma\,d\nu&\le\overline C,\\
\nu\{\|\Sigma\|_{\rm op}>R\}&\le\overline C/R\qquad(R>0).
\end{aligned}\end{gathered}$}
\ifdim\wd\equationmeasure>\dimexpr\columnwidth-36pt\relax
\typeout{MANUSCRIPT-WIDE 6.86}\begin{manuscriptwide}\eqanchor{6.86}\begin{equation*}\tag{6.86}\begin{gathered}\begin{aligned}
\int\operatorname{tr}\Sigma\,d\nu&\le\overline C,\\
\nu\{\|\Sigma\|_{\rm op}>R\}&\le\overline C/R\qquad(R>0).
\end{aligned}\end{gathered}\end{equation*}\end{manuscriptwide}
\else \eqanchor{6.86}\begin{equation*}\tag{6.86}\begin{gathered}\begin{aligned}
\int\operatorname{tr}\Sigma\,d\nu&\le\overline C,\\
\nu\{\|\Sigma\|_{\rm op}>R\}&\le\overline C/R\qquad(R>0).
\end{aligned}\end{gathered}\end{equation*}\fi

For smooth retained $f$, define $\ell_\infty(f)=\|\nabla_Af\|_\infty$, $Rf=KJf-JAf$ and $Lf=D^{-1}Rf$. Let $f_0=s$, with $f_z$ as in Theorem \hyperlink{res6.36}{6.8}. For $z\ge0$,
\sbox{\equationmeasure}{$\displaystyle\begin{gathered}\begin{aligned}
\|Rf\|^2&\le\overline C\ell_\infty(f)^2,\\
0&\le q_A[f]+z\|f\|^2-f_z[f]\\
&\le\frac{\overline C}{\delta+z}\ell_\infty(f)^2,\\
\|Lf\|^2&\le\frac{\overline C}{\delta^2}\ell_\infty(f)^2,\\
0&\le f_z[f]-s[f]-z\|f\|^2\\
&\le\frac{z\overline C}{\delta(\delta+z)}\ell_\infty(f)^2.
\end{aligned}\end{gathered}$}
\ifdim\wd\equationmeasure>\dimexpr\columnwidth-36pt\relax
\typeout{MANUSCRIPT-WIDE 6.87}\begin{manuscriptwide}\eqanchor{6.87}\begin{equation*}\tag{6.87}\begin{gathered}\begin{aligned}
\|Rf\|^2&\le\overline C\ell_\infty(f)^2,\\
0&\le q_A[f]+z\|f\|^2-f_z[f]\\
&\le\frac{\overline C}{\delta+z}\ell_\infty(f)^2,\\
\|Lf\|^2&\le\frac{\overline C}{\delta^2}\ell_\infty(f)^2,\\
0&\le f_z[f]-s[f]-z\|f\|^2\\
&\le\frac{z\overline C}{\delta(\delta+z)}\ell_\infty(f)^2.
\end{aligned}\end{gathered}\end{equation*}\end{manuscriptwide}
\else \eqanchor{6.87}\begin{equation*}\tag{6.87}\begin{gathered}\begin{aligned}
\|Rf\|^2&\le\overline C\ell_\infty(f)^2,\\
0&\le q_A[f]+z\|f\|^2-f_z[f]\\
&\le\frac{\overline C}{\delta+z}\ell_\infty(f)^2,\\
\|Lf\|^2&\le\frac{\overline C}{\delta^2}\ell_\infty(f)^2,\\
0&\le f_z[f]-s[f]-z\|f\|^2\\
&\le\frac{z\overline C}{\delta(\delta+z)}\ell_\infty(f)^2.
\end{aligned}\end{gathered}\end{equation*}\fi

The energy-minimizing lift is $Jf-Lf$. For smooth $f,g$ and $t\ge0$, the discarded response also obeys
\sbox{\equationmeasure}{$\displaystyle\begin{gathered}|\langle Rf,e^{-tD}Rg\rangle|
\le\overline C e^{-\delta t}
\ell_\infty(f)\ell_\infty(g).\end{gathered}$}
\ifdim\wd\equationmeasure>\dimexpr\columnwidth-36pt\relax
\typeout{MANUSCRIPT-WIDE 6.88}\begin{manuscriptwide}\eqanchor{6.88}\begin{equation*}\tag{6.88}\begin{gathered}|\langle Rf,e^{-tD}Rg\rangle|
\le\overline C e^{-\delta t}
\ell_\infty(f)\ell_\infty(g).\end{gathered}\end{equation*}\end{manuscriptwide}
\else \eqanchor{6.88}\begin{equation*}\tag{6.88}\begin{gathered}|\langle Rf,e^{-tD}Rg\rangle|
\le\overline C e^{-\delta t}
\ell_\infty(f)\ell_\infty(g).\end{gathered}\end{equation*}\fi

These statements hold on the full spaces and their physical gauge-invariant restrictions. Theorem \hyperlink{res6.7}{6.2} gives $s_i\le2B_i/(a_ir_G)$, so uniform local data make the constants independent of link count. Equation \hyperlink{eq6.86}{(6.86)} averages over surrounding retained configurations: it gives neither a uniform bound after fixing the surrounding retained configuration nor an essential-supremum bound on $\Sigma$. Thus \hyperlink{eq6.87}{(6.87)} controls specified smooth probes, not every unit-energy physical vector; \hyperlink{eq6.88}{(6.88)} is a weak memory form on that core. Here $\overline C,\delta,z$ have energy units and $t$ inverse-energy units. Uniform refinement and a retained infrared gap remain unproved.\par

Proof in the Supplement, Appendix \crosspdflink{Entropy_Geometry_and_Lattice_Gaps_Supplement.pdf}{appC.38}{C.13}.\par

\subsubsection*{Localization and infrared stiffness}

A pinching or tube-energy comparison can provide the independent infrared input only if it covers every retained physical direction, including local low-energy components and localization errors. The following criterion states these requirements after exact relaxation.\par

\hypertarget{res6.42}{}\noindent \textbf{Theorem \hyperlink{res6.42}{6.11} (complete-channel infrared certificate).} Let $s$ be the densely defined closed nonnegative infrared form on the nonzero centered physical space $\mathcal H_R^0$, after all hidden variables in this comparison have been exactly relaxed, and let $S$ be its operator. Let $\mathcal D$ be a complex linear form core for $s$, and let $V:\mathcal H_R^0\to\bigoplus_{i=1}^N\mathcal K_i$ be an isometry, with components $V_i$. For each channel choose a closed nonnegative form $\ell_i$, an orthogonal projection $P_i$, and $\beta_i>0$. Suppose $V_i\mathcal D\subset D(\ell_i)$ and, for $z\in D(\ell_i)$ and $x\in\mathcal D$,
\sbox{\equationmeasure}{$\displaystyle\begin{gathered}\begin{aligned}
\ell_i[z]&\ge\beta_i\|(I-P_i)z\|^2,\\
\sum_i\ell_i[V_ix]&\le s[x]+\varepsilon\|x\|^2,
\qquad\varepsilon\ge0.
\end{aligned}\end{gathered}$}
\ifdim\wd\equationmeasure>\dimexpr\columnwidth-36pt\relax
\typeout{MANUSCRIPT-WIDE 6.92}\begin{manuscriptwide}\eqanchor{6.92}\begin{equation*}\tag{6.92}\begin{gathered}\begin{aligned}
\ell_i[z]&\ge\beta_i\|(I-P_i)z\|^2,\\
\sum_i\ell_i[V_ix]&\le s[x]+\varepsilon\|x\|^2,
\qquad\varepsilon\ge0.
\end{aligned}\end{gathered}\end{equation*}\end{manuscriptwide}
\else \eqanchor{6.92}\begin{equation*}\tag{6.92}\begin{gathered}\begin{aligned}
\ell_i[z]&\ge\beta_i\|(I-P_i)z\|^2,\\
\sum_i\ell_i[V_ix]&\le s[x]+\varepsilon\|x\|^2,
\qquad\varepsilon\ge0.
\end{aligned}\end{gathered}\end{equation*}\fi

The second inequality is one additive comparison for the relaxed form. Suppose also that the local low-energy components obey
\sbox{\equationmeasure}{$\displaystyle\begin{gathered}\sum_i\|P_iV_ix\|^2\le\eta\|x\|^2,
\\  0\le\eta<1.\end{gathered}$}
\ifdim\wd\equationmeasure>\dimexpr\columnwidth-36pt\relax
\typeout{MANUSCRIPT-WIDE 6.93}\begin{manuscriptwide}\eqanchor{6.93}\begin{equation*}\tag{6.93}\begin{gathered}\sum_i\|P_iV_ix\|^2\le\eta\|x\|^2,
\qquad 0\le\eta<1.\end{gathered}\end{equation*}\end{manuscriptwide}
\else \eqanchor{6.93}\begin{equation*}\tag{6.93}\begin{gathered}\sum_i\|P_iV_ix\|^2\le\eta\|x\|^2,
\\  0\le\eta<1.\end{gathered}\end{equation*}\fi

Set $\beta=\min_i\beta_i$. If $m_0=(1-\eta)\beta-\varepsilon>0$, then $S\ge m_0I$. If $s=q_0$ is the base of the exactly compatible centered physical tower of Theorem \hyperlink{res6.38}{6.7}, with its full vertical bounds $\delta_j>0$, then
\sbox{\equationmeasure}{$\displaystyle\begin{gathered}\begin{aligned}
S&\ge m_0I,\qquad m_0=(1-\eta)\beta-\varepsilon,\\
K_n&\ge\left(m_0^{-1}+\sum_{j=1}^n\delta_j^{-1}\right)^{-1}I.
\end{aligned}\end{gathered}$}
\ifdim\wd\equationmeasure>\dimexpr\columnwidth-36pt\relax
\typeout{MANUSCRIPT-WIDE 6.94}\begin{manuscriptwide}\eqanchor{6.94}\begin{equation*}\tag{6.94}\begin{gathered}\begin{aligned}
S&\ge m_0I,\qquad m_0=(1-\eta)\beta-\varepsilon,\\
K_n&\ge\left(m_0^{-1}+\sum_{j=1}^n\delta_j^{-1}\right)^{-1}I.
\end{aligned}\end{gathered}\end{equation*}\end{manuscriptwide}
\else \eqanchor{6.94}\begin{equation*}\tag{6.94}\begin{gathered}\begin{aligned}
S&\ge m_0I,\qquad m_0=(1-\eta)\beta-\varepsilon,\\
K_n&\ge\left(m_0^{-1}+\sum_{j=1}^n\delta_j^{-1}\right)^{-1}I.
\end{aligned}\end{gathered}\end{equation*}\fi

Proof in the Supplement, Appendix \crosspdflink{Entropy_Geometry_and_Lattice_Gaps_Supplement.pdf}{appC.41}{C.14}. The constants $\beta_i$, $\varepsilon$, $m_0$ and $\delta_j$ have physical energy units; $\eta$ is dimensionless. Any size or tube estimate used here must establish the channel inequalities \hyperlink{eq6.92}{(6.92)} on their stated domains, with the same vacuum subtraction.\par

For finite-rank $P_i$, the optimal $\eta$ is the norm of the Gram matrix of adjoint images of local low modes. This is the exact leakage norm on the entire centered space, not a variational excitation-energy estimate. Appendix \crosspdflink{Entropy_Geometry_and_Lattice_Gaps_Supplement.pdf}{appC.41}{C.14} gives the actual ground-law localization identity and vacuum obstruction. Channels and low modes may proliferate with volume; common positive $m_0$, exact hierarchy compatibility and a summable inverse vertical-energy budget remain unproved Yang-Mills inputs.\par

\subsubsection*{Actual terminal hierarchies}

\hypertarget{res6.45}{}\noindent \textbf{Theorem \hyperlink{res6.45}{6.12} (actual terminal regulator and exact relaxed hierarchy).} Fix a finite spatial compact-link $SU(3)$ Hamiltonian of Theorem \hyperlink{res2.2}{2.2}, with actual vacuum $\Omega_N$, energy $E_{0,N}$ and ground-transformed form $q_N$ on the complete physical vacuum complement $\mathcal H_N$. Its operator is $K_N=\Omega_N^{-1}(H_N-E_{0,N})\Omega_N$. Let $\mathcal H_0\subset\cdots\subset\mathcal H_N$ be closed retained holonomy subspaces in this actual ground law, with orthogonal projections $P_j$ and $P_N=I$. For example, nested gauge-stable holonomy sigma-algebras define these spaces by conditional expectation. Omit any zero retained space. Put
\sbox{\equationmeasure}{$\displaystyle\begin{gathered}\begin{aligned}
D(s_j^{(N)})&=P_jD(q_N),\\
s_j^{(N)}[x]&=\min_{\substack{g\in D(q_N)\\P_jg=x}}q_N[g].
\end{aligned}\end{gathered}$}
\ifdim\wd\equationmeasure>\dimexpr\columnwidth-36pt\relax
\typeout{MANUSCRIPT-WIDE 6.95}\begin{manuscriptwide}\eqanchor{6.95}\begin{equation*}\tag{6.95}\begin{gathered}\begin{aligned}
D(s_j^{(N)})&=P_jD(q_N),\\
s_j^{(N)}[x]&=\min_{\substack{g\in D(q_N)\\P_jg=x}}q_N[g].
\end{aligned}\end{gathered}\end{equation*}\end{manuscriptwide}
\else \eqanchor{6.95}\begin{equation*}\tag{6.95}\begin{gathered}\begin{aligned}
D(s_j^{(N)})&=P_jD(q_N),\\
s_j^{(N)}[x]&=\min_{\substack{g\in D(q_N)\\P_jg=x}}q_N[g].
\end{aligned}\end{gathered}\end{equation*}\fi

These are densely defined closed forms with compact resolvents. Their minima are unique, their linear minimizing lifts compose exactly, and adjacent relaxation is compatible including domains. If $\iota_j:\mathcal H_j\to\mathcal H_N$ denotes inclusion, then
\sbox{\equationmeasure}{$\displaystyle\begin{gathered}\begin{aligned}
(S_j^{(N)})^{-1}&=\iota_j^*K_N^{-1}\iota_j,\\
s_{j-1}^{(N)}[x]&=\min_{\substack{y\in D(s_j^{(N)})\\P_{j-1}y=x}}s_j^{(N)}[y],\\
D(s_{j-1}^{(N)})&=P_{j-1}D(s_j^{(N)}).
\end{aligned}\end{gathered}$}
\ifdim\wd\equationmeasure>\dimexpr\columnwidth-36pt\relax
\typeout{MANUSCRIPT-WIDE 6.96}\begin{manuscriptwide}\eqanchor{6.96}\begin{equation*}\tag{6.96}\begin{gathered}\begin{aligned}
(S_j^{(N)})^{-1}&=\iota_j^*K_N^{-1}\iota_j,\\
s_{j-1}^{(N)}[x]&=\min_{\substack{y\in D(s_j^{(N)})\\P_{j-1}y=x}}s_j^{(N)}[y],\\
D(s_{j-1}^{(N)})&=P_{j-1}D(s_j^{(N)}).
\end{aligned}\end{gathered}\end{equation*}\end{manuscriptwide}
\else \eqanchor{6.96}\begin{equation*}\tag{6.96}\begin{gathered}\begin{aligned}
(S_j^{(N)})^{-1}&=\iota_j^*K_N^{-1}\iota_j,\\
s_{j-1}^{(N)}[x]&=\min_{\substack{y\in D(s_j^{(N)})\\P_{j-1}y=x}}s_j^{(N)}[y],\\
D(s_{j-1}^{(N)})&=P_{j-1}D(s_j^{(N)}).
\end{aligned}\end{gathered}\end{equation*}\fi

No locality or Markov contraction property of a relaxed form is assumed. The proof supplies explicit terminal path-product fiber bounds, including their eliminated-block cost. A terminal estimate $q_N[g]\ge d_{N,j}\|(P_j-P_{j-1})g\|^2$ for every $g\in D(q_N)$ descends to the full vertical estimate for $s_j^{(N)}$, with the same $d_{N,j}>0$. Consequently, if one separately establishes $s_0^{(N)}\ge m_{N,0}I$ with $m_{N,0}>0$, then
\sbox{\equationmeasure}{$\displaystyle\begin{gathered}\begin{aligned}
s_j^{(N)}[x]&\ge d_{N,j}\|(I-P_{j-1})x\|^2,\\
K_N&\ge\left(m_{N,0}^{-1}
+\sum_{j=1}^{N}d_{N,j}^{-1}\right)^{-1}I.
\end{aligned}\end{gathered}$}
\ifdim\wd\equationmeasure>\dimexpr\columnwidth-36pt\relax
\typeout{MANUSCRIPT-WIDE 6.97}\begin{manuscriptwide}\eqanchor{6.97}\begin{equation*}\tag{6.97}\begin{gathered}\begin{aligned}
s_j^{(N)}[x]&\ge d_{N,j}\|(I-P_{j-1})x\|^2,\\
K_N&\ge\left(m_{N,0}^{-1}
+\sum_{j=1}^{N}d_{N,j}^{-1}\right)^{-1}I.
\end{aligned}\end{gathered}\end{equation*}\end{manuscriptwide}
\else \eqanchor{6.97}\begin{equation*}\tag{6.97}\begin{gathered}\begin{aligned}
s_j^{(N)}[x]&\ge d_{N,j}\|(I-P_{j-1})x\|^2,\\
K_N&\ge\left(m_{N,0}^{-1}
+\sum_{j=1}^{N}d_{N,j}^{-1}\right)^{-1}I.
\end{aligned}\end{gathered}\end{equation*}\fi

A uniform upper bound on the parenthesis gives a common numerical gap, even if different terminal regulators have different marginal laws and Hilbert spaces. In particular, $m_{N,0}\ge m_0>0$ and $d_{N,j}\ge d_j$ with $\sum_jd_j^{-1}<\infty$ suffice. These uniform estimates are not proved here. The construction uses the actual terminal vacuum; it does not establish compatibility between different terminal cutoffs, observable survival, or continuum dynamics. Proof in Appendix \crosspdflink{Entropy_Geometry_and_Lattice_Gaps_Supplement.pdf}{appC.44}{C.16}.\par

Retaining a spatial separator does not justify independent-interior estimates. Proposition \crosspdflink{Entropy_Geometry_and_Lattice_Gaps_Supplement.pdf}{res6.47}{6.31} (Supplement, Appendix \crosspdflink{Entropy_Geometry_and_Lattice_Gaps_Supplement.pdf}{appC.46}{C.17}) fixes the shared link of two Wilson plaquettes: the remaining bare potential splits, but the actual conditional law is nonfactorizing for every positive magnetic coefficient. At small $b/a$, its log density has a nonzero mixed term of order $(b/a)^2$ with analytic remainder. Uniform hierarchies must therefore control generated conditional interactions; approximate factorization is not excluded and continuum locality is not established.\par

\subsubsection*{Generated interactions}

Exact elimination must control generated interactions, the omitted remainder and transformed probes. Theorem \crosspdflink{Entropy_Geometry_and_Lattice_Gaps_Supplement.pdf}{res6.21}{6.32} gives a convergent expansion under a specified local conditional-cumulant decomposition with a common volume- and regulator-independent exponentially weighted interaction norm. Yang-Mills use still requires local dependence, integrability or small-/large-field control, gauge covariance and boundary representations. One-scale control allows a gapless retained field, and a finite-range Gaussian decomposition [\hyperlink{ref39}{29}] supplies no non-Abelian fluctuation construction. The cumulant, survival and complete-sector contraction estimates remain open.\par

\hypertarget{sec6.4}{}\subsection{Regulator and vacuum comparisons}

A hierarchy constructed within one terminal vacuum need not agree with a hierarchy from another cutoff. The following comparisons state the electric-symbol, dynamical and actual-vacuum estimates needed to relate those constructions.\par

\subsubsection*{Compatibility of the electric metric}

Exact static marginalization also generates an electric form. Its conditional mean need not describe fluctuations along the discarded fibers, which can invalidate a first-order generator estimate.\par

\hypertarget{res6.2}{}\noindent \textbf{Theorem \hyperlink{res6.2}{6.13} (electric-symbol requirement).} Let $\pi:M\to N$ be a smooth surjective submersion of compact manifolds without boundary, $\mu$ a smooth positive probability, and $K$ a nonnegative reversible elliptic diffusion Hamiltonian with smooth coefficients, symbol $a_f$, and $K1=0$. Put $\nu=\pi_*\mu$, $Jf=f\circ\pi$, and let $A$ be the self-adjoint operator of the compressed form $q_A(f)=q_K(Jf)$. Define
\sbox{\equationmeasure}{$\displaystyle\begin{gathered}\begin{aligned}
a_x^\pi(\xi,\eta)&=a_f(\pi^*\xi,\pi^*\eta),\\
\bar a_y&=\mathbb E_\mu[a_x^\pi\mid \pi(x)=y].
\end{aligned}\end{gathered}$}
\ifdim\wd\equationmeasure>\dimexpr\columnwidth-36pt\relax
\typeout{MANUSCRIPT-WIDE 6.4}\begin{manuscriptwide}\eqanchor{6.4}\begin{equation*}\tag{6.4}\begin{gathered}\begin{aligned}
a_x^\pi(\xi,\eta)&=a_f(\pi^*\xi,\pi^*\eta),\\
\bar a_y&=\mathbb E_\mu[a_x^\pi\mid \pi(x)=y].
\end{aligned}\end{gathered}\end{equation*}\end{manuscriptwide}
\else \eqanchor{6.4}\begin{equation*}\tag{6.4}\begin{gathered}\begin{aligned}
a_x^\pi(\xi,\eta)&=a_f(\pi^*\xi,\pi^*\eta),\\
\bar a_y&=\mathbb E_\mu[a_x^\pi\mid \pi(x)=y].
\end{aligned}\end{gathered}\end{equation*}\fi

If $\|(KJ-JA)f\|^2\le c q_A(f)$ for every smooth $f$ and one finite $c$, then $a_x^\pi=\bar a_{\pi(x)}$ everywhere. For an estimate restricted to gauge-invariant probes, equality is forced on the cotangent directions tested by invariant phases. Conversely, once these symbols agree, a conditional covariance bound on the remaining drift fluctuation suffices for the first-order estimate.\par

Proof in the Supplement, Appendix \crosspdflink{Entropy_Geometry_and_Lattice_Gaps_Supplement.pdf}{appC.2}{C.19}. The necessity is visible in a physical $SU(3)$ example. Three edge-disjoint triangles meeting at one root, with edge weights $a/3$, $a>0$, reduce to loop variables $x,y,z$. Retaining $u=xy$ and $v=yz$ gives exactly matching Haar marginals in the pure-electric case. Nevertheless, for the invariant phases $f_n=\exp(in[F(u)+F(v)])$, $F=\operatorname{Re}\operatorname{tr}/3$,
\sbox{\equationmeasure}{$\displaystyle\begin{gathered}\frac{\|(KJ-JA)f_n\|^2}{q_A(f_n)}
=\frac{a n^2}{108}\longrightarrow\infty.\end{gathered}$}
\ifdim\wd\equationmeasure>\dimexpr\columnwidth-36pt\relax
\typeout{MANUSCRIPT-WIDE 6.5}\begin{manuscriptwide}\eqanchor{6.5}\begin{equation*}\tag{6.5}\begin{gathered}\frac{\|(KJ-JA)f_n\|^2}{q_A(f_n)}
=\frac{a n^2}{108}\longrightarrow\infty.\end{gathered}\end{equation*}\end{manuscriptwide}
\else \eqanchor{6.5}\begin{equation*}\tag{6.5}\begin{gathered}\frac{\|(KJ-JA)f_n\|^2}{q_A(f_n)}
=\frac{a n^2}{108}\longrightarrow\infty.\end{gathered}\end{equation*}\fi

The obstruction persists for every smooth invariant finite potential on this graph; the displayed constant concerns the Haar case. A change of density cannot remove a second-order symbol fluctuation. Theorem \hyperlink{res6.30}{6.16} and Proposition \hyperlink{res6.31}{6.17} give a positive-time comparison for such fluctuations; their coefficient bounds still require control under refinement. For maps with projectable electric symbol, the required inputs are exact marginal and form compatibility, compatible operator domains, and summable centered drift covariances in physical units, uniformly in the retained and exterior configurations.\par

\subsubsection*{Controlled holonomy elimination}

Conditional-expectation closures and time-marginal entropy bounds are established in [\hyperlink{ref82}{30}, Sections 2.3 and 3.1] and [\hyperlink{ref83}{31}, Theorem 2.15]. Here the electric conditional Fisher tensor controls retained correlations and observed path laws; compression cancellation improves the correlation bound.\par

\hypertarget{res6.6}{}\noindent \textbf{Theorem \hyperlink{res6.6}{6.14} (quadratic error in retained dynamics).} Let $K\ge0$ and $A\ge0$ be self-adjoint operators on Hilbert spaces $\mathcal H$ and $\mathcal K$, and let $J:\mathcal K\to\mathcal H$ be an isometry. Suppose $JD(A)\subset D(K)$ and, on $D(A)$,
\sbox{\equationmeasure}{$\displaystyle\begin{gathered}\begin{aligned}
R&=KJ-JA,\qquad J^*R=0,\\
\|Rf\|^2&\le c\|A^{1/2}f\|^2,\qquad c\ge0.
\end{aligned}\end{gathered}$}
\ifdim\wd\equationmeasure>\dimexpr\columnwidth-36pt\relax
\typeout{MANUSCRIPT-WIDE 6.16}\begin{manuscriptwide}\eqanchor{6.16}\begin{equation*}\tag{6.16}\begin{gathered}\begin{aligned}
R&=KJ-JA,\qquad J^*R=0,\\
\|Rf\|^2&\le c\|A^{1/2}f\|^2,\qquad c\ge0.
\end{aligned}\end{gathered}\end{equation*}\end{manuscriptwide}
\else \eqanchor{6.16}\begin{equation*}\tag{6.16}\begin{gathered}\begin{aligned}
R&=KJ-JA,\qquad J^*R=0,\\
\|Rf\|^2&\le c\|A^{1/2}f\|^2,\qquad c\ge0.
\end{aligned}\end{gathered}\end{equation*}\fi

Then, for every $t\ge0$,
\sbox{\equationmeasure}{$\displaystyle\begin{gathered}\begin{aligned}
\|e^{-tK}J-Je^{-tA}\|&\le\sqrt{ct/2},\\
\|J^*e^{-tK}J-e^{-tA}\|&\le ct/2.
\end{aligned}\end{gathered}$}
\ifdim\wd\equationmeasure>\dimexpr\columnwidth-36pt\relax
\typeout{MANUSCRIPT-WIDE 6.17}\begin{manuscriptwide}\eqanchor{6.17}\begin{equation*}\tag{6.17}\begin{gathered}\begin{aligned}
\|e^{-tK}J-Je^{-tA}\|&\le\sqrt{ct/2},\\
\|J^*e^{-tK}J-e^{-tA}\|&\le ct/2.
\end{aligned}\end{gathered}\end{equation*}\end{manuscriptwide}
\else \eqanchor{6.17}\begin{equation*}\tag{6.17}\begin{gathered}\begin{aligned}
\|e^{-tK}J-Je^{-tA}\|&\le\sqrt{ct/2},\\
\|J^*e^{-tK}J-e^{-tA}\|&\le ct/2.
\end{aligned}\end{gathered}\end{equation*}\fi

No gap on the discarded space is assumed. The constant $c$ has energy units and $t$ has inverse-energy units.\par

Proof in the Supplement, Appendix \crosspdflink{Entropy_Geometry_and_Lattice_Gaps_Supplement.pdf}{appC.8}{C.2}.\par

For a simple-loop holonomy, $Jf=f(W)$ is isometric under the actual vacuum law and its exact marginal. Its compressed form is the electric diffusion of that marginal. Appendix \crosspdflink{Entropy_Geometry_and_Lattice_Gaps_Supplement.pdf}{appC.8}{C.2} identifies $R$ with a centered conditional score, differentiated along the electric horizontal lift. Its covariance supplies $c$ in \hyperlink{eq6.16}{(6.16)}. This lift matters: after eliminating gauge variables, the electric form generally has mixed derivatives, so an ordinary parameter Fisher tensor cannot simply replace the horizontal score.\par

Proposition \crosspdflink{Entropy_Geometry_and_Lattice_Gaps_Supplement.pdf}{res6.17}{6.28} verifies this comparison for two bare-Wilson $SU(3)$ plaquettes sharing one edge. All seven edge weights equal $a>0$, the magnetic coefficient is $b\ge0$, and $\varepsilon=b/a$. With $F(w)=\operatorname{Re}\operatorname{tr}w/3$, the actual outer-loop marginal has Haar density
\sbox{\equationmeasure}{$\displaystyle\begin{gathered}p_\varepsilon(w)
=1+\frac7{768}\varepsilon^2F(w)
+O_{C^q}(\varepsilon^3),\\  q\ge0.\end{gathered}$}
\ifdim\wd\equationmeasure>\dimexpr\columnwidth-36pt\relax
\typeout{MANUSCRIPT-WIDE 6.18}\begin{manuscriptwide}\eqanchor{6.18}\begin{equation*}\tag{6.18}\begin{gathered}p_\varepsilon(w)
=1+\frac7{768}\varepsilon^2F(w)
+O_{C^q}(\varepsilon^3),\qquad q\ge0.\end{gathered}\end{equation*}\end{manuscriptwide}
\else \eqanchor{6.18}\begin{equation*}\tag{6.18}\begin{gathered}p_\varepsilon(w)
=1+\frac7{768}\varepsilon^2F(w)
+O_{C^q}(\varepsilon^3),\\  q\ge0.\end{gathered}\end{equation*}\fi

The remainder is controlled in each fixed smoothness norm on this fixed block. Let $K$ be its actual ground-state diffusion Hamiltonian, $A$ the exact marginal form operator and $M_t=J^*e^{-tK}J$. For some finite $C$ and all sufficiently small $\varepsilon\ge0$,
\sbox{\equationmeasure}{$\displaystyle\begin{gathered}\begin{aligned}
\|M_t-e^{-tA}\|
&\le\left(\frac9{128}+C\varepsilon\right)at\varepsilon^2,\\
D(\mathbb P_W^T\Vert\overline{\mathbb P}^{\,T})
&\le\left(\frac1{128}+C\varepsilon\right)aT\varepsilon^2.
\end{aligned}\end{gathered}$}
\ifdim\wd\equationmeasure>\dimexpr\columnwidth-36pt\relax
\typeout{MANUSCRIPT-WIDE 6.19}\begin{manuscriptwide}\eqanchor{6.19}\begin{equation*}\tag{6.19}\begin{gathered}\begin{aligned}
\|M_t-e^{-tA}\|
&\le\left(\frac9{128}+C\varepsilon\right)at\varepsilon^2,\\
D(\mathbb P_W^T\Vert\overline{\mathbb P}^{\,T})
&\le\left(\frac1{128}+C\varepsilon\right)aT\varepsilon^2.
\end{aligned}\end{gathered}\end{equation*}\end{manuscriptwide}
\else \eqanchor{6.19}\begin{equation*}\tag{6.19}\begin{gathered}\begin{aligned}
\|M_t-e^{-tA}\|
&\le\left(\frac9{128}+C\varepsilon\right)at\varepsilon^2,\\
D(\mathbb P_W^T\Vert\overline{\mathbb P}^{\,T})
&\le\left(\frac1{128}+C\varepsilon\right)aT\varepsilon^2.
\end{aligned}\end{gathered}\end{equation*}\fi

Here $\mathbb P_W^T$ is the actual stationary outer-holonomy path law; $\overline{\mathbb P}^{\,T}$ is the stationary Markov law generated by the marginal form. Their dynamics differ at nonzero small coupling, despite using the same marginal. The complete statement and proof are in the Supplement, Appendix \crosspdflink{Entropy_Geometry_and_Lattice_Gaps_Supplement.pdf}{appC.9}{C.3}.\par

This interacting-loop elimination gives a finite-block error with leading coefficients, but no certified coupling interval or uniform iterated bound. Theorem \hyperlink{res6.12}{6.27} transfers retained contraction when the error stays below its margin and convergent renormalized reflected probes span the limiting physical sector. Appendix \crosspdflink{Entropy_Geometry_and_Lattice_Gaps_Supplement.pdf}{appC.8}{C.2} also gives a probe-dependent criterion. Reconstruction, observable survival, boundary-sector control and weak-bare-coupling extension remain required.\par

\subsubsection*{Actual edge blocking and high-energy errors}

Ordinary edge blocking uses disjoint serial paths and has a projectable electric symbol, even in the full interacting fine lattice. It therefore permits a joint first-order bound under the actual vacuum. This controls a summable high-energy remainder at fixed physical times; it does not control the retained low-energy error.\par

\hypertarget{res6.33}{}\noindent \textbf{Proposition \hyperlink{res6.33}{6.15} (actual edge blocking and a summable ultraviolet remainder).} Take any finite Wilson Hamiltonian \hyperlink{eq2.11}{(2.11)}, with at most $\nu$ incident plaquettes per link. Embed an ordinary coarse graph with $N_c\ge1$ edges by connected oriented paths of two distinct fine links in the fine graph, with no fine link shared by two paths. Let each coarse variable be the ordered path product. Let $\pi$ be this product map, $\mu=\Omega^2dm$ its actual fine ground law, $\bar\mu=\pi_*\mu$, $Jf=f\circ\pi$, $K=\Omega^{-1}(H-E_0)\Omega$, and $A$ the operator of the exact restricted form. Then, on the full spaces and their gauge-invariant restrictions,
\sbox{\equationmeasure}{$\displaystyle\begin{gathered}\begin{aligned}
q_A(f)&=2a\int\sum_{e=1}^{N_c}|\nabla_e f|^2\,d\bar\mu,\\
R&=KJ-JA,\qquad J^*R=0,\\
\|Rf\|^2&\le c q_A(f),\quad f\in D(A),\\
c&=\frac{64\nu^2N_c}{27}\frac{b^2}{a}.
\end{aligned}\end{gathered}$}
\ifdim\wd\equationmeasure>\dimexpr\columnwidth-36pt\relax
\typeout{MANUSCRIPT-WIDE 6.65}\begin{manuscriptwide}\eqanchor{6.65}\begin{equation*}\tag{6.65}\begin{gathered}\begin{aligned}
q_A(f)&=2a\int\sum_{e=1}^{N_c}|\nabla_e f|^2\,d\bar\mu,\\
R&=KJ-JA,\qquad J^*R=0,\\
\|Rf\|^2&\le c q_A(f),\quad f\in D(A),\\
c&=\frac{64\nu^2N_c}{27}\frac{b^2}{a}.
\end{aligned}\end{gathered}\end{equation*}\end{manuscriptwide}
\else \eqanchor{6.65}\begin{equation*}\tag{6.65}\begin{gathered}\begin{aligned}
q_A(f)&=2a\int\sum_{e=1}^{N_c}|\nabla_e f|^2\,d\bar\mu,\\
R&=KJ-JA,\qquad J^*R=0,\\
\|Rf\|^2&\le c q_A(f),\quad f\in D(A),\\
c&=\frac{64\nu^2N_c}{27}\frac{b^2}{a}.
\end{aligned}\end{gathered}\end{equation*}\fi

Here $JD(A)\subset D(K)$. For any sequence of these actual finite-regulator blockings, indexed by $j\ge0$, fix physical preparation time $\eta>0$ and reference energy $E_{\rm ref}>0$. Define, for $t\ge0$,
\sbox{\equationmeasure}{$\displaystyle\begin{gathered}\begin{aligned}
\mathsf E_j(t)&=e^{-\eta A_j}
 (J_j^*e^{-tK_j}J_j-e^{-tA_j})e^{-\eta A_j},\\
\Lambda_j&=\eta^{-1}\bigl[\log(1+c_j/E_{\rm ref})
 +(j+1)\log2\bigr],\\
P_j&=\mathbf1_{[0,\Lambda_j]}(A_j).
\end{aligned}\end{gathered}$}
\ifdim\wd\equationmeasure>\dimexpr\columnwidth-36pt\relax
\typeout{MANUSCRIPT-WIDE 6.66}\begin{manuscriptwide}\eqanchor{6.66}\begin{equation*}\tag{6.66}\begin{gathered}\begin{aligned}
\mathsf E_j(t)&=e^{-\eta A_j}
 (J_j^*e^{-tK_j}J_j-e^{-tA_j})e^{-\eta A_j},\\
\Lambda_j&=\eta^{-1}\bigl[\log(1+c_j/E_{\rm ref})
 +(j+1)\log2\bigr],\\
P_j&=\mathbf1_{[0,\Lambda_j]}(A_j).
\end{aligned}\end{gathered}\end{equation*}\end{manuscriptwide}
\else \eqanchor{6.66}\begin{equation*}\tag{6.66}\begin{gathered}\begin{aligned}
\mathsf E_j(t)&=e^{-\eta A_j}
 (J_j^*e^{-tK_j}J_j-e^{-tA_j})e^{-\eta A_j},\\
\Lambda_j&=\eta^{-1}\bigl[\log(1+c_j/E_{\rm ref})
 +(j+1)\log2\bigr],\\
P_j&=\mathbf1_{[0,\Lambda_j]}(A_j).
\end{aligned}\end{gathered}\end{equation*}\fi

In common physical units, without a running-coupling assumption,
\sbox{\equationmeasure}{$\displaystyle\begin{gathered}\begin{aligned}
\|\mathsf E_j(t)-P_j\mathsf E_j(t)P_j\|
&\le tE_{\rm ref}2^{-j-1},\\
\sum_{j\ge n}\|\mathsf E_j(t)-P_j\mathsf E_j(t)P_j\|
&\le tE_{\rm ref}2^{-n}.
\end{aligned}\end{gathered}$}
\ifdim\wd\equationmeasure>\dimexpr\columnwidth-36pt\relax
\typeout{MANUSCRIPT-WIDE 6.67}\begin{manuscriptwide}\eqanchor{6.67}\begin{equation*}\tag{6.67}\begin{gathered}\begin{aligned}
\|\mathsf E_j(t)-P_j\mathsf E_j(t)P_j\|
&\le tE_{\rm ref}2^{-j-1},\\
\sum_{j\ge n}\|\mathsf E_j(t)-P_j\mathsf E_j(t)P_j\|
&\le tE_{\rm ref}2^{-n}.
\end{aligned}\end{gathered}\end{equation*}\end{manuscriptwide}
\else \eqanchor{6.67}\begin{equation*}\tag{6.67}\begin{gathered}\begin{aligned}
\|\mathsf E_j(t)-P_j\mathsf E_j(t)P_j\|
&\le tE_{\rm ref}2^{-j-1},\\
\sum_{j\ge n}\|\mathsf E_j(t)-P_j\mathsf E_j(t)P_j\|
&\le tE_{\rm ref}2^{-n}.
\end{aligned}\end{gathered}\end{equation*}\fi

Proof in the Supplement, Appendix \crosspdflink{Entropy_Geometry_and_Lattice_Gaps_Supplement.pdf}{appC.32}{C.20}. These numerical norm bounds can be summed without a common cutoff Hilbert space. The $A_j$ are actual generated marginal parents, generally not native coarse Wilson Hamiltonians; thresholds adapt to error coefficients and retained volume. Spectral preparation gives the summation, while \hyperlink{eq6.65}{(6.65)} supplies the Yang-Mills joint drift bound. Control of $P_j\mathsf E_j(t)P_j$, compatibility across terminal cutoffs, a surviving local field and complete-sector infrared contraction remain unproved (Sections \hyperlink{sec6.5}{VI E} and \hyperlink{sec6.6}{VI F}).\par

\subsubsection*{Positive-time comparison for overlapping holonomies}

Positive-time preparation permits a comparison even when retained electric symbols fluctuate along eliminated fibers. The following estimate accommodates the resulting second-order defect. It does not require a gap on discarded modes.\par

\hypertarget{res6.30}{}\noindent \textbf{Theorem \hyperlink{res6.30}{6.16} (mixed-order defects on positive-time probes).} Let $K\ge0$ and $A\ge0$ be self-adjoint on Hilbert spaces $\mathcal H$ and $\mathcal K$, and let $J:\mathcal K\to\mathcal H$ be an isometry. Suppose $JD(A)\subset D(K)$ and, for every $f\in D(A)$,
\sbox{\equationmeasure}{$\displaystyle\begin{gathered}\begin{aligned}
R&=KJ-JA,\qquad J^*R=0,\\
\|Rf\|^2&\le\alpha\|Af\|^2
 +\beta\|A^{1/2}f\|^2+\gamma\|f\|^2,
\end{aligned}\end{gathered}$}
\ifdim\wd\equationmeasure>\dimexpr\columnwidth-36pt\relax
\typeout{MANUSCRIPT-WIDE 6.60}\begin{manuscriptwide}\eqanchor{6.60}\begin{equation*}\tag{6.60}\begin{gathered}\begin{aligned}
R&=KJ-JA,\qquad J^*R=0,\\
\|Rf\|^2&\le\alpha\|Af\|^2
 +\beta\|A^{1/2}f\|^2+\gamma\|f\|^2,
\end{aligned}\end{gathered}\end{equation*}\end{manuscriptwide}
\else \eqanchor{6.60}\begin{equation*}\tag{6.60}\begin{gathered}\begin{aligned}
R&=KJ-JA,\qquad J^*R=0,\\
\|Rf\|^2&\le\alpha\|Af\|^2
 +\beta\|A^{1/2}f\|^2+\gamma\|f\|^2,
\end{aligned}\end{gathered}\end{equation*}\fi

where $\alpha,\beta,\gamma\ge0$ are finite. Set $D_t=J^*e^{-tK}J-e^{-tA}$. For every $t\ge0$ and $\eta>0$,
\sbox{\equationmeasure}{$\displaystyle\begin{gathered}\begin{aligned}
\|e^{-\eta A}D_te^{-\eta A}\|
&\le\frac{t^2}{2}
\left(\frac{\alpha}{e^2\eta^2}
 +\frac{\beta}{2e\eta}+\gamma\right).
\end{aligned}\end{gathered}$}
\ifdim\wd\equationmeasure>\dimexpr\columnwidth-36pt\relax
\typeout{MANUSCRIPT-WIDE 6.61}\begin{manuscriptwide}\eqanchor{6.61}\begin{equation*}\tag{6.61}\begin{gathered}\begin{aligned}
\|e^{-\eta A}D_te^{-\eta A}\|
&\le\frac{t^2}{2}
\left(\frac{\alpha}{e^2\eta^2}
 +\frac{\beta}{2e\eta}+\gamma\right).
\end{aligned}\end{gathered}\end{equation*}\end{manuscriptwide}
\else \eqanchor{6.61}\begin{equation*}\tag{6.61}\begin{gathered}\begin{aligned}
\|e^{-\eta A}D_te^{-\eta A}\|
&\le\frac{t^2}{2}
\left(\frac{\alpha}{e^2\eta^2}
 +\frac{\beta}{2e\eta}+\gamma\right).
\end{aligned}\end{gathered}\end{equation*}\fi

Here $t,\eta$ have inverse-energy units; $\alpha$ is dimensionless, $\beta$ has energy units and $\gamma$ has energy-squared units. Appendix \crosspdflink{Entropy_Geometry_and_Lattice_Gaps_Supplement.pdf}{appC.29}{C.21} proves a sharper spectral bound and a probe-dependent integral criterion. Preparation uses the specified operator $A$. Neither its identification with fine physical translation nor any Yang-Mills coefficient estimate, continuum construction or spectral gap is asserted.\par

\hypertarget{res6.31}{}\noindent \textbf{Proposition \hyperlink{res6.31}{6.17} (an actual joint-holonomy block).} On Theorem \hyperlink{res6.2}{6.13}'s three-triangle $SU(3)$ carrier, use $H=-a(\Delta_x+\Delta_y+\Delta_z)+V$, $a>0$, with any smooth real simultaneous-conjugation-invariant potential $V$. Let $\rho=\Omega^2$ be its actual normalized ground density, $K=\Omega^{-1}(H-E_0)\Omega$, $\pi(x,y,z)=(xy,yz)$, $Jf=f\circ\pi$, and $A$ the exact marginal-form operator. Put $s=\max_i\|\nabla_i\log\rho\|_\infty$ and $B=\max_i\|\nabla_iV\|_\infty$, for $i=x,y,z$. Then $s\le8B/(3a)$, $JD(A)\subset D(K)$ and $J^*(KJ-JA)=0$. For every $f\in D(A)$,
\sbox{\equationmeasure}{$\displaystyle\begin{gathered}\begin{aligned}
\|(KJ-JA)f\|^2
&\le512\|Af\|^2+\beta_s q_A(f),\\
\beta_s&=2a\bigl[8\sqrt6
+(28\sqrt5+2\sqrt2)s\bigr]^2 .
\end{aligned}\end{gathered}$}
\ifdim\wd\equationmeasure>\dimexpr\columnwidth-36pt\relax
\typeout{MANUSCRIPT-WIDE 6.62}\begin{manuscriptwide}\eqanchor{6.62}\begin{equation*}\tag{6.62}\begin{gathered}\begin{aligned}
\|(KJ-JA)f\|^2
&\le512\|Af\|^2+\beta_s q_A(f),\\
\beta_s&=2a\bigl[8\sqrt6
+(28\sqrt5+2\sqrt2)s\bigr]^2 .
\end{aligned}\end{gathered}\end{equation*}\end{manuscriptwide}
\else \eqanchor{6.62}\begin{equation*}\tag{6.62}\begin{gathered}\begin{aligned}
\|(KJ-JA)f\|^2
&\le512\|Af\|^2+\beta_s q_A(f),\\
\beta_s&=2a\bigl[8\sqrt6
+(28\sqrt5+2\sqrt2)s\bigr]^2 .
\end{aligned}\end{gathered}\end{equation*}\fi

These statements hold on the full based-variable spaces and their physical invariant subspaces. For $V=0$, put $L_u=-\Delta_u$, $L_v=-\Delta_v$. Then $A=2a(L_u+L_v)$ and
\sbox{\equationmeasure}{$\displaystyle\begin{gathered}\begin{aligned}
\|(KJ-JA)f\|^2
&=\frac{a^2}{2}\|L_u^{1/2}L_v^{1/2}f\|^2\\
&\le\frac1{32}\|Af\|^2,\qquad f\in D(A).
\end{aligned}\end{gathered}$}
\ifdim\wd\equationmeasure>\dimexpr\columnwidth-36pt\relax
\typeout{MANUSCRIPT-WIDE 6.63}\begin{manuscriptwide}\eqanchor{6.63}\begin{equation*}\tag{6.63}\begin{gathered}\begin{aligned}
\|(KJ-JA)f\|^2
&=\frac{a^2}{2}\|L_u^{1/2}L_v^{1/2}f\|^2\\
&\le\frac1{32}\|Af\|^2,\qquad f\in D(A).
\end{aligned}\end{gathered}\end{equation*}\end{manuscriptwide}
\else \eqanchor{6.63}\begin{equation*}\tag{6.63}\begin{gathered}\begin{aligned}
\|(KJ-JA)f\|^2
&=\frac{a^2}{2}\|L_u^{1/2}L_v^{1/2}f\|^2\\
&\le\frac1{32}\|Af\|^2,\qquad f\in D(A).
\end{aligned}\end{gathered}\end{equation*}\fi

The constant $1/32$ is sharp even on physical functions. Thus the very map obstructing the first-order comparison admits Theorem \hyperlink{res6.30}{6.16}'s positive-time estimate. Proof in the Supplement, Appendix \crosspdflink{Entropy_Geometry_and_Lattice_Gaps_Supplement.pdf}{appC.30}{C.22}, including an explicitly coupled Wilson-loop potential, exterior-data qualifications and physical scaling. Its interacting coefficient bound still deteriorates as $1/(hg_h^6)$ under the stated refinement; no summable continuum error follows.\par

\subsubsection*{Entropy comparisons before reconstruction}

Entropy comparisons can establish convergence of renormalized correlations before a physical Hilbert space is available. They must compare actual regulator laws, with the field normalization and counterterms included in the observables.\par

\hypertarget{res6.4}{}\noindent \textbf{Theorem \hyperlink{res6.4}{6.18} (summable entropy comparisons).} Set $S_{j,0}=1$. For each $n\ge1$, let $\mathscr T_n$ be a complex locally convex test space with continuous seminorm $p_n$, and let $S_{j,n}$ be linear correlation functionals, with $S_{0,n}$ continuous. For each $j$, suppose probability laws $\mu_j,\nu_j$ share a measurable space. For each $n,f$, write $F=F_{j,n,f}$ for a complex representative and assume, for $\lambda\in\{\mu_j,\nu_j\}$,
\sbox{\equationmeasure}{$\displaystyle\begin{gathered}\begin{aligned}
S_{j+1,n}(f)&=\mathbb E_{\mu_j}F,\\
|S_{j,n}(f)-\mathbb E_{\nu_j}F|&\le d_{j,n}p_n(f),\\
\mathbb E_\lambda|F|^2&\le M_{j,n}^2p_n(f)^2,\\
D(\mu_j\Vert\nu_j)&\le\epsilon_j.
\end{aligned}\end{gathered}$}
\ifdim\wd\equationmeasure>\dimexpr\columnwidth-36pt\relax
\typeout{MANUSCRIPT-WIDE 6.9}\begin{manuscriptwide}\eqanchor{6.9}\begin{equation*}\tag{6.9}\begin{gathered}\begin{aligned}
S_{j+1,n}(f)&=\mathbb E_{\mu_j}F,\\
|S_{j,n}(f)-\mathbb E_{\nu_j}F|&\le d_{j,n}p_n(f),\\
\mathbb E_\lambda|F|^2&\le M_{j,n}^2p_n(f)^2,\\
D(\mu_j\Vert\nu_j)&\le\epsilon_j.
\end{aligned}\end{gathered}\end{equation*}\end{manuscriptwide}
\else \eqanchor{6.9}\begin{equation*}\tag{6.9}\begin{gathered}\begin{aligned}
S_{j+1,n}(f)&=\mathbb E_{\mu_j}F,\\
|S_{j,n}(f)-\mathbb E_{\nu_j}F|&\le d_{j,n}p_n(f),\\
\mathbb E_\lambda|F|^2&\le M_{j,n}^2p_n(f)^2,\\
D(\mu_j\Vert\nu_j)&\le\epsilon_j.
\end{aligned}\end{gathered}\end{equation*}\fi

All constants are finite and nonnegative. Suppose
\sbox{\equationmeasure}{$\displaystyle\begin{gathered}\begin{aligned} R_{J,n}&:=\sum_{j\ge J}\bigl(2M_{j,n}\sqrt{\epsilon_j}+d_{j,n}\bigr),\\ R_{0,n}&<\infty\quad(n\ge1).\end{aligned}\end{gathered}$}
\ifdim\wd\equationmeasure>\dimexpr\columnwidth-36pt\relax
\typeout{MANUSCRIPT-WIDE 6.10}\begin{manuscriptwide}\eqanchor{6.10}\begin{equation*}\tag{6.10}\begin{gathered}\begin{aligned} R_{J,n}&:=\sum_{j\ge J}\bigl(2M_{j,n}\sqrt{\epsilon_j}+d_{j,n}\bigr),\\ R_{0,n}&<\infty\quad(n\ge1).\end{aligned}\end{gathered}\end{equation*}\end{manuscriptwide}
\else \eqanchor{6.10}\begin{equation*}\tag{6.10}\begin{gathered}\begin{aligned} R_{J,n}&:=\sum_{j\ge J}\bigl(2M_{j,n}\sqrt{\epsilon_j}+d_{j,n}\bigr),\\ R_{0,n}&<\infty\quad(n\ge1).\end{aligned}\end{gathered}\end{equation*}\fi

Then the full sequence converges to continuous linear functionals $S_n$, with
\sbox{\equationmeasure}{$\displaystyle\begin{gathered}|S_n(f)-S_{J,n}(f)|\le p_n(f)R_{J,n}.\end{gathered}$}
\ifdim\wd\equationmeasure>\dimexpr\columnwidth-36pt\relax
\typeout{MANUSCRIPT-WIDE 6.11}\begin{manuscriptwide}\eqanchor{6.11}\begin{equation*}\tag{6.11}\begin{gathered}|S_n(f)-S_{J,n}(f)|\le p_n(f)R_{J,n}.\end{gathered}\end{equation*}\end{manuscriptwide}
\else \eqanchor{6.11}\begin{equation*}\tag{6.11}\begin{gathered}|S_n(f)-S_{J,n}(f)|\le p_n(f)R_{J,n}.\end{gathered}\end{equation*}\fi

For Schwartz spaces of arbitrary test kernels, the limits form a tempered correlation hierarchy. Order-by-order continuity does not give the growth control across orders required for reconstruction [\hyperlink{ref14}{25}]. This conclusion supplies neither the full reconstruction hypotheses nor a mass gap.\par

Proof in the Supplement, Appendix \crosspdflink{Entropy_Geometry_and_Lattice_Gaps_Supplement.pdf}{appC.4}{C.23}. That appendix also gives a quantitative reflected-survival test. Lemma \crosspdflink{Entropy_Geometry_and_Lattice_Gaps_Supplement.pdf}{res6.15}{6.33} expresses the required entropy increment through a source variance. These comparison and moment estimates remain unproved for four-dimensional Yang-Mills.\par

\subsubsection*{Comparing vacua in one electric form}

The missing connection is control of the vacuum ratio in the actual electric form. Unlike an averaged information cost, a pointwise gradient bound controls every physical mode.\par

\hypertarget{res6.5}{}\noindent \textbf{Theorem \hyperlink{res6.5}{6.19} (vacuum-gradient comparison).} On the same finite link manifold and physical space as Theorem \hyperlink{res2.2}{2.2}, let $H_i=H_{\rm el}+V_i$, $i=0,1$, with the same positive electric weights and smooth real invariant potentials. Write $\Omega_i,E_i,\Delta_i$ for their normalized positive vacua, ground energies and physical gaps. Set
\sbox{\equationmeasure}{$\displaystyle\begin{gathered}\begin{aligned}
r&=\Omega_1/\Omega_0,\qquad d\mu_i=\Omega_i^2dm,\\
\ell&=\|\sqrt{\Gamma_a(\log r,\log r)}\|_\infty,\\
J&=\int\Gamma_a(\log r,\log r)\,d\mu_1.
\end{aligned}\end{gathered}$}
\ifdim\wd\equationmeasure>\dimexpr\columnwidth-36pt\relax
\typeout{MANUSCRIPT-WIDE 6.12}\begin{manuscriptwide}\eqanchor{6.12}\begin{equation*}\tag{6.12}\begin{gathered}\begin{aligned}
r&=\Omega_1/\Omega_0,\qquad d\mu_i=\Omega_i^2dm,\\
\ell&=\|\sqrt{\Gamma_a(\log r,\log r)}\|_\infty,\\
J&=\int\Gamma_a(\log r,\log r)\,d\mu_1.
\end{aligned}\end{gathered}\end{equation*}\end{manuscriptwide}
\else \eqanchor{6.12}\begin{equation*}\tag{6.12}\begin{gathered}\begin{aligned}
r&=\Omega_1/\Omega_0,\qquad d\mu_i=\Omega_i^2dm,\\
\ell&=\|\sqrt{\Gamma_a(\log r,\log r)}\|_\infty,\\
J&=\int\Gamma_a(\log r,\log r)\,d\mu_1.
\end{aligned}\end{gathered}\end{equation*}\fi

Let $\mathbb P_i^T$ be their stationary full-configuration diffusion laws on an interval of length $T>0$, with generators $L_i=-\Omega_i^{-1}(H_i-E_i)\Omega_i$. Then
\sbox{\equationmeasure}{$\displaystyle\begin{gathered}\begin{aligned}
|\sqrt{\Delta_1}-\sqrt{\Delta_0}|&\le\ell,\\
D(\mathbb P_1^T\Vert\mathbb P_0^T)
 &=D(\mu_1\Vert\mu_0)+TJ,\\
J&=\|(H_0-E_0)^{1/2}\Omega_1\|^2.
\end{aligned}\end{gathered}$}
\ifdim\wd\equationmeasure>\dimexpr\columnwidth-36pt\relax
\typeout{MANUSCRIPT-WIDE 6.13}\begin{manuscriptwide}\eqanchor{6.13}\begin{equation*}\tag{6.13}\begin{gathered}\begin{aligned}
|\sqrt{\Delta_1}-\sqrt{\Delta_0}|&\le\ell,\\
D(\mathbb P_1^T\Vert\mathbb P_0^T)
 &=D(\mu_1\Vert\mu_0)+TJ,\\
J&=\|(H_0-E_0)^{1/2}\Omega_1\|^2.
\end{aligned}\end{gathered}\end{equation*}\end{manuscriptwide}
\else \eqanchor{6.13}\begin{equation*}\tag{6.13}\begin{gathered}\begin{aligned}
|\sqrt{\Delta_1}-\sqrt{\Delta_0}|&\le\ell,\\
D(\mathbb P_1^T\Vert\mathbb P_0^T)
 &=D(\mu_1\Vert\mu_0)+TJ,\\
J&=\|(H_0-E_0)^{1/2}\Omega_1\|^2.
\end{aligned}\end{gathered}\end{equation*}\fi

If $\ell<\sqrt{\Delta_0}$ and $d=(\sqrt{\Delta_0}-\ell)^2$, then
\sbox{\equationmeasure}{$\displaystyle\begin{gathered}\begin{aligned}
\Delta_1&\ge d>0,\\
D(\mu_1\Vert\mu_0)&\le 2\ell^2/d,\\
D(\mathbb P_1^T\Vert\mathbb P_0^T)&\le(T+2/d)\ell^2.
\end{aligned}\end{gathered}$}
\ifdim\wd\equationmeasure>\dimexpr\columnwidth-36pt\relax
\typeout{MANUSCRIPT-WIDE 6.14}\begin{manuscriptwide}\eqanchor{6.14}\begin{equation*}\tag{6.14}\begin{gathered}\begin{aligned}
\Delta_1&\ge d>0,\\
D(\mu_1\Vert\mu_0)&\le 2\ell^2/d,\\
D(\mathbb P_1^T\Vert\mathbb P_0^T)&\le(T+2/d)\ell^2.
\end{aligned}\end{gathered}\end{equation*}\end{manuscriptwide}
\else \eqanchor{6.14}\begin{equation*}\tag{6.14}\begin{gathered}\begin{aligned}
\Delta_1&\ge d>0,\\
D(\mu_1\Vert\mu_0)&\le 2\ell^2/d,\\
D(\mathbb P_1^T\Vert\mathbb P_0^T)&\le(T+2/d)\ell^2.
\end{aligned}\end{gathered}\end{equation*}\fi

Here $T$ has inverse-energy units. These entropies compare classical configuration and path laws; they are not quantum relative entropies between distinct pure vacua. Common measurable observations obey the same entropy upper bound by data processing.\par

Proof in the Supplement, Appendix \crosspdflink{Entropy_Geometry_and_Lattice_Gaps_Supplement.pdf}{appC.6}{C.25}.\par

The theorem supplies a conditional refinement strategy. At step $j$, a reference Hamiltonian on the next carrier must share the target's electric operator, have at least the preceding certified gap, and reproduce the preceding correlations up to Theorem \hyperlink{res6.4}{6.18}'s representative errors. If its comparison loss is $\ell_j$ and the following bounds hold, the gaps stay at least $m$ and the path comparisons supply that theorem at each fixed physical time horizon:
\sbox{\equationmeasure}{$\displaystyle\begin{gathered}\begin{aligned}
L&=\sum_j\ell_j<\sqrt{\Delta_{\rm start}},\\
m&=(\sqrt{\Delta_{\rm start}}-L)^2>0,\\
\sum_j(M_{j,n}\ell_j+d_{j,n})&<\infty\quad(n\ge1).
\end{aligned}\end{gathered}$}
\ifdim\wd\equationmeasure>\dimexpr\columnwidth-36pt\relax
\typeout{MANUSCRIPT-WIDE 6.15}\begin{manuscriptwide}\eqanchor{6.15}\begin{equation*}\tag{6.15}\begin{gathered}\begin{aligned}
L&=\sum_j\ell_j<\sqrt{\Delta_{\rm start}},\\
m&=(\sqrt{\Delta_{\rm start}}-L)^2>0,\\
\sum_j(M_{j,n}\ell_j+d_{j,n})&<\infty\quad(n\ge1).
\end{aligned}\end{gathered}\end{equation*}\end{manuscriptwide}
\else \eqanchor{6.15}\begin{equation*}\tag{6.15}\begin{gathered}\begin{aligned}
L&=\sum_j\ell_j<\sqrt{\Delta_{\rm start}},\\
m&=(\sqrt{\Delta_{\rm start}}-L)^2>0,\\
\sum_j(M_{j,n}\ell_j+d_{j,n})&<\infty\quad(n\ge1).
\end{aligned}\end{gathered}\end{equation*}\fi

The maps, reference gaps for added degrees of freedom and moment bounds remain to be constructed. Reconstruction, relativistic covariance, Yang-Mills identification and reflected survival require Sections \hyperlink{sec6.1}{VI A}, \hyperlink{sec6.5}{VI E}, \hyperlink{sec6.6}{VI F}. Even independent product comparisons can have $\ell$ growing as the square root of volume, motivating Section \hyperlink{sec4}{IV}'s local and conditional estimates.\par

Proposition \crosspdflink{Entropy_Geometry_and_Lattice_Gaps_Supplement.pdf}{res6.16}{6.34} gives the complementary obstruction: static entropy, average Fisher information and every fixed-horizon path entropy can tend to zero while regulator gaps collapse. Its slow modes concentrate in shrinking regions. Thus convergence of information costs alone does not supply a uniform physical decay estimate. Appendix \crosspdflink{Entropy_Geometry_and_Lattice_Gaps_Supplement.pdf}{appC.6}{C.25} also explains why changing electric coefficients prevents direct full-path comparison.\par

\subsubsection*{Comparing local vacuum marginals}

\hypertarget{res6.49}{}\noindent \textbf{Theorem \hyperlink{res6.49}{6.20} (local comparison of actual changing vacua).} Let $X=G^r$, with $G$ compact and connected, a fixed bi-invariant product metric, Haar probability $dw$, and $d=r\dim G>0$. For each actual finite compact-link Hamiltonian $H_j=-\sum_ea_{j,e}\Delta_e+V_j$, $a_{j,e}>0$, with smooth real potential and normalized positive vacuum $\Omega_j$, retain $r$ path holonomies on $X$. Each path $b$ must have a nonempty set $E_{j,b}$ of edges occurring once in it and in no other retained path. Other edges may be shared. Let $p_j\,dw$ be the pushforward of $\Omega_j^2dm_j$. Put
\sbox{\equationmeasure}{$\displaystyle\begin{gathered}\begin{aligned}
M_{j,e}&=\sup_{\xi_e}\operatorname{osc}_{u_e}V_j(u_e,\xi_e),\\
C_j&=\sum_{b=1}^{r}\min_{e\in E_{j,b}}M_{j,e}/a_{j,e},\\
\eta_j^3&=\frac{27}{4}\sqrt d\,(\sqrt{C_j}+\sqrt{C_{j+1}})\\
&\quad\times W_1(p_j\,dw,p_{j+1}\,dw).
\end{aligned}\end{gathered}$}
\ifdim\wd\equationmeasure>\dimexpr\columnwidth-36pt\relax
\typeout{MANUSCRIPT-WIDE 6.108}\begin{manuscriptwide}\eqanchor{6.108}\begin{equation*}\tag{6.108}\begin{gathered}\begin{aligned}
M_{j,e}&=\sup_{\xi_e}\operatorname{osc}_{u_e}V_j(u_e,\xi_e),\\
C_j&=\sum_{b=1}^{r}\min_{e\in E_{j,b}}M_{j,e}/a_{j,e},\\
\eta_j^3&=\frac{27}{4}\sqrt d\,(\sqrt{C_j}+\sqrt{C_{j+1}})\\
&\quad\times W_1(p_j\,dw,p_{j+1}\,dw).
\end{aligned}\end{gathered}\end{equation*}\end{manuscriptwide}
\else \eqanchor{6.108}\begin{equation*}\tag{6.108}\begin{gathered}\begin{aligned}
M_{j,e}&=\sup_{\xi_e}\operatorname{osc}_{u_e}V_j(u_e,\xi_e),\\
C_j&=\sum_{b=1}^{r}\min_{e\in E_{j,b}}M_{j,e}/a_{j,e},\\
\eta_j^3&=\frac{27}{4}\sqrt d\,(\sqrt{C_j}+\sqrt{C_{j+1}})\\
&\quad\times W_1(p_j\,dw,p_{j+1}\,dw).
\end{aligned}\end{gathered}\end{equation*}\fi

where $W_1$ uses the geodesic distance on $X$. Then
\sbox{\equationmeasure}{$\displaystyle\begin{gathered}\begin{aligned}
\int_X|\nabla\sqrt{p_j}|^2\,dw&\le C_j,\\
\|\sqrt{p_{j+1}}-\sqrt{p_j}\|_2&\le\eta_j.
\end{aligned}\end{gathered}$}
\ifdim\wd\equationmeasure>\dimexpr\columnwidth-36pt\relax
\typeout{MANUSCRIPT-WIDE 6.109}\begin{manuscriptwide}\eqanchor{6.109}\begin{equation*}\tag{6.109}\begin{gathered}\begin{aligned}
\int_X|\nabla\sqrt{p_j}|^2\,dw&\le C_j,\\
\|\sqrt{p_{j+1}}-\sqrt{p_j}\|_2&\le\eta_j.
\end{aligned}\end{gathered}\end{equation*}\end{manuscriptwide}
\else \eqanchor{6.109}\begin{equation*}\tag{6.109}\begin{gathered}\begin{aligned}
\int_X|\nabla\sqrt{p_j}|^2\,dw&\le C_j,\\
\|\sqrt{p_{j+1}}-\sqrt{p_j}\|_2&\le\eta_j.
\end{aligned}\end{gathered}\end{equation*}\fi

If $\sum_j\eta_j<\infty$, the square roots converge in $L^2(dw)$ to a nonnegative unit vector $\psi$, and $p_j\to p=\psi^2$ in $L^1(dw)$. Every bounded retained configuration observable has convergent centered vacuum vectors and mixed Gram matrices in this representation. A nonconstant real analytic real observable has strictly positive limiting variance.\par

This comparison uses unconditional actual marginals and averaged Fisher information, without global vacuum fidelity, a full-cutoff Hilbert embedding or bounds at every exterior value. Gauge-invariant uses retain compatible endpoint gauge data and invariant observables. The static vectors represent a commuting configuration algebra; they do not construct a quantum reduced state, local field or physical dynamics. The Yang-Mills transport-error budget remains unproved. Proof and tail estimates are in the Supplement, Appendix \crosspdflink{Entropy_Geometry_and_Lattice_Gaps_Supplement.pdf}{appC.48}{C.27}.\par

\hypertarget{sec6.5}{}\subsection{Limiting dynamics and surviving observables}

We now assemble compatible one-step estimates and control the observables that must survive regulator removal. A limiting semigroup and a surviving vector still leave complete-sector decay and local relativistic field reconstruction to be proved.\par

\subsubsection*{Summable errors construct limiting dynamics}

Exact compatibility allows a stronger conclusion than unrelated comparisons at successive scales. The following theorem constructs the limiting time evolution instead of assuming it.\par

\hypertarget{res6.8}{}\noindent \textbf{Theorem \hyperlink{res6.8}{6.21} (nested reductions and limiting dynamics).} Let $A_j\ge0$ be self-adjoint on $\mathcal H_j$, and let $J_j:\mathcal H_j\to\mathcal H_{j+1}$ be isometries. Suppose $J_jD(A_j)\subset D(A_{j+1})$ and there are constants $c_j\ge0$ such that
\sbox{\equationmeasure}{$\displaystyle\begin{gathered}\begin{aligned}
R_j&=A_{j+1}J_j-J_jA_j,&J_j^*R_j&=0,\\
\|R_jf\|^2&\le c_j\|A_j^{1/2}f\|^2,&
\sum_{j=0}^\infty c_j&<\infty.
\end{aligned}\end{gathered}$}
\ifdim\wd\equationmeasure>\dimexpr\columnwidth-36pt\relax
\typeout{MANUSCRIPT-WIDE 6.24}\begin{manuscriptwide}\eqanchor{6.24}\begin{equation*}\tag{6.24}\begin{gathered}\begin{aligned}
R_j&=A_{j+1}J_j-J_jA_j,&J_j^*R_j&=0,\\
\|R_jf\|^2&\le c_j\|A_j^{1/2}f\|^2,&
\sum_{j=0}^\infty c_j&<\infty.
\end{aligned}\end{gathered}\end{equation*}\end{manuscriptwide}
\else \eqanchor{6.24}\begin{equation*}\tag{6.24}\begin{gathered}\begin{aligned}
R_j&=A_{j+1}J_j-J_jA_j,&J_j^*R_j&=0,\\
\|R_jf\|^2&\le c_j\|A_j^{1/2}f\|^2,&
\sum_{j=0}^\infty c_j&<\infty.
\end{aligned}\end{gathered}\end{equation*}\fi

The defect identities and inequality hold for $f\in D(A_j)$. All generators and $c_j$ use common physical energy units. Form the Hilbert direct limit $\mathcal H_\infty$, with compatible isometries $I_j$ and dense union of their ranges. Then $I_je^{-tA_j}I_j^*$ converges strongly, uniformly on compact time intervals, to $e^{-tA_\infty}$ for a unique nonnegative self-adjoint $A_\infty$. With $C_n=\sum_{j\ge n}c_j$ and $t\ge0$,
\sbox{\equationmeasure}{$\displaystyle\begin{gathered}\begin{aligned}
\|e^{-tA_\infty}I_n-I_ne^{-tA_n}\|
 &\le\sqrt{tC_n/2},\\
\|I_n^*e^{-tA_\infty}I_n-e^{-tA_n}\|
 &\le tC_n/2.
\end{aligned}\end{gathered}$}
\ifdim\wd\equationmeasure>\dimexpr\columnwidth-36pt\relax
\typeout{MANUSCRIPT-WIDE 6.25}\begin{manuscriptwide}\eqanchor{6.25}\begin{equation*}\tag{6.25}\begin{gathered}\begin{aligned}
\|e^{-tA_\infty}I_n-I_ne^{-tA_n}\|
 &\le\sqrt{tC_n/2},\\
\|I_n^*e^{-tA_\infty}I_n-e^{-tA_n}\|
 &\le tC_n/2.
\end{aligned}\end{gathered}\end{equation*}\end{manuscriptwide}
\else \eqanchor{6.25}\begin{equation*}\tag{6.25}\begin{gathered}\begin{aligned}
\|e^{-tA_\infty}I_n-I_ne^{-tA_n}\|
 &\le\sqrt{tC_n/2},\\
\|I_n^*e^{-tA_\infty}I_n-e^{-tA_n}\|
 &\le tC_n/2.
\end{aligned}\end{gathered}\end{equation*}\fi

The embedded operator domains form an operator core, and the energy forms are exactly preserved. If normalized vacua satisfy $A_j\Omega_j=0$ and $J_j\Omega_j=\Omega_{j+1}$, they define $\Omega_\infty\in\ker A_\infty$. Every centered nonzero $f\in D(A_n^{1/2})$ survives: writing $v=I_nf$, $\sigma^2=\|f\|^2$ and $E=\|A_n^{1/2}f\|^2$, for $0\le t<\infty$,
\sbox{\equationmeasure}{$\displaystyle\begin{gathered}\begin{aligned}
\|v\|^2&=\sigma^2,\qquad
\|A_\infty^{1/2}v\|^2=E,\\
\langle v,e^{-tA_\infty}v\rangle
 &\ge\sigma^2e^{-tE/\sigma^2}>0.
\end{aligned}\end{gathered}$}
\ifdim\wd\equationmeasure>\dimexpr\columnwidth-36pt\relax
\typeout{MANUSCRIPT-WIDE 6.26}\begin{manuscriptwide}\eqanchor{6.26}\begin{equation*}\tag{6.26}\begin{gathered}\begin{aligned}
\|v\|^2&=\sigma^2,\qquad
\|A_\infty^{1/2}v\|^2=E,\\
\langle v,e^{-tA_\infty}v\rangle
 &\ge\sigma^2e^{-tE/\sigma^2}>0.
\end{aligned}\end{gathered}\end{equation*}\end{manuscriptwide}
\else \eqanchor{6.26}\begin{equation*}\tag{6.26}\begin{gathered}\begin{aligned}
\|v\|^2&=\sigma^2,\qquad
\|A_\infty^{1/2}v\|^2=E,\\
\langle v,e^{-tA_\infty}v\rangle
 &\ge\sigma^2e^{-tE/\sigma^2}>0.
\end{aligned}\end{gathered}\end{equation*}\fi

Proof in the Supplement, Appendix \crosspdflink{Entropy_Geometry_and_Lattice_Gaps_Supplement.pdf}{appC.11}{C.28}. Its exact orthogonal-defect identity makes the squared constants $c_j$ add; no independence of spatial blocks or gap on discarded modes is assumed. Uniform bounds $c_j(\xi)\le c_j$ give the same conclusions for every compatible exterior datum $\xi$. The appendix also gives a conditional-score sum for direct path-entropy comparison; relative entropy is not treated as satisfying a triangle inequality.\par

The theorem constructs a Hilbert space, vacuum and time evolution for a specified compatible tower. Yang-Mills use requires compatible actual marginal laws and generated electric forms; a few matching correlations do not suffice. The tower, renormalized local fields, locality and relativistic covariance remain to be constructed. Even $R_j=0$ allows new orthogonal modes with energies tending to zero; a common strict contraction on the complete limiting physical sector is still necessary.\par

\hypertarget{res6.32}{}\noindent \textbf{Proposition \hyperlink{res6.32}{6.22} (limiting dynamics with mixed-order defects).} Use the Hilbert spaces, compatible isometries, operator-domain inclusions and exact cancellation of Theorem \hyperlink{res6.8}{6.21}. Replace its first-order inequality by
$\|R_jf\|^2\le\alpha_j\|A_jf\|^2+\beta_jq_j(f)+\gamma_j\|f\|^2$
on $D(A_j)$, with nonnegative coefficients. For one fixed reference energy $E_{\rm ref}>0$, assume
\sbox{\equationmeasure}{$\displaystyle\begin{gathered}\sum_{j=0}^\infty
\left(\alpha_j+\frac{\beta_j}{E_{\rm ref}}
+\frac{\gamma_j}{E_{\rm ref}^2}\right)<\infty.\end{gathered}$}
\ifdim\wd\equationmeasure>\dimexpr\columnwidth-36pt\relax
\typeout{MANUSCRIPT-WIDE 6.64}\begin{manuscriptwide}\eqanchor{6.64}\begin{equation*}\tag{6.64}\begin{gathered}\sum_{j=0}^\infty
\left(\alpha_j+\frac{\beta_j}{E_{\rm ref}}
+\frac{\gamma_j}{E_{\rm ref}^2}\right)<\infty.\end{gathered}\end{equation*}\end{manuscriptwide}
\else \eqanchor{6.64}\begin{equation*}\tag{6.64}\begin{gathered}\sum_{j=0}^\infty
\left(\alpha_j+\frac{\beta_j}{E_{\rm ref}}
+\frac{\gamma_j}{E_{\rm ref}^2}\right)<\infty.\end{gathered}\end{equation*}\fi

Then the embedded semigroups converge strongly, uniformly on compact time intervals, to $e^{-tA_\infty}$ for a nonnegative self-adjoint operator. The union of embedded operator domains is an operator core, and the energy forms are exactly preserved. Compatible normalized vacua define a limiting vacuum; every centered nonzero finite-energy seed retains its norm and energy and satisfies \hyperlink{eq6.26}{(6.26)}.\par

Proof in the Supplement, Appendix \crosspdflink{Entropy_Geometry_and_Lattice_Gaps_Supplement.pdf}{appC.31}{C.29}. The coefficient sums use common physical units. This extends the allowed comparison beyond projectable electric symbols; it does not establish these sums for the Yang-Mills hierarchy or supply a gap for newly retained modes.\par

\subsubsection*{Exact relaxed towers and spectral survival}

Let $\mathcal H_0\subset\mathcal H_1\subset\cdots$ be closed subspaces of their Hilbert completion $\mathcal H$, with orthogonal projections $P_j\to I$ strongly. Let $q_j$ be densely defined closed nonnegative forms on $\mathcal H_j$, with generators $K_j$. Suppose minimizing $q_j$ over the fibers of $P_{j-1}|_{\mathcal H_j}$ gives exactly $q_{j-1}$, including the domain image, and
\sbox{\equationmeasure}{$\displaystyle\begin{gathered}q_j[g]\ge\delta_j\|(P_j-P_{j-1})g\|^2
\\ (g\in D(q_j)),\\  \delta_j>0.\end{gathered}$}
\ifdim\wd\equationmeasure>\dimexpr\columnwidth-36pt\relax
\typeout{MANUSCRIPT-WIDE 6.98}\begin{manuscriptwide}\eqanchor{6.98}\begin{equation*}\tag{6.98}\begin{gathered}q_j[g]\ge\delta_j\|(P_j-P_{j-1})g\|^2
\quad(g\in D(q_j)),\qquad \delta_j>0.\end{gathered}\end{equation*}\end{manuscriptwide}
\else \eqanchor{6.98}\begin{equation*}\tag{6.98}\begin{gathered}q_j[g]\ge\delta_j\|(P_j-P_{j-1})g\|^2
\\ (g\in D(q_j)),\\  \delta_j>0.\end{gathered}\end{equation*}\fi

For $m>n$, define
\sbox{\equationmeasure}{$\displaystyle\begin{gathered}T_{n,m}=\sum_{j=n+1}^{m}\delta_j^{-1},
\\  R_j(z)=I_j(K_j+z)^{-1}I_j^*,\\  z>0,\end{gathered}$}
\ifdim\wd\equationmeasure>\dimexpr\columnwidth-36pt\relax
\typeout{MANUSCRIPT-WIDE 6.99}\begin{manuscriptwide}\eqanchor{6.99}\begin{equation*}\tag{6.99}\begin{gathered}T_{n,m}=\sum_{j=n+1}^{m}\delta_j^{-1},
\qquad R_j(z)=I_j(K_j+z)^{-1}I_j^*,\quad z>0,\end{gathered}\end{equation*}\end{manuscriptwide}
\else \eqanchor{6.99}\begin{equation*}\tag{6.99}\begin{gathered}T_{n,m}=\sum_{j=n+1}^{m}\delta_j^{-1},
\\  R_j(z)=I_j(K_j+z)^{-1}I_j^*,\\  z>0,\end{gathered}\end{equation*}\fi

where $I_j:\mathcal H_j\hookrightarrow\mathcal H$ is the inclusion. Then
\sbox{\equationmeasure}{$\displaystyle\begin{gathered}\begin{aligned}
\|(I-P_n)g\|^2&\le T_{n,m}q_m[g],\quad g\in D(q_m),\\
\|R_m(z)-R_n(z)\|
&\le \sqrt{T_{n,m}/z}+T_{n,m}/4.
\end{aligned}\end{gathered}$}
\ifdim\wd\equationmeasure>\dimexpr\columnwidth-36pt\relax
\typeout{MANUSCRIPT-WIDE 6.100}\begin{manuscriptwide}\eqanchor{6.100}\begin{equation*}\tag{6.100}\begin{gathered}\begin{aligned}
\|(I-P_n)g\|^2&\le T_{n,m}q_m[g],\quad g\in D(q_m),\\
\|R_m(z)-R_n(z)\|
&\le \sqrt{T_{n,m}/z}+T_{n,m}/4.
\end{aligned}\end{gathered}\end{equation*}\end{manuscriptwide}
\else \eqanchor{6.100}\begin{equation*}\tag{6.100}\begin{gathered}\begin{aligned}
\|(I-P_n)g\|^2&\le T_{n,m}q_m[g],\quad g\in D(q_m),\\
\|R_m(z)-R_n(z)\|
&\le \sqrt{T_{n,m}/z}+T_{n,m}/4.
\end{aligned}\end{gathered}\end{equation*}\fi

Consequently $\sum_j\delta_j^{-1}<\infty$ implies norm convergence $R_n(z)\to R(z)$, uniformly for $z\ge z_0>0$, with error at most $\sqrt{T_n/z}+T_n/4$, where $T_n=\sum_{j>n}\delta_j^{-1}$. No positive infrared gap is required.\par

Convergence of forms, resolvents and semigroups on varying Hilbert spaces has a systematic framework [\hyperlink{ref102}{32}, Sections 2.2-2.6]. The following theorem supplies a spectral-tightness criterion for the specified nested embeddings.\par

\hypertarget{res6.46}{}\noindent \textbf{Theorem \hyperlink{res6.46}{6.23} (spectral tightness and a surviving limiting Hamiltonian).} Let $\mathcal H_n$ be increasing closed subspaces with dense union in $\mathcal H$, let $I_n$ be their inclusions, and let $K_n\ge0$ be self-adjoint on $\mathcal H_n$. Write $q_n$ for the closed form associated with $K_n$. Suppose the extended resolvents $R_n(z)=I_n(K_n+z)^{-1}I_n^*$ converge in operator norm for one $z>0$. This holds under the coherent exact-tower hypotheses and summable inverse shell energies of \hyperlink{eq6.98}{(6.98)}-\hyperlink{eq6.100}{(6.100)}. For a total family $\mathcal D\subset\mathcal H$, choose $u_n\in\mathcal H_n$ with $I_nu_n\to u$ for each $u\in\mathcal D$, and suppose
\sbox{\equationmeasure}{$\displaystyle\begin{gathered}\lim_{L\to\infty}\limsup_{n\to\infty}
 \|\mathbf1_{(L,\infty)}(K_n)u_n\|^2=0.\end{gathered}$}
\ifdim\wd\equationmeasure>\dimexpr\columnwidth-36pt\relax
\typeout{MANUSCRIPT-WIDE 6.101}\begin{manuscriptwide}\eqanchor{6.101}\begin{equation*}\tag{6.101}\begin{gathered}\lim_{L\to\infty}\limsup_{n\to\infty}
 \|\mathbf1_{(L,\infty)}(K_n)u_n\|^2=0.\end{gathered}\end{equation*}\end{manuscriptwide}
\else \eqanchor{6.101}\begin{equation*}\tag{6.101}\begin{gathered}\lim_{L\to\infty}\limsup_{n\to\infty}
 \|\mathbf1_{(L,\infty)}(K_n)u_n\|^2=0.\end{gathered}\end{equation*}\fi

Then there is a unique nonnegative self-adjoint $K$ on the whole $\mathcal H$ such that, for every $z>0$,
\sbox{\equationmeasure}{$\displaystyle\begin{gathered}R_n(z)\longrightarrow (K+z)^{-1}
\\ \hbox{in operator norm}.\end{gathered}$}
\ifdim\wd\equationmeasure>\dimexpr\columnwidth-36pt\relax
\typeout{MANUSCRIPT-WIDE 6.102}\begin{manuscriptwide}\eqanchor{6.102}\begin{equation*}\tag{6.102}\begin{gathered}R_n(z)\longrightarrow (K+z)^{-1}
\quad\hbox{in operator norm}.\end{gathered}\end{equation*}\end{manuscriptwide}
\else \eqanchor{6.102}\begin{equation*}\tag{6.102}\begin{gathered}R_n(z)\longrightarrow (K+z)^{-1}
\\ \hbox{in operator norm}.\end{gathered}\end{equation*}\fi

The extended Euclidean semigroups converge strongly, uniformly on compact $t\ge0$ intervals, to $e^{-tK}$; their convergence is also in operator norm on compact intervals bounded away from zero. The extended unitary groups converge strongly to $e^{-itK/\hbar}$, uniformly on compact real-time intervals.\par

Condition \hyperlink{eq6.101}{(6.101)} is also necessary for these resolvent limits to be the resolvents of a densely defined self-adjoint operator on all of $\mathcal H$. It suffices to verify it on the stated total family, with no form-domain requirement. Uniform finite-energy bounds for convergent representatives imply \hyperlink{eq6.101}{(6.101)}, but finite first moments are not required.\par

If the regulators have one common normalized vacuum $\Omega$ and $K_n\ge m_*(I-|\Omega\rangle\langle\Omega|)$, $m_*>0$, the limit has the same unique vacuum and lower bound. A convergent centered probe $I_nu_n\to u\ne0$ then has
\sbox{\equationmeasure}{$\displaystyle\begin{gathered}\langle u_n,e^{-tK_n}u_n\rangle
 \longrightarrow\langle u,e^{-tK}u\rangle>0
\\ (0\le t<\infty).\end{gathered}$}
\ifdim\wd\equationmeasure>\dimexpr\columnwidth-36pt\relax
\typeout{MANUSCRIPT-WIDE 6.103}\begin{manuscriptwide}\eqanchor{6.103}\begin{equation*}\tag{6.103}\begin{gathered}\langle u_n,e^{-tK_n}u_n\rangle
 \longrightarrow\langle u,e^{-tK}u\rangle>0
\qquad(0\le t<\infty).\end{gathered}\end{equation*}\end{manuscriptwide}
\else \eqanchor{6.103}\begin{equation*}\tag{6.103}\begin{gathered}\langle u_n,e^{-tK_n}u_n\rangle
 \longrightarrow\langle u,e^{-tK}u\rangle>0
\\ (0\le t<\infty).\end{gathered}\end{equation*}\fi

The common gap is an independent input; it may be supplied by Theorem \hyperlink{res6.38}{6.7}'s base-gap and shell-budget estimate. This constructs Hilbert-space quantum dynamics and surviving vectors under the stated hypotheses. Local renormalized fields, relativistic covariance and identification with four-dimensional Yang-Mills remain separate requirements. Proof in the Supplement, Appendix \crosspdflink{Entropy_Geometry_and_Lattice_Gaps_Supplement.pdf}{appC.45}{C.32}.\par

For finite-volume compact regulators this is a cutoff limit: norm limits of their compact resolvents remain compact. A thermodynamic limit and its local relativistic interpretation require separate arguments.\par

Proposition \crosspdflink{Entropy_Geometry_and_Lattice_Gaps_Supplement.pdf}{res6.51}{6.37} (Supplement, Appendix \crosspdflink{Entropy_Geometry_and_Lattice_Gaps_Supplement.pdf}{appC.50}{C.33}) shows that exact relaxation, summable inverse shell energies and a common finite-level gap can coexist with a lost limiting direction. Spectral tightness is therefore a separate requirement.\par

\subsubsection*{Bare plaquettes at weak coupling}

\hypertarget{res6.10}{}\noindent \textbf{Proposition \hyperlink{res6.10}{6.24} (bare plaquette variance collapses at weak bare coupling).} Consider the cubic periodic $SU(3)$ Hamiltonians of Section \hyperlink{sec2}{II}, with side at least four, and let $\gamma=b/a>0$. There are explicit constants $C,\gamma_0$ determined by one fixed smooth coordinate test function on $SU(3)$, independent of volume, for which every elementary $F_p=\operatorname{Re}\operatorname{Tr}(U_p)/3$ satisfies
\sbox{\equationmeasure}{$\displaystyle\begin{gathered}\omega(1-F_p)\le C\gamma^{-1/2},\\  \operatorname{Var}_\omega(F_p)\le\frac32C\gamma^{-1/2},\\  \gamma\ge\gamma_0.\end{gathered}$}
\ifdim\wd\equationmeasure>\dimexpr\columnwidth-36pt\relax
\typeout{MANUSCRIPT-WIDE 6.30}\begin{manuscriptwide}\eqanchor{6.30}\begin{equation*}\tag{6.30}\begin{gathered}\omega(1-F_p)\le C\gamma^{-1/2},\qquad \operatorname{Var}_\omega(F_p)\le\frac32C\gamma^{-1/2},\qquad \gamma\ge\gamma_0.\end{gathered}\end{equation*}\end{manuscriptwide}
\else \eqanchor{6.30}\begin{equation*}\tag{6.30}\begin{gathered}\omega(1-F_p)\le C\gamma^{-1/2},\\  \operatorname{Var}_\omega(F_p)\le\frac32C\gamma^{-1/2},\\  \gamma\ge\gamma_0.\end{gathered}\end{equation*}\fi

The same bounds hold in local state limits at each fixed finite $\gamma$. Along a sequence $\gamma_h\to\infty$, a detector covariance matrix containing $F_p$ as a bare coordinate cannot have a uniform strictly positive lower eigenvalue.\par

Proof and explicit admissible constants $C,\gamma_0$ in the Supplement, Appendix \crosspdflink{Entropy_Geometry_and_Lattice_Gaps_Supplement.pdf}{appC.15}{C.34}, equation \crosspdflink{Entropy_Geometry_and_Lattice_Gaps_Supplement.pdf}{eqC.72}{(C.72)}.\par

In this coefficient convention, \hyperlink{eq6.30}{(6.30)} bounds bare plaquette variance by $O(g_h^2)$. Renormalized observables may survive through rescaling and mixing; spatial averages still require separate correlation estimates. More precisely, if $\operatorname{Var}(F_{p,j})\le C\gamma_j^{-1/2}$ and $O_j=Z_j(F_{p,j}-\omega_j(F_{p,j}))$, then $|Z_j|^2\gamma_j^{-1/2}\to0$ forces $\|O_j\Omega_j\|\to0$. A surviving variance therefore requires at least this normalization scale; it gives no sufficient renormalization or reflected-survival bound.\par

\subsubsection*{Positive-time observables}

Observable survival can be studied before exponential decay. Spatial smearing alone need not produce a finite time-zero vector for a continuum composite field. Use separated positive physical times, retaining one normalization and mixing prescription across the probe family.\par

Corollary \crosspdflink{Entropy_Geometry_and_Lattice_Gaps_Supplement.pdf}{res6.18}{6.35} gives survival for renormalized representatives with orthogonal refinement innovations. Under Theorem \hyperlink{res6.8}{6.21}'s first-order defect hypotheses, Proposition \crosspdflink{Entropy_Geometry_and_Lattice_Gaps_Supplement.pdf}{res6.19}{6.36} bounds their summed energy by terminal norm and energy and the sum of defect constants. For a coherent infinite family with a bounded defect sum, uniform norm and energy bounds replace separate increment estimates. Filtering a bounded terminal probe by $e^{-\tau A_N}$ gives energy at most $\|F_N\|^2/(2e\tau)$, but its norm and coarse overlap may vanish; static compatibility ensures neither coherence nor nonzero overlap. Appendix \crosspdflink{Entropy_Geometry_and_Lattice_Gaps_Supplement.pdf}{appC.1}{C.1} shows when rescaling a fixed loop cannot give a finite-energy seed; Theorem \hyperlink{res6.9}{6.25} tests survival through positive-time pairings. Surviving spectral measures have Proposition \hyperlink{res2.15}{2.7}'s Laplace representation, but their support need not stay away from zero.\par

\hypertarget{res6.9}{}\noindent \textbf{Theorem \hyperlink{res6.9}{6.25} (two positive times prevent ultraviolet spectral escape).} For every regulator index $h$, let $H_h\ge0$ be self-adjoint on a Hilbert space with normalized $\Omega_h$ satisfying $H_h\Omega_h=0$, and let $u_{h,i}\perp\Omega_h$ be vectors indexed by a countable set. Fix $\tau>0$. Suppose that constants $M_i<\infty$, $c>0$, and one index $i_0$ satisfy, uniformly in the chosen cutoff and volume sequence,
\sbox{\equationmeasure}{$\displaystyle\begin{gathered}G_h^{ii}(\tau):=\langle u_{h,i},e^{-\tau H_h}u_{h,i}\rangle\le M_i,\\  G_h^{i_0i_0}(2\tau)\ge c.\end{gathered}$}
\ifdim\wd\equationmeasure>\dimexpr\columnwidth-36pt\relax
\typeout{MANUSCRIPT-WIDE 6.27}\begin{manuscriptwide}\eqanchor{6.27}\begin{equation*}\tag{6.27}\begin{gathered}G_h^{ii}(\tau):=\langle u_{h,i},e^{-\tau H_h}u_{h,i}\rangle\le M_i,\qquad G_h^{i_0i_0}(2\tau)\ge c.\end{gathered}\end{equation*}\end{manuscriptwide}
\else \eqanchor{6.27}\begin{equation*}\tag{6.27}\begin{gathered}G_h^{ii}(\tau):=\langle u_{h,i},e^{-\tau H_h}u_{h,i}\rangle\le M_i,\\  G_h^{i_0i_0}(2\tau)\ge c.\end{gathered}\end{equation*}\fi

Set $v_{h,i}=e^{-\tau H_h}u_{h,i}$ and $\mu_h^{ij}(B)=\langle v_{h,i},\mathbf1_B(H_h)v_{h,j}\rangle$. Then, for $R\ge0$,
\sbox{\equationmeasure}{$\displaystyle\begin{gathered}\mu_h^{ii}([0,\infty))\le M_i,\\  \mu_h^{ii}([R,\infty))\le M_i e^{-\tau R},\\  \int E^n\,d\mu_h^{ii}(E)\le M_i\left(\frac{n}{e\tau}\right)^n\\ (n\ge1).\end{gathered}$}
\ifdim\wd\equationmeasure>\dimexpr\columnwidth-36pt\relax
\typeout{MANUSCRIPT-WIDE 6.28}\begin{manuscriptwide}\eqanchor{6.28}\begin{equation*}\tag{6.28}\begin{gathered}\mu_h^{ii}([0,\infty))\le M_i,\qquad \mu_h^{ii}([R,\infty))\le M_i e^{-\tau R},\qquad \int E^n\,d\mu_h^{ii}(E)\le M_i\left(\frac{n}{e\tau}\right)^n\quad(n\ge1).\end{gathered}\end{equation*}\end{manuscriptwide}
\else \eqanchor{6.28}\begin{equation*}\tag{6.28}\begin{gathered}\mu_h^{ii}([0,\infty))\le M_i,\\  \mu_h^{ii}([R,\infty))\le M_i e^{-\tau R},\\  \int E^n\,d\mu_h^{ii}(E)\le M_i\left(\frac{n}{e\tau}\right)^n\\ (n\ge1).\end{gathered}\end{equation*}\fi

Every regulator sequence has a subsequence on which all entries converge weakly to measures $\mu^{ij}$ whose every finite submatrix is a positive matrix-valued measure, with $\mu^{i_0i_0}([0,\infty))\ge c$. Consequently all the pairings
\sbox{\equationmeasure}{$\displaystyle\begin{gathered}\langle v_{h,i},e^{-tH_h}v_{h,j}\rangle\longrightarrow\int_{[0,\infty)}e^{-tE}\,d\mu^{ij}(E),\\  t\ge0,\end{gathered}$}
\ifdim\wd\equationmeasure>\dimexpr\columnwidth-36pt\relax
\typeout{MANUSCRIPT-WIDE 6.29}\begin{manuscriptwide}\eqanchor{6.29}\begin{equation*}\tag{6.29}\begin{gathered}\langle v_{h,i},e^{-tH_h}v_{h,j}\rangle\longrightarrow\int_{[0,\infty)}e^{-tE}\,d\mu^{ij}(E),\qquad t\ge0,\end{gathered}\end{equation*}\end{manuscriptwide}
\else \eqanchor{6.29}\begin{equation*}\tag{6.29}\begin{gathered}\langle v_{h,i},e^{-tH_h}v_{h,j}\rangle\longrightarrow\int_{[0,\infty)}e^{-tE}\,d\mu^{ij}(E),\\  t\ge0,\end{gathered}\end{equation*}\fi

converge, uniformly on compact time intervals. The limiting diagonal pairing for $i_0$ is nonzero at $t=0$ and continuous there.\par

Proof in the Supplement, Appendix \crosspdflink{Entropy_Geometry_and_Lattice_Gaps_Supplement.pdf}{appC.14}{C.35}.\par

The earlier-time bound in \hyperlink{eq6.27}{(6.27)} is essential. On $\mathbb C\Omega\oplus\mathbb Ce_h$, take $H_h\Omega=0$, $H_he_h=N_he_h$, $N_h\to\infty$, and $u_h=e^{\tau N_h}e_h$. Then $G_h(2\tau)=1$, but the translated correlation $e^{-tN_h}$ vanishes for every $t>0$; the estimate at $\tau$ fails. Conversely, \hyperlink{eq6.27}{(6.27)} permits limiting spectral weight at zero: it controls ultraviolet escape, not the infrared edge. A local quantum field, a continuum-dense probe family, and Yang-Mills identification still require reconstruction and renormalization [\hyperlink{ref14}{25}, \hyperlink{ref13}{26}].\par

\subsubsection*{Smoothing and reflection positivity}

Four-dimensional heat flow can mix the two sides of the reflection plane. Theorem \crosspdflink{Entropy_Geometry_and_Lattice_Gaps_Supplement.pdf}{res6.20}{6.38} exhibits a free gauge-invariant Maxwell violation of reflection positivity for observables classified by their time labels; the unflowed field remains reflection positive. Spatial smoothing preserves half-space support where defined, but needs separate ultraviolet estimates. Positive-flow-time finiteness [\hyperlink{ref23}{33}, Section 7] is perturbative and assumes renormalization of the underlying theory; it constructs no nonperturbative law.\par

\hypertarget{sec6.6}{}\subsection{Spectral transfer on the physical sector}

Under the reconstruction data of Section \hyperlink{sec6.1}{VI A}, a common contraction on the complete centered physical space converts the limiting dynamics into a spectral gap.\par

\subsubsection*{Contraction at one physical time}

Retain the reconstructed data and \hyperlink{eq6.2}{(6.2)}, without assuming vacuum uniqueness. Write $P_\Omega=|\Omega\rangle\langle\Omega|$, $\mathcal K=\Omega^\perp$, and $v_F=q(F)-S(F)\Omega$. The centered reflected forms are
\sbox{\equationmeasure}{$\displaystyle\begin{gathered}B_t(F,G)=S((\theta F)^*\tau_tG)-\overline{S(F)}S(G),\\  B_t(F,G)=\langle v_F,e^{-tH}v_G\rangle,\\  B_0(F,F)=\|v_F\|^2.\end{gathered}$}
\ifdim\wd\equationmeasure>\dimexpr\columnwidth-36pt\relax
\typeout{MANUSCRIPT-WIDE 6.31}\begin{manuscriptwide}\eqanchor{6.31}\begin{equation*}\tag{6.31}\begin{gathered}B_t(F,G)=S((\theta F)^*\tau_tG)-\overline{S(F)}S(G),\qquad B_t(F,G)=\langle v_F,e^{-tH}v_G\rangle,\qquad B_0(F,F)=\|v_F\|^2.\end{gathered}\end{equation*}\end{manuscriptwide}
\else \eqanchor{6.31}\begin{equation*}\tag{6.31}\begin{gathered}B_t(F,G)=S((\theta F)^*\tau_tG)-\overline{S(F)}S(G),\\  B_t(F,G)=\langle v_F,e^{-tH}v_G\rangle,\\  B_0(F,F)=\|v_F\|^2.\end{gathered}\end{equation*}\fi

These positive forms satisfy $B_t(F,F)=\|e^{-tH/2}v_F\|^2$. Testing \hyperlink{eq6.2}{(6.2)} against the dense range of $q$ gives $q(\tau_tF)=e^{-tH}q(F)$; hence $S(\tau_tF)=S(F)$. Thus reflected separation $t$ gives squared-norm evolution through $t/2$. Times use the physical normalization of $H$; with $\hbar=1$, time has inverse-energy units.\par

\hypertarget{res6.11}{}\noindent \textbf{Theorem \hyperlink{res6.11}{6.26} (exact single-time gap criterion).} Fix $t_*>0$ and $m>0$. Under \hyperlink{eq6.2}{(6.2)}, the following are equivalent: (i) $H\ge m(I-P_\Omega)$ as forms on $D(H^{1/2})$; (ii)
\sbox{\equationmeasure}{$\displaystyle\begin{gathered}B_{t_*}(F,F)\le\rho B_0(F,F),\\  \rho=e^{-mt_*}<1;\end{gathered}$}
\ifdim\wd\equationmeasure>\dimexpr\columnwidth-36pt\relax
\typeout{MANUSCRIPT-WIDE 6.32}\begin{manuscriptwide}\eqanchor{6.32}\begin{equation*}\tag{6.32}\begin{gathered}B_{t_*}(F,F)\le\rho B_0(F,F),\qquad \rho=e^{-mt_*}<1;\end{gathered}\end{equation*}\end{manuscriptwide}
\else \eqanchor{6.32}\begin{equation*}\tag{6.32}\begin{gathered}B_{t_*}(F,F)\le\rho B_0(F,F),\\  \rho=e^{-mt_*}<1;\end{gathered}\end{equation*}\fi

for every $F\in\mathcal A_+$; (iii) $\|e^{-t_*H}|_{\mathcal K}\|\le\rho$. In (ii) it suffices to use a complex linear observable subspace whose centered quotient images are dense in $\mathcal K$. These conditions imply $\ker H=\mathbb C\Omega$. If $\mathcal K\ne\{0\}$, then $m\le\Delta:=\inf\operatorname{spec}(H|_{\mathcal K})<\infty$, and the optimal contraction factor is
\sbox{\equationmeasure}{$\displaystyle\begin{gathered}r(t_*)=\sup_{B_0(F,F)>0}\frac{B_{t_*}(F,F)}{B_0(F,F)}=e^{-t_*\Delta},\\  \Delta=-t_*^{-1}\log r(t_*).\end{gathered}$}
\ifdim\wd\equationmeasure>\dimexpr\columnwidth-36pt\relax
\typeout{MANUSCRIPT-WIDE 6.33}\begin{manuscriptwide}\eqanchor{6.33}\begin{equation*}\tag{6.33}\begin{gathered}r(t_*)=\sup_{B_0(F,F)>0}\frac{B_{t_*}(F,F)}{B_0(F,F)}=e^{-t_*\Delta},\qquad \Delta=-t_*^{-1}\log r(t_*).\end{gathered}\end{equation*}\end{manuscriptwide}
\else \eqanchor{6.33}\begin{equation*}\tag{6.33}\begin{gathered}r(t_*)=\sup_{B_0(F,F)>0}\frac{B_{t_*}(F,F)}{B_0(F,F)}=e^{-t_*\Delta},\\  \Delta=-t_*^{-1}\log r(t_*).\end{gathered}\end{equation*}\fi

Proof in the Supplement, Appendix \crosspdflink{Entropy_Geometry_and_Lattice_Gaps_Supplement.pdf}{appC.19}{C.37}.\par

The bound also gives $B_s(F,F)\le e^{-ms}B_0(F,F)$ and $\|e^{-sH}v_F\|\le e^{-ms}\|v_F\|$ for $s\ge0$. Thus $\rho$ contracts squared norm at $t_*/2$ and norm at $t_*$. This equivalence assumes reconstruction; its Yang-Mills estimate remains unproved.\par

For the regulator pairings, let $\mathcal A_n$ be analogous unital complex test algebras with reflections $\theta_n$ and unital linear allowed translations $\tau_{n,t}$, $t\in\mathcal T_n\subset[0,\infty)$, where $0\in\mathcal T_n$ and $\tau_{n,0}=I$. The normalized linear functionals are $S_n$, and the observable representatives $i_n:\mathcal A\to\mathcal A_n$ are unital linear maps preserving $*$ and intertwining reflection; they include all field renormalizations, without requiring multiplicativity. Positivity, convergence and the common sequence $t_n\in\mathcal T_n$ are specified in the following theorem.\par

\hypertarget{res6.12}{}\noindent \textbf{Theorem \hyperlink{res6.12}{6.27} (conditional mass gap from regulator pairings at one time).} Assume \hyperlink{eq6.2}{(6.2)} and a complex linear $\mathcal V\subset\mathcal A_+$ whose centered quotient images are dense in $\mathcal K$. For each regulator choose unital linear observable representatives $i_n$, with all renormalizations included, a normalized functional $S_n$, a reflection $\theta_n$, and one common sequence of allowed physical times $t_n\to t_*>0$. Put $s_n(F)=S_n(i_nF)$ and
\sbox{\equationmeasure}{$\displaystyle\begin{gathered}B_{n,0}(F,G)\\ =S_n((\theta_n i_nF)^*i_nG)\\ {}-\overline{s_n(F)}s_n(G),\\  B_{n,*}(F,G)\\ =S_n((\theta_n i_nF)^*\tau_{n,t_n}i_nG)\\ {}-\overline{s_n(F)}s_n(G).\end{gathered}$}
\ifdim\wd\equationmeasure>\dimexpr\columnwidth-36pt\relax
\typeout{MANUSCRIPT-WIDE 6.34}\begin{manuscriptwide}\eqanchor{6.34}\begin{equation*}\tag{6.34}\begin{gathered}B_{n,0}(F,G)=S_n((\theta_n i_nF)^*i_nG)-\overline{s_n(F)}s_n(G),\qquad B_{n,*}(F,G)=S_n((\theta_n i_nF)^*\tau_{n,t_n}i_nG)-\overline{s_n(F)}s_n(G).\end{gathered}\end{equation*}\end{manuscriptwide}
\else \eqanchor{6.34}\begin{equation*}\tag{6.34}\begin{gathered}B_{n,0}(F,G)\\ =S_n((\theta_n i_nF)^*i_nG)\\ {}-\overline{s_n(F)}s_n(G),\\  B_{n,*}(F,G)\\ =S_n((\theta_n i_nF)^*\tau_{n,t_n}i_nG)\\ {}-\overline{s_n(F)}s_n(G).\end{gathered}\end{equation*}\fi

Assume these forms are positive semidefinite and converge, for each fixed $F,G\in\mathcal V$, to $B_0(F,G)$ and $B_{t_*}(F,G)$. Convergence of one-point expectations and uncentered reflected pairings is sufficient. Suppose $\rho_n\ge0$, $\rho_n\to\rho\in[0,1)$ and, for every fixed $F\in\mathcal V$, there are $\eta_n(F)\ge0$ tending to zero such that, eventually,
\sbox{\equationmeasure}{$\displaystyle\begin{gathered}0\le B_{n,*}(F,F)\le\rho_n B_{n,0}(F,F)+\eta_n(F).\end{gathered}$}
\ifdim\wd\equationmeasure>\dimexpr\columnwidth-36pt\relax
\typeout{MANUSCRIPT-WIDE 6.35}\begin{manuscriptwide}\eqanchor{6.35}\begin{equation*}\tag{6.35}\begin{gathered}0\le B_{n,*}(F,F)\le\rho_n B_{n,0}(F,F)+\eta_n(F).\end{gathered}\end{equation*}\end{manuscriptwide}
\else \eqanchor{6.35}\begin{equation*}\tag{6.35}\begin{gathered}0\le B_{n,*}(F,F)\le\rho_n B_{n,0}(F,F)+\eta_n(F).\end{gathered}\end{equation*}\fi

Finally, require a surviving reflected vector: for some $F_0\in\mathcal V$ and $v_0>0$,
\sbox{\equationmeasure}{$\displaystyle\begin{gathered}\liminf_n B_{n,0}(F_0,F_0)\ge v_0.\end{gathered}$}
\ifdim\wd\equationmeasure>\dimexpr\columnwidth-36pt\relax
\typeout{MANUSCRIPT-WIDE 6.36}\begin{manuscriptwide}\eqanchor{6.36}\begin{equation*}\tag{6.36}\begin{gathered}\liminf_n B_{n,0}(F_0,F_0)\ge v_0.\end{gathered}\end{equation*}\end{manuscriptwide}
\else \eqanchor{6.36}\begin{equation*}\tag{6.36}\begin{gathered}\liminf_n B_{n,0}(F_0,F_0)\ge v_0.\end{gathered}\end{equation*}\fi

Then $0<\rho<1$, $\dim\mathcal H>1$, $\ker H=\mathbb C\Omega$, and $0<-\log\rho/t_*\le\Delta<\infty$.\par

Proof in the Supplement, Appendix \crosspdflink{Entropy_Geometry_and_Lattice_Gaps_Supplement.pdf}{appC.20}{C.38}.\par

For example, physical regulator gaps $m_n\to m>0$ give $\rho_n=e^{-m_nt_n}$ and hence $\Delta\ge m$, provided the stated physical pairings converge. An error $\eta_n(F)\le\varepsilon_n B_{n,0}(F,F)$ with $\varepsilon_n\to0$ is also allowed. This transfer uses only zero time and one positive time; physical reconstruction remains an independent input. Ordinary Euclidean variance cannot replace \hyperlink{eq6.36}{(6.36)}, by Remark \crosspdflink{Entropy_Geometry_and_Lattice_Gaps_Supplement.pdf}{res6.27}{6.42}.\par

Theorem \crosspdflink{Entropy_Geometry_and_Lattice_Gaps_Supplement.pdf}{res6.13}{6.39} gives the equivalent Gram-matrix criterion, including its direction of finite-dimensional approximation; Corollary \crosspdflink{Entropy_Geometry_and_Lattice_Gaps_Supplement.pdf}{res6.14}{6.40} gives the integrated-response criterion (Supplement, Appendices \crosspdflink{Entropy_Geometry_and_Lattice_Gaps_Supplement.pdf}{appC.21}{C.39} and \crosspdflink{Entropy_Geometry_and_Lattice_Gaps_Supplement.pdf}{appC.23}{C.40}). Both require control of every complex combination in a family dense in the complete centered physical sector, with the same physical normalization of observables and time. A finite set of diagonal correlations or separately normalized time slices cannot replace these requirements.\par

\subsubsection*{Relativistic interpretation}

For a relativistic mass interpretation, additionally supply a strongly continuous unitary representation of the proper orthochronous Poincare group, an invariant vacuum, strongly commuting translation generators $(H,\mathbf P)$ transforming covariantly under Lorentz transformations, and joint spectrum in the closed forward cone. For $m>0$, Proposition \crosspdflink{Entropy_Geometry_and_Lattice_Gaps_Supplement.pdf}{res6.24}{6.41} identifies $H\ge m(I-P_\Omega)$ with joint spectral support on $\Omega^\perp$ satisfying $p^0\ge0$ and $(p^0)^2-|\mathbf p|^2\ge m^2$. This asserts neither an isolated one-particle state nor attainment of the bound.\par

For four-dimensional Yang-Mills, the remaining tasks are to construct the intended local theory with \hyperlink{eq6.2}{(6.2)} and its relativistic and short-distance identification, establish convergence of renormalized reflected pairs with a surviving centered vector, and prove a common strict contraction on the complete generated physical sector at one fixed physical time. Theorems \hyperlink{res6.11}{6.26} and \hyperlink{res6.12}{6.27} and Proposition \crosspdflink{Entropy_Geometry_and_Lattice_Gaps_Supplement.pdf}{res6.24}{6.41} then imply the mass gap. The complete-form comparison of Theorem \crosspdflink{Entropy_Geometry_and_Lattice_Gaps_Supplement.pdf}{res3.11}{3.9} and the complete-domain entropy criterion of Theorem \crosspdflink{Entropy_Geometry_and_Lattice_Gaps_Supplement.pdf}{res3.8}{3.7} provide alternative sufficient inputs, not estimates already established for that target theory.\par

\hypertarget{sec7}{}\section{Remaining Yang-Mills estimates}

The target has three spatial dimensions and relativistic time. Theorem \hyperlink{res6.45}{6.12} constructs complete relaxed hierarchies within the actual finite regulators. A common numerical gap requires a uniform base infrared estimate and an inverse-shell-energy budget, in physical units across cutoff and volume. Different terminal vacua need not first be identified for that numerical bound; their comparison is required to construct continuum dynamics.\par

Proposition \hyperlink{res6.48}{6.3} improves the actual pointwise score estimate to square-root growth in $b/a$, uniformly in volume and every surrounding configuration. Its loop-error certificate improves from $O(|C|b^2/a)$ to $O(|C|b)$, but still grows on the proposed weak-bare-coupling refinement. The path-product fiber estimates in Appendix \crosspdflink{Entropy_Geometry_and_Lattice_Gaps_Supplement.pdf}{appC.44}{C.16} retain a cost from all varied links. Generated relaxed forms need not be local Markov forms, and Proposition \crosspdflink{Entropy_Geometry_and_Lattice_Gaps_Supplement.pdf}{res6.47}{6.31} proves actual conditional nonfactorization across a spatial separator. Uniform conditional Poincare or direct full vertical-energy estimates must control these generated interactions.\par

Positive conditional weighted curvature is a sufficient route, not a necessary condition for a gap. Proposition \crosspdflink{Entropy_Geometry_and_Lattice_Gaps_Supplement.pdf}{res6.50}{6.29} gives actual SU(3) conditional tensors that become negative on open sets at large magnetic/electric ratio. Restricting only retained holonomies near identity does not remove the eliminated configurations causing this effect. A small-field proof must also control the excluded configurations through their energy or capacity; small probability alone does not bound every physical vector. The finite-graph calculation does not show that the full cubic regulator loses its gap.\par

Volume independence and cutoff uniformity are separate issues. Large-$N$ reduction concerns an appropriate neutral observable sector under its symmetry-realization conditions, including unbroken center symmetry [\hyperlink{ref97}{34}]; it supplies neither a fixed-$SU(3)$ complete-sector estimate nor bounds at every conditioned exterior. Exponentially small stable-particle mass shifts at large box size, expressed through forward scattering amplitudes, describe an already massive theory [\hyperlink{ref98}{35}]. They can guide volume comparisons after that spectral input is established, but cannot supply it by assuming a positive mass. The pinching comparison of Section \hyperlink{sec5}{V} must include the actual vacuum subtraction, every physical channel and the effect of its surroundings to meet Theorem \hyperlink{res6.42}{6.11}.\par

Theorem \hyperlink{res6.49}{6.20} weakens the comparison needed for static local vacuum convergence: unconditional local transport errors, together with averaged Fisher estimates, suffice. The explicit bound allows Fisher information to grow under refinement. Proving those actual transport errors could compare changing vacua without global fidelity or exact full-cutoff embeddings. The conclusion concerns commuting configuration observables; it does not identify their physical time evolution or construct renormalized local field operators. Uniform spectral tightness and nonzero physical-time probe limits remain necessary, as in Theorems \hyperlink{res6.9}{6.25} and \hyperlink{res6.46}{6.23}.\par

For an actually coherent infinite hierarchy, summable inverse shell energies give full-resolvent convergence, and Theorem \hyperlink{res6.46}{6.23} constructs a limiting Hamiltonian if physical probes do not escape to infinite energy. Finite-volume cutoff removal still requires a separate thermodynamic and relativistic local-field construction. Neither the coefficient diagnostics, Balaban's ultraviolet stability [\hyperlink{ref53}{36}], the multiscale functional-inequality framework [\hyperlink{ref51}{22}], nor geometric confinement benchmarks provide the remaining complete infrared estimate. Once reconstruction, nontrivial survival and a common complete-sector decay bound are established, Section \hyperlink{sec6.6}{VI F} transfers the gap.\par

\noindent \textbf{References.} The bibliography includes the sources cited in the article and essential Supplemental Material.\par


\begin{thebibliography}{99}

\interlinepenalty=10000

\bibitem[1]{ref2}\hypertarget{ref2}{}Gauvin, S. Controlled Completion of Dual-Affine Entropy Geometry: Input-Output Born Structures, Compact Holonomy, and Relative-Entropy Generating Functions. Preprints 2026, 2026090123. \url{https://doi.org/10.20944/preprints202609.0123.v1}

\bibitem[2]{ref37}\hypertarget{ref37}{}S. Bravyi, D. DiVincenzo, and D. Loss, Polynomial-time algorithm for simulation of weakly interacting quantum spin systems, Communications in Mathematical Physics 284, 481-507 (2008). \url{https://arxiv.org/abs/0707.1894}

\bibitem[3]{ref58}\hypertarget{ref58}{}D. A. Yarotsky, Ground states in relatively bounded quantum perturbations of classical lattice systems, Communications in Mathematical Physics 261, 799-819 (2006), Theorem 1. \url{https://doi.org/10.1007/s00220-005-1456-9} Preprint: \url{https://arxiv.org/abs/math-ph/0412040}

\bibitem[4]{ref59}\hypertarget{ref59}{}K. Osterwalder and E. Seiler, Gauge field theories on a lattice, Annals of Physics 110, 440-471 (1978). \url{https://doi.org/10.1016/0003-4916(78)90039-8}

\bibitem[5]{ref96}\hypertarget{ref96}{}J. Thompson, Derivatives of Feynman-Kac Semigroups, Journal of Theoretical Probability 32, 950-973 (2019). Version-specific locator: arXiv:1611.04373v4 (2020), Theorem 2.2. \url{https://doi.org/10.1007/s10959-018-0824-2}; \url{https://arxiv.org/abs/1611.04373}

\bibitem[6]{ref90}\hypertarget{ref90}{}M. Griesemer and D. Hasler, On the Smooth Feshbach-Schur Map, arXiv:0704.3244 (2007), Section 2. \url{https://arxiv.org/abs/0704.3244}

\bibitem[7]{ref101}\hypertarget{ref101}{}W. N. Anderson, Jr., and G. E. Trapp, Shorted Operators. II, SIAM Journal on Applied Mathematics 28(1), 60-71 (1975). \url{https://doi.org/10.1137/0128007}

\bibitem[8]{ref104}\hypertarget{ref104}{}Yu. M. Arlinskii, Shorting, parallel addition and form sums of nonnegative selfadjoint linear relations, Linear Algebra and its Applications 599, 156-200 (2020), Section 3. \url{https://doi.org/10.1016/j.laa.2020.04.007}; \url{https://arxiv.org/abs/1912.00910}

\bibitem[9]{ref79}\hypertarget{ref79}{}\crosspdflink{Entropy_Geometry_and_Lattice_Gaps_Supplement.pdf}{appA}{Essential Supplemental Material for this article, accompanying PDF, containing complete proofs of the physical electric, entropy and response estimates, screening and scale comparisons, compatible dynamics, renormalized-observable criteria and conditional spectral transfer (Appendices A-C).}

\bibitem[10]{ref4}\hypertarget{ref4}{}J. Kogut and L. Susskind, Hamiltonian formulation of Wilson's lattice gauge theories, Physical Review D 11, 395-408 (1975). \url{https://doi.org/10.1103/PhysRevD.11.395}

\bibitem[11]{ref93}\hypertarget{ref93}{}Y. Chen et al., Glueball Spectrum and Matrix Elements on Anisotropic Lattices, Physical Review D 73, 014516 (2006), Sections II and III. \url{https://doi.org/10.1103/PhysRevD.73.014516}; \url{https://arxiv.org/abs/hep-lat/0510074}

\bibitem[12]{ref99}\hypertarget{ref99}{}H. Lehmann, Über Eigenschaften von Ausbreitungsfunktionen und Renormierungskonstanten quantisierter Felder, Il Nuovo Cimento 11, 342-357 (1954), Section 1(a), Eqs. (5)-(10). \url{https://doi.org/10.1007/BF02783624}

\bibitem[13]{ref65}\hypertarget{ref65}{}S. Deffner and E. Lutz, Generalized Clausius inequality for nonequilibrium quantum processes, Physical Review Letters 105, 170402 (2010). \url{https://doi.org/10.1103/PhysRevLett.105.170402}; arXiv:1005.4495. \url{https://arxiv.org/abs/1005.4495}

\bibitem[14]{ref105}\hypertarget{ref105}{}H. Shen, R. Zhu, and X. Zhu, A Stochastic Analysis Approach to Lattice Yang-Mills at Strong Coupling, Communications in Mathematical Physics 400, 805-851 (2023), Theorem 1.4 and Corollary 1.6. \url{https://doi.org/10.1007/s00220-022-04609-1}; \url{https://arxiv.org/abs/2204.12737}

\bibitem[15]{ref8}\hypertarget{ref8}{}N. Madras and D. Randall, Markov chain decomposition for convergence rate analysis, Annals of Applied Probability 12, 581-606 (2002). \url{https://doi.org/10.1214/aoap/1026915617}

\bibitem[16]{ref9}\hypertarget{ref9}{}N. Grunewald, F. Otto, C. Villani, and M. G. Westdickenberg, A two-scale approach to logarithmic Sobolev inequalities and the hydrodynamic limit, Annales de l'Institut Henri Poincaré, Probabilités et Statistiques 45, 302-351 (2009). \url{https://doi.org/10.1214/07-AIHP200}

\bibitem[17]{ref60}\hypertarget{ref60}{}M. Mosonyi, G. Bunth, and P. Vrana, Geometric relative entropies and barycentric Renyi divergences, Linear Algebra and its Applications 699, 159-276 (2024). \url{https://doi.org/10.1016/j.laa.2024.06.005} Preprint: \url{https://arxiv.org/abs/2207.14282}

\bibitem[18]{ref62}\hypertarget{ref62}{}I. Frerot and T. Roscilde, Quantum variance: a measure of quantum coherence and quantum correlations for many-body systems, Physical Review B 94, 075121 (2016), Appendices B and D. \url{https://doi.org/10.1103/PhysRevB.94.075121} Preprint: \url{https://arxiv.org/abs/1509.06741}

\bibitem[19]{ref50}\hypertarget{ref50}{}P. Caputo and D. Parisi, Block Factorization of the Relative Entropy via Spatial Mixing, Communications in Mathematical Physics 388, 793-818 (2021). \url{https://doi.org/10.1007/s00220-021-04237-1}

\bibitem[20]{ref100}\hypertarget{ref100}{}P. Caputo and J. Salez, Entropy factorization via curvature, Journal of Functional Analysis 291(1), 111498 (2026), Theorem 2. \url{https://doi.org/10.1016/j.jfa.2026.111498}; \url{https://arxiv.org/abs/2407.13457}

\bibitem[21]{ref57}\hypertarget{ref57}{}P. Cattiaux and A. Guillin, Functional Inequalities via Lyapunov conditions, arXiv:1001.1822 (2010). \url{https://arxiv.org/abs/1001.1822}

\bibitem[22]{ref51}\hypertarget{ref51}{}R. Bauerschmidt, T. Bodineau, and B. Dagallier, Stochastic dynamics and the Polchinski equation: an introduction, Probability Surveys 21, 200-290 (2024). \url{https://doi.org/10.1214/24-PS27}

\bibitem[23]{ref103}\hypertarget{ref103}{}C. Bierlich, S. Chakraborty, G. Gustafson, and L. Lönnblad, Setting the string shoving picture in a new frame, Journal of High Energy Physics 2021, no. 3, 270 (2021), Section 2.3, Eq. (2.3). \url{https://doi.org/10.1007/JHEP03(2021)270}; \url{https://arxiv.org/abs/2010.07595}

\bibitem[24]{ref94}\hypertarget{ref94}{}G. S. Bali, Casimir scaling of SU(3) static potentials, Physical Review D 62, 114503 (2000), Section V. \url{https://doi.org/10.1103/PhysRevD.62.114503}; \url{https://arxiv.org/abs/hep-lat/0006022}

\bibitem[25]{ref14}\hypertarget{ref14}{}K. Osterwalder and R. Schrader, Axioms for Euclidean Green's functions II, Communications in Mathematical Physics 42, 281-305 (1975). \url{https://doi.org/10.1007/BF01608978}

\bibitem[26]{ref13}\hypertarget{ref13}{}K. Osterwalder and R. Schrader, Axioms for Euclidean Green's functions, Communications in Mathematical Physics 31, 83-112 (1973). \url{https://doi.org/10.1007/BF01645738}

\bibitem[27]{ref84}\hypertarget{ref84}{}A. Ashtekar and J. Lewandowski, Differential geometry on the space of connections via graphs and projective limits, Journal of Geometry and Physics 17(3), 191-230 (1995), Sections 2 and 6.1. \url{https://doi.org/10.1016/0393-0440(95)00028-G}; arXiv:hep-th/9412073. \url{https://arxiv.org/abs/hep-th/9412073}

\bibitem[28]{ref91}\hypertarget{ref91}{}P. Li and S.-T. Yau, On the parabolic kernel of the Schrodinger operator, Acta Mathematica 156, 153-201 (1986), Theorem 2.3. \url{https://doi.org/10.1007/BF02399203}

\bibitem[29]{ref39}\hypertarget{ref39}{}D. C. Brydges, G. Guadagni, and P. K. Mitter, Finite range decomposition of Gaussian processes, Journal of Statistical Physics 115, 415-449 (2004). \url{https://doi.org/10.1023/B:JOSS.0000019818.81237.66}; \url{https://arxiv.org/abs/math-ph/0303013}

\bibitem[30]{ref82}\hypertarget{ref82}{}F. Legoll and T. Lelievre, Effective dynamics using conditional expectations, Nonlinearity 23(9), 2131-2163 (2010). \url{https://doi.org/10.1088/0951-7715/23/9/006}

\bibitem[31]{ref83}\hypertarget{ref83}{}M. H. Duong, A. Lamacz, M. A. Peletier, A. Schlichting, and U. Sharma, Quantification of coarse-graining error in Langevin and overdamped Langevin dynamics, Nonlinearity 31(10), 4517-4566 (2018). \url{https://doi.org/10.1088/1361-6544/aaced5}

\bibitem[32]{ref102}\hypertarget{ref102}{}K. Kuwae and T. Shioya, Convergence of spectral structures: a functional analytic theory and its applications to spectral geometry, Communications in Analysis and Geometry 11(4), 599-673 (2003), Sections 2.2-2.6. \url{https://doi.org/10.4310/CAG.2003.v11.n4.a1}

\bibitem[33]{ref23}\hypertarget{ref23}{}M. Luscher and P. Weisz, Perturbative analysis of the gradient flow in non-abelian gauge theories, Journal of High Energy Physics 2011, no. 2, 051 (2011). \url{https://arxiv.org/abs/1101.0963}

\bibitem[34]{ref97}\hypertarget{ref97}{}P. Kovtun, M. Unsal, and L. G. Yaffe, Volume independence in large Nc QCD-like gauge theories, Journal of High Energy Physics 2007, no. 6, 019 (2007). \url{https://doi.org/10.1088/1126-6708/2007/06/019}; \url{https://arxiv.org/abs/hep-th/0702021}

\bibitem[35]{ref98}\hypertarget{ref98}{}M. Luscher, Volume dependence of the energy spectrum in massive quantum field theories. I. Stable particle states, Communications in Mathematical Physics 104, 177-206 (1986). \url{https://doi.org/10.1007/BF01211589}

\bibitem[36]{ref53}\hypertarget{ref53}{}T. Balaban, Large field renormalization. II. Localization, exponentiation, and bounds for the R operation, Communications in Mathematical Physics 122, 355-392 (1989). \url{https://doi.org/10.1007/BF01238433}

\bibitem[37]{ref5}\hypertarget{ref5}{}J. C. Baez, Spin Networks in Gauge Theory, Advances in Mathematics 117, 253-272 (1996); preprint Spin Network States in Gauge Theory, arXiv:gr-qc/9411007. \url{https://doi.org/10.1006/aima.1996.0012}

\bibitem[38]{ref7}\hypertarget{ref7}{}B. C. Hall, Harmonic analysis with respect to heat kernel measure, arXiv:quant-ph/0006037 (2000). \url{https://arxiv.org/abs/quant-ph/0006037}

\bibitem[39]{ref38}\hypertarget{ref38}{}B. Nachtergaele and R. Sims, On the dynamics of lattice systems with unbounded on-site terms in the Hamiltonian, arXiv:1410.8174 (2014), especially Theorem 4.1. \url{https://arxiv.org/abs/1410.8174}

\bibitem[40]{ref87}\hypertarget{ref87}{}T. Kato, Perturbation Theory for Linear Operators, corrected printing of the second edition (Springer-Verlag, Berlin, 1980; reprinted 1995), Chapter VII, Section 3.5, Theorem 3.9. \url{https://doi.org/10.1007/978-3-642-66282-9}

\bibitem[41]{ref61}\hypertarget{ref61}{}M. S. Lin and M. Tomamichel, Investigating Properties of a Family of Quantum Renyi Divergences, Quantum Information Processing 14, 1501-1512 (2015), Theorem 5. \url{https://doi.org/10.1007/s11128-015-0935-y} Preprint: \url{https://arxiv.org/abs/1408.6897}

\bibitem[42]{ref17}\hypertarget{ref17}{}D. Petz and C. Sudar, Geometries of quantum states, ESI preprint 204 (1995); Journal of Mathematical Physics 37, 2662-2673 (1996). \url{https://www.esi.ac.at/preprints/esi204.pdf}

\bibitem[43]{ref85}\hypertarget{ref85}{}G. Lindblad, Completely positive maps and entropy inequalities, Communications in Mathematical Physics 40, 147-151 (1975). \url{https://doi.org/10.1007/BF01609396}

\bibitem[44]{ref81}\hypertarget{ref81}{}I. V. Girsanov, On transforming a certain class of stochastic processes by absolutely continuous substitution of measures, Theory of Probability and Its Applications 5(3), 285-301 (1960). \url{https://doi.org/10.1137/1105027}

\bibitem[45]{ref95}\hypertarget{ref95}{}M. Foster and C. Michael, Hadrons with a heavy colour-adjoint particle, Physical Review D 59, 094509 (1999), Section 2. \url{https://doi.org/10.1103/PhysRevD.59.094509}; \url{https://arxiv.org/abs/hep-lat/9811010}

\bibitem[46]{ref6}\hypertarget{ref6}{}B. Simon, Schrödinger semigroups, Bulletin of the American Mathematical Society 7, 447-526 (1982). \url{https://doi.org/10.1090/S0273-0979-1982-15040-6}

\bibitem[47]{ref22}\hypertarget{ref22}{}M. Luscher, Properties and uses of the Wilson flow in lattice QCD, Journal of High Energy Physics 2010, no. 8, 071 (2010). \url{https://arxiv.org/abs/1006.4518}

\end{thebibliography}
\end{document}